\documentclass[10pt,journal,compsoc]{IEEEtran}
\usepackage{amsmath,amsfonts}
\usepackage{algorithm}
\usepackage{amsmath,amssymb,bm}   
\usepackage{algpseudocode}        
\usepackage{array}
\usepackage{textcomp}
\usepackage{stfloats}
\usepackage{url}
\usepackage{verbatim}
\usepackage{graphicx}
\usepackage{cite}
\usepackage{scalerel}
\usepackage{color}
\usepackage[english]{babel}
\usepackage{hyperref}
\usepackage{mathtools}
\usepackage{multirow}
\usepackage{makecell}
\usepackage{caption}
\usepackage{subcaption}
\usepackage{rotating}
\usepackage[capitalize]{cleveref}
\usepackage{mathtools}            

\usepackage{mathtools}
\usepackage{diagbox}\usepackage{tikz}
\usepackage{pgfplots}
\usetikzlibrary{calc,positioning}
\usetikzlibrary{decorations.markings,arrows.meta}

\tikzset{
  vtx/.style={circle,fill=black,inner sep=1.8pt},
  rvt/.style={circle,fill=red!80!black,inner sep=2pt},
  ovt/.style={circle,fill=orange!90!black,inner sep=2.4pt},
  edge/.style={line width=1.0pt},
  boldedge/.style={line width=1.5pt, draw=blue!70},
  innerseg/.style={line width=1.3pt, draw=cyan!70!black},
  auxline/.style={line width=1.6pt, draw=cyan!60!blue},
  lab/.style={font=\large},
  midarrow/.style={
    postaction={decorate},
    decoration={markings, mark=at position 0.5 with {\arrow{Stealth}}}
  }
}

\title{Semi-sparsity Generalization for Variational Mesh Denoising}

\author{Junqing Huang~\IEEEmembership{Student Member,~IEEE,}
	Haihui Wang~\IEEEmembership{Member,~IEEE,}
	Michael Ruzhansky
	\thanks{Manuscript received XX, XXXX, 2025; revised XX, XXXX, XX and accepted XX, XXXX, XX. Date of publication XX, XXXX, XX; date of current version XX, XXXX, XX. This work was supported in part by the FWO Odysseus 1 grant G.0H94.18N: Analysis and Partial Differential Equations and by the Methusalem programme of the Ghent University Special Research Fund (BOF) (Grant number 01M01021); and in part by the National Science and Technology Major Project, China, under Grant J2019-I-0001-0001 and Grant J2019-I-0019-0018. Michael Ruzhansky is also supported by EPSRC grant EP/R003025/2. (Corresponding author: Michael Ruzhansky.)}
	\thanks{Junqing Huang, Michael Ruzhansky are with Department of Mathematics: Analysis, Logic and Discrete Mathematics, Ghent University, Belgium; Michael Ruzhansky is also with School of Mathematical Sciences, Queen Mary University of London, UK. (E-mail: \{Junqing.Huang, Michael.Ruzhansky\} @UGent.be).}
	\thanks{Haihui Wang is with the School of Mathematical Sciences, Beihang University (BUAA), China (e-mail: whhmath@buaa.edu.cn).}
	\thanks{Junqing Huang and Haihui Wang contributed equally to this work.}
	\thanks{Digital Object Identifier no. XX.XXXX/TIP.XXXX.XXXXXXX.}}
\begin{document}

\IEEEtitleabstractindextext{
\begin{abstract}
In this paper, we propose a new variational framework for 3D surface denoising over triangulated meshes, which is inspired by the success of semi-sparse regularization in image processing. Differing from the uniformly sampled image data, mesh surfaces are typically represented by irregular, non-uniform structures, which thus complicate the direct application of the standard formulation and pose challenges in both model design and numerical implementation. To bridge this gap, we first introduce the discrete approximations of higher-order differential operators over triangle meshes and then develop a semi-sparsity regularized minimization model for mesh denoising. This new model is efficiently solved by using a multi-block alternating direction method of multipliers (ADMM) and achieves high-quality simultaneous fitting performance --- preserving sharp features while promoting piecewise-polynomial smoothing surfaces. To verify its effectiveness, we also present a series of experimental results on both synthetic and real scanning data, showcasing the competitive and superior results compared to state-of-the-art methods, both visually and quantitatively.
\end{abstract}

}
\maketitle

\IEEEraisesectionheading{\section{Introduction}}\label{sec:introduction}

\IEEEPARstart{W}{ith} the rapid advancement of depth cameras and 3D scanning technologies, the acquisition of real-world 3D scenes has become increasingly easier and more efficient in recent years. However, the reconstructed surfaces are often far from perfection, even coupled with high-quality scanners. In many scenarios, it is inevitable to introduce various noises that may be caused by measurement errors during the scanning process and/or computational errors in reconstruction and resampling algorithms. The notorious noise not only degrades the precision of geometrical representation but also propagates and amplifies errors in downstream geometry processing (e.g., reconstruction, simulation, and visualization), posing potential limitations for practical applications. In this context, 3D surface denoising, particularly over triangulated meshes, has long constituted a fundamental problem in geometry processing. The goal is to suppress noise while restoring intrinsic geometric structures and preserving sharp features—an endeavor that continues to attract significant attention in both theoretical modeling and computational methodologies.

The core challenge in geometric surface denoising lies in the intrinsic difficulty of distinguishing the interest of geometric features, such as corners, edges, creases, and fine textures, from high-frequency stochastic noise, as both are characterized by similar high-frequency behavior. This ambiguity is further exacerbated under severe noise corruption, non-uniform sampling, or complex topological configurations. To address this challenge, numerous approaches have been proposed in the literature. Existing mesh denoising methods can be broadly classified into the following categories: diffusion-based methods~\cite{desbrun1999implicit, meyer2003discrete, bajaj2003anisotropic, tasdizen2002geometric, wu2008diffusion, clarenz2000anisotropic}, filtering-based methods~\cite{fleishman2003bilateral, yagou2002mesh, jones2003non, shen2004fuzzy, wang2006bilateral, sun2007fast, zheng2010bilateral, wang2012cascaded, zhang2015guided, wang2016mesh, wei2016tensor, chen2019structure, wei2018mesh, li2018non, yadav2017mesh}, variational-based methods~\cite{sorkine2006differential, wu2009scale, zhang2015variational, lu2015robust, wu2015mesh, liu2020mesh, he2013mesh, zhao2018robust, zhong2019robust, wei2014bi, zhang2018static} and their higher-order extensions~\cite{liu2019novel, liu2019triangulated, liu2021mesh, zhang2022novel, baumgartner2023total}, as well as data-driven approaches~\cite{wang2019data, li2020normalf, li2020dnf}. Despite the remarkable progress, there remains considerable interest in developing more powerful, theoretically justified, and computationally efficient denoising methods.

In this paper, we propose a simple yet effective variational framework for triangulated mesh denoising, which takes advantage of semi-sparsity priors to achieve high-quality denoising performance. This new method is primarily inspired by recent advances of semi-sparse regularization for edge-preserving image filtering~\cite{huang2023semi}, where such priors have been demonstrated to be particularly effective for fitting piecewise-polynomial smoothing surfaces while preserving sparse features such as edges and singularities. From a variational perspective, the denoising task can be formulated as a variational-based regularization model that enforces geometric fidelity while promoting smoothness under suitable regularity constraints. In particular, it is essential to preserve sharpening edges and locally smooth regions (e.g., approximately quadratic patches) without introducing artificial discontinuities or spurious features. On this premise, the proposed method integrates semi-sparse regularization with an $L^2$-norm data fidelity term, providing a balance between geometric smoothness and feature preservation. Similar to many existing methods~\cite{sun2007fast, zheng2010bilateral, zhang2015variational, zhang2015guided, zhang2018static, liu2019novel, liu2021mesh}, we employ a two-stage optimization scheme for surface denoising, in which surface normals are first refined via the proposed semi-sparsity model and vertex positions are then updated accordingly. Within this framework, the proposed method establishes a new paradigm for triangulated mesh denoising, offering an effective balance between feature preservation and noise suppression. Our contributions are summarized as follows:

\begin{itemize}

\item We design a simple and effective semi-sparsity-based minimization model for triangulated mesh denoising, providing a variational framework that preserves sharp features while suppressing staircase artifacts in smooth regions.
 
\item We present an efficient numerical scheme based on a multi-block alternating direction method of multipliers (ADMM) for solving the resulting non-convex and non-smooth optimization problem.
 
\item We conduct extensive numerical experiments on synthetic and real scanned datasets, demonstrating the versatility and superior denoising performance of the proposed method compared with state-of-the-art methods.
		
\end{itemize}

The remainder of this paper is organized as follows. A brief survey of existing mesh denoising methods is presented in~\cref{related_work}. The discrete differential operators and their higher-order generalizations on triangulated meshes are introduced in~\cref{spaces_and_operators}. The proposed semi-sparsity-based denoising model is formulated in~\cref{semi_sparsity_model}, where we introduce the regularization strategy for face-normal filtering and the multi-block ADMM algorithm for optimization. In addition, a short discussion on vertex position updating is also provided. In~\cref{experimental_results}, we present the results of our semi-sparsity regularization method and further compare it to the state-of-the-art methods visually and quantitatively. Finally, we conclude with remarks and discuss directions for future work in~\cref{conclusion}.


\section{Related Work}
\label{related_work}


We now review representative methods for triangulated mesh denoising, which, as discussed earlier, can be broadly categorized into diffusion-based methods, filtering-based methods, variational-based methods and their higher-order extensions, and data-driven methods. It is worth noting that many of them are inspired by feature-preserving image filtering/denoising methods, because meshes and images --- from signal processing perspectives --- share similar intrinsic geometric characteristics despite differences in their data structures and representations. 

\textbf{Diffusion-based Methods:} Early studies of triangulated mesh denoising can be traced back to isotropic diffusion methods~\cite{desbrun1999implicit, meyer2003discrete}, in which vertex positions or surface normals are evolved under a discrete heat flow to smooth high-frequency noise while retaining geometric structure. Due to its isotropic nature, the diffusion process treats local geometric features and noise equally, resulting in denoising at the expense of feature blurring or, in extreme cases, complete loss of surface features. The idea was later extended to anisotropic diffusion~\cite{bajaj2003anisotropic,tasdizen2002geometric, wu2008diffusion, clarenz2000anisotropic}, in which directional and data-dependent conductivities, for example, normal jumps, dihedral angles, curvature indicators, or structure-tensor cues,  were employed to better adapt to complex local geometry. While anisotropic schemes generally outperform isotropic smoothing under moderate noise, they may still struggle to preserve sharp creases on heavily corrupted meshes and often require careful parameter tuning. In addition, multi-scale variants~\cite{wu2009scale} advocate multi-resolution strategies to separate noise from features across scales, which tend to reduce staircase artifacts and preserve smoothly varying curvature by suppressing oscillations. 

\textbf{Filter-based Methods:} Similar to filtering techniques in image processing, filtering-based mesh denoising methods can be broadly categorized into local and global filters. Local filters aggregate information from a neighborhood via data-dependent weights to suppress noise while preserving features. The key difference lies in their dynamic weights designed in these methods. For example, median/mean-type schemes~\cite{yagou2002mesh, shen2004fuzzy, sun2007fast} update face normals using robust neighborhood statistics (e.g., vector median or area-weighted averages) followed by a vertex update; they perform well under mild noise but tend to weaken sharp features at higher noise levels.  Bilateral mesh filters~\cite{fleishman2003bilateral, jones2003non, wang2006bilateral, zheng2010bilateral} incorporate both spatial proximity (e.g., triangle centroids) and range similarity (face normals) in their weighting, which improves the performance in recovering strong features but can not avoid over-smoothing limitations at the sharp and corner features. Guided normal filtering~\cite{zhang2015guided, wei2016tensor} and cascaded schemes~\cite{wang2012cascaded, wang2016mesh} further introduce carefully constructed guidance fields to enhance robustness and preserve geometric details. Despite their limitations, local filters remain popular in practice due to simplicity and ease of implementation. 

In contrast to local filters, non-local methods have also been studied in the context of triangulated mesh denoising. For example, nonlocal filtering methods~\cite{chen2019structure, wei2018mesh, li2018non, yadav2017mesh} generalize local ideas to patch-based neighborhoods that extend beyond the one-ring, in which similar mesh patches are first identified and grouped, and then collaborative filtering (e.g., weighted aggregation or related operations) is performed to exploit pattern similarity in the underlying surface. Nonlocal graph methods~\cite{arvanitis2018feature, zhao2019graph} construct data-adaptive graphs and perform Laplacian-like smoothing. Nonlocal methods are generally more effective at recovering repeated structures and subtle structures, while they are sensitive to patch matching (mis-registration across sharp ridges can blur features) and computationally intensive due to patch construction, matching, and group-wise processing in practical applications. 

\textbf{Variational and Higher-order Models:} Variational modeling is also a core framework for triangular mesh denoising, which has also greatly benefited from the success of variational methods in image processing. Variational methods typically formulate the denoising problem as an energy minimization problem with data fidelity and regularization terms. Variational models can be analyzed in the continuous setting via Euler–Lagrange equations within the framework of PDEs or directly solved in the discrete setting with modern optimization algorithms. A canonical example is total variation (TV) regularization, which employs an $L_1$-norm penalty on first-order differences, and provides strong edge-preserving behavior. A systematic study of diffusion methods and their relationship to TV models, together with a comparison of their effects on mesh denoising, is presented in~\cite{wu2009scale}. Building on this line, \cite{zhang2015variational} develops an efficient augmented-Lagrangian (ALM) solver for the variational TV-based regularization. Despite its advantages, TV regularization is prone to staircasing in smoothly varying regions. To mitigate this, higher-order models have been proposed~\cite{liu2019novel, liu2019triangulated}; while they reduce staircasing, they may blur fine structures or round sharp edges under severe noise. Hybrid formulations that combine TV with higher-order terms~\cite{liu2021mesh, baumgartner2023total, zhang2022novel} offer partial improvements but can still produce artifacts near salient features.

In parallel, optimization-based methods impose explicit regularization to achieve robust, high-fidelity denoising. For example, the bi-normal filtering strategy of~\cite{wei2014bi} refines the facet-normal field by exploiting piecewise-consistency within an optimization framework. The Static/Dynamic (SD) filter of~\cite{zhang2018static} formulates a nonlinear optimization that promotes signal smoothness while preserving variations associated with features at selected scales. In addition, nonconvex regularizers have also been explored for mesh denoising --- for instance, sparsity-inducing $\ell_0$-based models with piecewise-constant priors~\cite{he2013mesh, zhao2018robust} --- which can strongly preserve sharp features but may introduce visual artifacts (e.g., feature distortion or over-sharpening) on texture-rich models or under mixed noise. More recently, a nonsmooth, nonconvex Mumford–Shah formulation~\cite{wang2022feature} has demonstrated the effectiveness of a shape-optimization routine for mesh denoising. We note, however, that nonconvex programs are generally more challenging to solve (e.g., susceptibility to local minima, heavier computation), which can limit their practicality and widespread adoption.

\textbf{Data-driven Methods:} Recent advances in machine learning and deep neural networks have also inspired data-driven approaches for triangulated mesh denoising. The cascaded normal regression (CNR) model~\cite{wang2016mesh}, for instance, learns a nonlinear mapping from noisy face normals to their ground-truth counterparts. Subsequent frameworks~\cite{wang2019data} typically adopt a two-stage architecture, where noisy normals are first smoothed via a learned regression network, followed by a secondary refinement stage to recover fine-scale geometric details. More sophisticated designs, such as NormalF-Net~\cite{li2020normalf}, employ dedicated denoising and refinement subnetworks, while DNF-Net~\cite{li2020dnf} introduces a convolutional framework that directly predicts clean normals from noisy inputs. Although these learning-based methods achieve state-of-the-art denoising performance, they generally depend on large and diverse training datasets. Moreover, their training process is computationally expensive, and their generalization ability can be sensitive to variations in geometry, noise characteristics, and scanning modalities. Numerous other deep-learning models with varying architectures have also been proposed; we refer the reader to the recent survey~\cite{chen2023geometric} for a comprehensive comparisons.


Despite distinct forms and generations, these methods are closely connected and can be understood within a general PDE/variational framework. For example, classical Laplace smoothing corresponds to an explicit Euler step for heat flow on a grid, while bilateral filtering is analogous to anisotropic diffusion with data-adaptive diffusivity. Many variational energies give rise to diffusion-type Euler–Lagrange equations—for example, total variation (TV) induces edge-preserving anisotropic diffusion reminiscent of the Perona–Malik model. Therefore, an explicit/implicit diffusion scheme can be formulated in the optimization framework for the underlying energy. Furthermore, many filters are equivalent to minimizing this energy in a small number of iterations. Nonlocal diffusion and filtering methods construct patch-similarity graphs whose Laplacians drive nonlocal smoothing—extending local diffusion beyond the one-ring. Low-rank patch models solve global variational problems with nuclear-norm (or factorized) regularizers on grouped patches, enabling collaborative denoising. In this sense, nonlocal filters can be seen either as diffusion models on a data-adaptive graph or as variational problems with nonlocal priors.

\section{Discrete Operators and Spaces}
\label{spaces_and_operators}



We now introduce the discrete differential operators over triangulated meshes, together with the higher-order analogues, which are central to geometric mesh processing. The definitions presented here follow recent studies~\cite{baumgartner2023total, liu2019triangulated, liu2021mesh, zhang2022novel}, to which the reader is referred for further details.

\subsection{Basic Notations} 

Let $\mathcal{M}=(V, E, T)$ denote a mesh surface with arbitrary topology with no degenerate triangles in $\mathbb{R}^{3}$, where $V=\left\{v_1, \ldots, v_I\right\}, E=\left\{e_1, \ldots, e_J\right\}$, and $T=\left\{\tau_1, \ldots, \tau_K\right\}$ are the set of vertices, edges, and triangles of $\mathcal{M}$, respectively. 

To derive the discrete differential operators, we here follow~\cite{liu2019triangulated} to show the relationships of $V, E, T$ in~\cref{fig_mesh_local_structures}. The 1-ring of the triangle $\tau_{i}$ in~\cref{fig_mesh_local_structures} (a) is denoted as $D_{1}\left(\tau_{i}\right)$, which is the set of the triangles sharing some common edges with $\tau_{i}$. In~\cref{fig_mesh_local_structures} (b), the set of lines connecting the barycenter and vertices of $\tau_{i}$ is denoted as $B_{1}\left(\tau_{i}\right)=\left\{l_{\tau_{i},j}: j=0,1,2\right\}$, where $j$ counterclockwise marks the vertex contained in $\tau_{i}$. Namely, $l_{\tau_{i}, j}$ is the line connecting the barycenter of $\tau_{i}$ and the vertex marked as $j$ in $\tau_{i}$. The set of lines connecting vertices of $\tau_{i}$ and barycenter of triangles in $D_{1}\left(\tau_{i}\right)$ is denoted as $B_{2}\left(\tau_{i}\right)$. In~\cref{fig_mesh_local_structures} (c), the 1-disk of the vertex $v_i$ is denoted as $M_{1}\left(v_i\right)$, which are the indices of triangles containing $v_i$. Furthermore, the 1-neighborhood of vertex $v_i$ in~\cref{fig_mesh_local_structures} (d) is denoted as $N_{1}\left(v_i\right)$, which is the set of vertices connecting to $v_i$.

\begin{figure*}[!t]
\centering
\begin{subfigure}[t]{.22\textwidth}
\centering
\resizebox{\linewidth}{!}
{     
    \begin{tikzpicture}[x=1cm,y=1cm]
      \coordinate (A) at (0,0);
      \coordinate (B) at (-4.8,0);
      \coordinate (C) at (-2.2,-4.2);
      \coordinate (E) at (-2.4,-0.2);
      \coordinate (F) at (-3.4,-2.2);
      \coordinate (G) at (-1.1,-2.1);
      \fill[green!60] (A) -- (E) -- (G) -- cycle;
      \fill[green!60] (B) -- (F) -- (E) -- cycle;
      \fill[green!60] (C) -- (G) -- (F) -- cycle;
      \draw[edge] (A) -- (E) -- (G) -- cycle;
      \draw[edge] (B) -- (F) -- (E) -- cycle;
      \draw[edge] (C) -- (G) -- (F) -- cycle;
      \foreach \p in {A,B,C,E,F,G}{\node[vtx] at (\p) {};}
      \node[lab, scale=1.25] at (barycentric cs:E=1,F=1,G=1) {$\tau_i$};
    \end{tikzpicture}
}
\caption{}
\end{subfigure}
\hfill
\begin{subfigure}[t]{.22\textwidth}
\centering
\resizebox{\linewidth}{!}
{ 
\begin{tikzpicture}[x=1cm,y=1cm]
      \coordinate (A) at (0,0);
      \coordinate (B) at (-4.8,0);
      \coordinate (C) at (-2.2,-4.2);
      \coordinate (E) at (-2.4,-0.2);
      \coordinate (F) at (-3.4,-2.2);
      \coordinate (G) at (-1.1,-2.1);
      \coordinate (C0) at ($ 0.3333*(E) + 0.3333*(F) + 0.3333*(G) $);
      \coordinate (C1) at ($ 0.3333*(A) + 0.3333*(E) + 0.3333*(G) $); 
      \coordinate (C2) at ($ 0.3333*(B) + 0.3333*(F) + 0.3333*(E) $); 
      \coordinate (C3) at ($ 0.3333*(C) + 0.3333*(G) + 0.3333*(F) $); 
      \draw[edge] (A) -- (E) -- (G) -- cycle;
      \draw[edge] (B) -- (F) -- (E) -- cycle;
      \draw[edge] (C) -- (G) -- (F) -- cycle;
      \draw[boldedge] (E) -- (C2) -- (F);
      \draw[boldedge] (E) -- (C1) -- (G);
      \draw[boldedge] (F) -- (C3) -- (G);
      \draw[innerseg,dashed] (C0)--(E);
      \draw[innerseg,dashed] (C0)--(F);
       \draw[innerseg,dashed] (C0)--(G);
      \foreach \p in {C0,C1,C2,C3}{\node[rvt] at (\p) {};}
      \foreach \p in {A,B,C,E,F,G}{\node[vtx] at (\p) {};}
      \node[lab, scale=1.0] at ($(C0)!0.5!(E)$) {$l_{\tau_i, 1}$};
      \node[lab, scale=1.0] at ($(C0)!0.5!(F)$) {$l_{\tau_i, 2}$};
      \node[lab, scale=1.0] at ($(C0)!0.5!(G)$) {$l_{\tau_i, 3}$};
      \node[lab, scale=1.25] at (barycentric cs:E=1,F=1,G=0.4) {$\tau_i$};
\end{tikzpicture}
}
\caption{}
\end{subfigure}
\hfill
\begin{subfigure}[t]{.22\textwidth}
\centering
\resizebox{\linewidth}{!}
{   
\begin{tikzpicture}[x=1cm,y=1cm]
  \coordinate (O) at (0,0);
  \foreach \i/\ang/\rad in {1/10/3.0,2/66/2.6,3/122/3.0,4/178/2.7,5/238/2.9,6/298/2.8,7/338/2.9}
  {
    \coordinate (V\i) at ({\rad*cos(\ang)},{\rad*sin(\ang)});
  }
  \fill[gray!70] (V1)--(V2)--(V3)--(V4)--(V5)--(V6)--(V7)--cycle;
  \draw[edge] (V1)--(V2)--(V3)--(V4)--(V5)--(V6)--(V7)--cycle;
  \foreach \i in {1,...,7}{ \draw[edge] (O)--(V\i); }
  \node[vtx] at (O) {};
  \foreach \i in {1,...,7}{ \node[vtx] at (V\i) {}; }
  \node[lab, scale=1.25] at (0.5,0.4) {$v_i$};
\end{tikzpicture}
}
\caption{}
\end{subfigure}
\hfill
\begin{subfigure}[t]{.22\textwidth}
\centering
\resizebox{\linewidth}{!}
{   
\begin{tikzpicture}[x=1cm,y=1cm]
  \coordinate (O) at (0,0);
  \foreach \i/\ang/\rad in {1/10/3.0,2/66/2.6,3/122/3.0,4/178/2.7,5/238/2.9,6/298/2.8,7/338/2.9}
  {
    \coordinate (W\i) at ({\rad*cos(\ang)},{\rad*sin(\ang)});
  }
  \draw[edge] (W1)--(W2)--(W3)--(W4)--(W5)--(W6)--(W7)--cycle;
  \foreach \i in {1,...,7}{ \draw[edge] (O)--(W\i); }
  \node[vtx] at (O) {};
  \foreach \i in {1,...,7}{ \node[ovt] at (W\i) {}; }
  \node[lab, scale=1.25] at (0.5,0.4) {$v_i$};
\end{tikzpicture}
}
\caption{}
\end{subfigure}
\caption{The illustration of $D_{1}\left(\tau_{i}\right), B_{1}\left(\tau_{i}\right), B_{2}\left(\tau_{i}\right), M_{1}\left(v_i\right)$, and $N_{1}\left(v_i\right)$(see also in~\cite{liu2019triangulated}). The elements contained in $D_{1}\left(\tau_{i}\right)$,$B_{1}\left(\tau_{i}\right)$, $B_{2}\left(\tau_{i}\right)$, $M_{1}\left(v_i\right)$, and $N_{1}\left(v_i\right)$ are plotted in green, cyan, blue, gray, and orange, respectively. (a) $D_{1}\left(\tau_{i}\right)$ is the 1-ring of the triangle $\tau_{i}$, which refers to three triangles; (b) $B_{1}\left(\tau_{i}\right)$ is the set of lines connecting the barycenter and vertices of $\tau_{i}$, which refers to three lines, and $B_{2}\left(\tau_{i}\right)$ is the set of lines connecting vertices of $\tau_{i}$ and barycenter of triangles contained in $D_{1}\left(\tau_{i}\right)$, which refers to six lines; (c) $M_{1}\left(v_i\right)$ is the 1-disk of the vertex $v_i$, which refers to seven triangles; (d) $N_{1}\left(v_i\right)$ is the 1-neighborhood of $v_i$, which refers to seven vertices.}
\label{fig_mesh_local_structures}
\end{figure*}

We further introduce the relative orientation of an edge $e$ to a triangle $\tau$, which is denoted by $\operatorname{sgn}(e, \tau)$, as follows. First, we assume that all triangles are with counterclockwise orientation and all edges are with randomly chosen fixed orientations. For an edge $e \prec \tau$, if the orientation of $e$ is consistent with the orientation of $\tau$, then $\operatorname{sgn}(e, \tau)=1$; otherwise, $\operatorname{sgn}(e, \tau)=-1$. Similarly, if $p$ is an endpoint of an edge $e$, then we write it as $v \prec e$. Similarly, $e \prec \tau$ denotes that $e$ is an edge of a triangle $\tau$; $v \prec \tau$ denotes that $v$ is a vertex of a triangle $\tau$.

\subsection{Discrete Operators and Generalizations}
\label{discrete_operators}


We follow~\cite{zhang2015variational, liu2019triangulated, liu2021mesh} and define the discrete differential operators over a piecewise constant triangulated surface $\mathcal{M}$. We denote the space $\mathcal{U}=\mathbb{R}^{|T|}$, which is isomorphic to the piecewise constant function space over $\mathcal{M}$. For example, $\mathrm{u} \in \mathcal{U}$ is restricted on the triangle $\tau$, which is written as $\mathrm{u}_\tau$ (or $\left.u\right|_\tau$). Analogically, we also introduce the function space $\mathcal{V}=\mathbb{R}^{|E|}$, in which $\mathrm{v} \in \mathcal{V}$ is restricted on the edge $e$ and denoted as $\mathrm{v}_e$ (or, $\mathrm{v}|_e$ equivalently).

With the above notations, the spaces $\mathcal{U}$ and $\mathcal{V}$ are assumed to be equipped with the area/length-weighted inner products and norms:
\begin{equation}
\begin{aligned}
\left(\mathrm{u}^1, \mathrm{u}^2\right)_\mathcal{U}=\sum_\tau \mathrm{u}_\tau^1 \mathrm{u}_\tau^2 s_\tau, \quad\|\mathrm{u}\|_\mathcal{U}=\sqrt{(\mathrm{u}, \mathrm{u})_\mathcal{U}},
\end{aligned}
\label{eq_inner_product_and_norm_u}
\end{equation}
\begin{equation}
\begin{aligned}
\left(\mathrm{v}^1, \mathrm{v}^2\right)_\mathcal{V}=\sum_e \mathrm{v}_e^1 \mathrm{v}_e^2 \operatorname{len}(e), \quad\|\mathrm{v}\|_\mathcal{V}=\sqrt{(\mathrm{v}, \mathrm{v})_\mathcal{V}},
\end{aligned}
\label{eq_inner_product_and_norm_v}
\end{equation}
where $s_\tau$ is the area of triangle $\tau$ and $\operatorname{len}(e)$ is the length of edge $e$.

As suggested in~\cite{zhang2015variational}, for any $\mathrm{u}\in \mathcal{U}$, it is possible to define the jump of $\mathrm{u}$ over an edge $e$ as,
\begin{equation}
\begin{aligned}
[\mathrm{u}]_e=\left\{\begin{array}{rr}
\displaystyle{\sum_{\tau, e \prec \tau}} \mathrm{u}|_\tau \operatorname{sgn}(e, \tau), & e \not \subset \partial \mathcal{M}, \\
0, & e \subset \partial \mathcal{M}.
\end{array}\right.
\end{aligned}
\label{eq_first_order_difference_over_e}
\end{equation}
Then, the first-order difference operator over the edge $e$, i.e., $D_\mathcal{M}:\mathcal{U} \rightarrow \mathcal{V}, \mathrm{u} \rightarrow D_\mathcal{M} \mathrm{u}$, is defined as, 
\begin{equation}
\begin{aligned}
D_\mathcal{M} \mathrm{u}|_e=[\mathrm{u}]_e \quad \forall e \; \text{for}\;\mathrm{u} \in \mathcal{U}.
\end{aligned}
\label{eq_first_order_differential_operator_over_e}
\end{equation}

As illustrated in~\cref{fig_mesh_local_structures} (a), there are three edges in each triangle $\tau_i$, i.e., $\{e_{\tau_i, 1}, e_{\tau_i, 2}, e_{\tau_i,3}\}$, which share with the triangles in $D_1(\tau_i)$. Thus, it is natural to define the gradient operator over the triangle $\tau$,
\begin{equation}
\begin{aligned}
\left.\nabla_{\mathcal{M}} \mathrm{u}\right|_\tau=\left(\left.D_{\mathcal{M}} \mathrm{u}\right|_{e_{\tau,1}},\left.D_{\mathcal{M}} \mathrm{u}\right|_{e_{\tau,2}},\left.D_{\mathcal{M}} \mathrm{u}\right|_{e_{\tau, 3}}\right),
\end{aligned}
\label{eq_gradient_operator_over_e}
\end{equation}
which is sometimes written as $\nabla \mathrm{u}=\left(\partial_1 \mathrm{u}, \partial_2 \mathrm{u}, \partial_3 \mathrm{u}\right)$ for simplicity. 

Let $\nabla:\mathcal{U}\to\mathcal{V}$ be the first–order difference operator with the weighted inner products $(\cdot,\cdot)_{\mathcal{U}}$ and $(\cdot,\cdot)_{\mathcal{V}}$. The adjoint $\nabla^\ast:\mathcal{V}\to\mathcal{U}$ is then characterized by
$$
(\mathrm{v},\,\nabla \mathrm{u})_{\mathcal{V}} \;=\; -(\nabla^\ast\mathrm{v},\,\mathrm{u})_{\mathcal{U}}.
$$
Based on the divergence theorem, the adjoint operator $\nabla^\ast\mathrm{v}|_\tau$ is given by:
\begin{equation}
\begin{aligned}
\nabla^\ast\mathrm{v}|_\tau = \frac{1}{s_\tau} \sum_{\substack{e\prec \tau\\ e \not\subset \partial\mathcal{M}}} \mathrm{v}_e \,\operatorname{sgn}(e,\tau)\, \operatorname{len}(e), \qquad \forall \tau\in T.
\end{aligned}
\label{eq_divergence operator}
\end{equation}

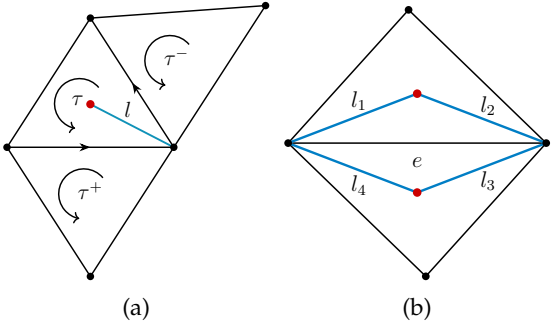
\begin{figure}[!t]
\centering
\begin{subfigure}[t]{0.2\textwidth}
\centering
\resizebox{\linewidth}{!}
{ 
 \begin{tikzpicture}[x=1cm,y=1cm]
  \coordinate (L) at (0,0);
  \coordinate (R) at (4,0);
  \coordinate (T) at (2,3.1);
  \coordinate (B) at (2,-3.1);
  \coordinate (P) at (6.2,3.4);
  \coordinate (C1) at ($ 0.3333*(L) + 0.3333*(B) + 0.3333*(R) $); 
  \coordinate (C2) at ($ 0.3333*(L) + 0.3333*(T) + 0.3333*(R) $); 
  \coordinate (C3) at ($ 0.3333*(T) + 0.3333*(P) + 0.3333*(R) $); 

  \draw[edge] (L)--(T)--(P)--(R)--(B)--cycle;
  \draw[edge, midarrow] (L)--(R);
  \draw[edge, midarrow] (R) -- (T);
  \draw[innerseg] (R)--(C2);
  \draw[->,edge] ($(C1)+(0.3,0.3)$) arc[start angle=-320,end angle=-100,radius=0.6];
  \draw[->,edge] ($(C2)+(0.2,0.4)$) arc[start angle=-320,end angle=-100,radius=0.6];
  \draw[->,edge] ($(C3)+(0.3,0.3)$) arc[start angle=-320,end angle=-100,radius=0.6];

  \node[lab,scale=1.25] at (C1) {$\tau^+$};
  \node[lab,scale=1.25] at ($(C2)+(-0.3,0.1)$)  {$\tau$};
  \node[lab,scale=1.25] at (C3) {$\tau^-$};
  \node[lab,scale=1.25] at ($(R)!0.55!(C2)+(0.0,0.35)$) {$l$};
  \foreach \p in {C2}{\node[rvt] at (\p) {};}
  \foreach \p in {L,R,P,T,B}{\node[vtx] at (\p) {};}
\end{tikzpicture}
}
\caption{}
\end{subfigure}
\begin{subfigure}[t]{0.2\textwidth}
\centering
\resizebox{\linewidth}{!}
{ 
 \begin{tikzpicture}[x=1cm,y=1cm]
  \coordinate (L) at (0,0);
  \coordinate (R) at (6,0);
  \coordinate (T) at (2.8,3.1);
  \coordinate (B) at (3.2,-3.1);
  \draw[edge] (L)--(T)--(R)--(B)--cycle;
  \draw[edge] (L)--(R); 
  \node[lab, scale=1.25] at ($(L)!0.5!(R)+(0,-0.45)$) {$e$};

  \coordinate (U) at (3,1.15);
  \draw[auxline] (L)--(U)--(R);
  \node[rvt] at (U) {};
  \node[lab,scale=1.25] at ($(L)!0.55!(U)+(0.0,0.35)$) {$l_1$};
  \node[lab,scale=1.25] at ($(U)!0.55!(R)+(0.0,0.35)$) {$l_2$};

  \coordinate (D) at (3,-1.15);
  \draw[auxline] (L)--(D)--(R);
  \node[rvt] at (D) {};
  \node[lab,scale=1.25] at ($(L)!0.55!(D)+(0.0,-0.35)$) {$l_4$};
  \node[lab,scale=1.25] at ($(D)!0.55!(R)+(0.0,-0.35)$) {$l_3$};

  \foreach \p in {L,R,T,B}{\node[vtx] at (\p) {};}
\end{tikzpicture}
}
\caption{}
\end{subfigure}
\caption{The illustration of (a) $[\mathrm{v}]_l$ (or $[[\mathrm{u}]]_l$) over the line $l$, and the associated $B_1(e)$ containing all the line $l_1, l_2$ and $l_3$ and $l_4$ in two adjacent oriented triangles sharing an edge $e$.}
\label{fig_mesh_second_order_differential_operator}
\end{figure}

By analogy, we can also define the first-order difference operator with respect to the variable $\mathrm{v} \in \mathrm{V}$. For this purpose, we introduce a line segment $l$ that connects the barycenter to a vertex of a triangle $\tau$ (see ~\cref{fig_mesh_local_structures} (b)). For $\mathrm{v} \in \mathcal{V}$, the $1$-form jump of $\mathrm{v}$ over $l$, under the Neumann boundary condition, is then defined as
\begin{equation}
\begin{aligned}
[\mathrm{v}]_l=\left\{\begin{array}{rr}
 \mathrm{v}_{e^+} \operatorname{sgn}(e^+, \tau) \!+\! \mathrm{v}_{e^-} \operatorname{sgn}(e^-, \tau), & e^-\, \text{and}\, e^- \not \!\subset\! \partial \mathcal{M}, \\
0, & e^+ \,\text{and}\, e^- \subset \partial \mathcal{M},
\end{array}\right.
\end{aligned}
\label{eq_first_order_difference_for_v}
\end{equation}
where $e^{+}$ and $e^{-}$ are two edges of $\tau$ incident to the vertex of touched by $l$. By convention, we assume that $e^+$ enters / $e^-$ leaves the common vertex in the anticlockwise direction. The two triangles $\tau^{+}$ and $\tau^{-}$ share the edges $e^{+}$ and $e^{-}$, respectively (see also~\cref{fig_mesh_second_order_differential_operator} (a)). 

Again, the first-order difference operator over the line $l$, i.e., $D_\mathcal{E}: \mathcal{V} \rightarrow \mathcal{W}$, is defined as
\begin{equation}
\begin{aligned}
D_\mathcal{E} \mathrm{v}|_l=[\mathrm{v}]_l, \quad \forall l \ \ \text{for}\; \mathrm{v} \in \mathcal{V},
\end{aligned}
\label{eq_edge_based_differential_operator}
\end{equation}
where $\mathcal{W}=R^{3 \times {|T|}}$. Moreover, the space $\mathcal{W}$ is equipped with the following inner product and norm;
\begin{equation}
\begin{aligned}
\left(\mathrm{w}^1, \mathrm{w}^2\right)_\mathcal{W}=\sum_l \mathrm{w}_l^1 \mathrm{w}_l^2 \operatorname{len} (l), \quad\|\mathrm{w}\|_\mathcal{W}=\sqrt{(\mathrm{w}, \mathrm{w})_\mathcal{W}},
\end{aligned}
\label{eq_inner_product_and_norm_w}
\end{equation}
where $\operatorname{len} (l)$ is the length of line segment $l$. The adjoint operator $\nabla^\ast\mathrm{w}|_e$ is given by:
\begin{equation}
\begin{aligned}
\nabla^\ast\mathrm{w}|_e = \frac{1}{\operatorname{len} (e)} \sum_{\substack{l \in B_1(e)}} \mathrm{w}_l \,\operatorname{sgn}(e,\tau_l)\, \operatorname{len}(l), \quad \forall e.
\end{aligned}
\label{eq_divergence operator1}
\end{equation}
Here, $B_1(e)$ is the set of lines associated with the edge $e$ (see~\cref{fig_mesh_second_order_differential_operator} (b)) and $\tau_l$ is the triangles containing the line $l$. More details for the derivation can be found in~\cite{liu2021mesh}.

On the other hand, \cref{eq_edge_based_differential_operator}, as demonstrated in~\cite{liu2019triangulated}, can be also understand as the second-order differential operator of $\mathrm{u}$ over the line $l$ in $\tau$, which can be verified by the following definition:
\begin{equation}
\begin{aligned}
[[\mathrm{u}]]_l 
= & [\mathrm{u}]_{e^+} \operatorname{sgn} (e^+, \tau^+) + [u]_{e^-} \operatorname{sgn} (e^-, \tau^-)\\
= & (\mathrm{u}_{\tau} \operatorname{sgn} (e^+, \tau) + \mathrm{u}_{\tau^+} \operatorname{sgn} (e^+, \tau^+)) \operatorname{sgn} (e^+, \tau^+) \\
 +&(\mathrm{u}_{\tau} \operatorname{sgn} (e^-, \tau) + \mathrm{u}_{\tau^-} \operatorname{sgn} (e^-, \tau^-)) \operatorname{sgn} (e^-, \tau^-)\\
= & (\mathrm{u}_{\tau^+}-\mathrm{u}_{\tau}) + (\mathrm{u}_{\tau^-}-\mathrm{u}_{\tau})\\
= & \mathrm{u}_{\tau^+}-2 \mathrm{u}_{\tau}+\mathrm{u}_{\tau^-},
\end{aligned}
\label{eq_face_based_second_order_differential_operator}
\end{equation}
which is sometimes written as $[[\mathrm{u}]]_{\tau,l}$. By this definition, we can see that $[[\mathrm{u}]]_l$ is invariant under the choice of the orientation of edge $e$. Then, the second-order differential operator $D_\mathcal{M}^2:\mathcal{U} \rightarrow \mathcal{W}, \mathrm{u} \rightarrow D_\mathcal{M}^2 \mathrm{u}$ is defined as, 
\begin{equation}
\begin{aligned}
D_\mathcal{M}^2 \mathrm{u}|_l=[[\mathrm{u}]]_l \quad \forall l, \quad \text{for}\,\mathrm{u} \in \mathcal{U}.
\end{aligned}
\label{eq_second_order_difference_over_for_u}
\end{equation}

As shown in~\cref{fig_mesh_local_structures} (b), there are three edges $\{l_{\tau_i, 1}, l_{\tau_i, 2}, l_{\tau_i,3}\}$ for each triangle $\tau_i$. As a result, it is natural to define the second-order gradient operator over the triangle $\tau$, i.e.,
\begin{equation}
\begin{aligned}
\left.\nabla_{\mathcal{M}}^2 \mathrm{u}\right|_\tau=\left(\left.D_{\mathcal{M}}^2 \mathrm{u}\right|_{l_{\tau,1}},\left.D_{\mathcal{M}}^2 \mathrm{u}\right|_{l_{\tau,2}},\left.D_{\mathcal{M}}^2 \mathrm{u}\right|_{l_{\tau,3}}\right),
\end{aligned}
\label{eq_second-order_gradient_operator}
\end{equation}
where $\left.\nabla_{\mathcal{M}} \mathrm{u}\right|_\tau$ is defined over the edges of $\tau$ and we sometimes write it as $\nabla^2 \mathrm{u}=\left(\partial_1^2 \mathrm{u}, \partial_2^2 \mathrm{u}, \partial_3^2 \mathrm{u}\right)$ for simplicity.

Again, let $\nabla^{2}:\mathcal{U}\to\mathcal{W}$ denote the second–order difference operator. The adjoint $(\nabla^{2})^\ast:\mathcal{W}\to\mathcal{U}$ is characterized by
$$
(\mathrm{w},\,\nabla^{2}\mathrm{u})_{\mathcal{W}} \;=\; \big((\nabla^{2})^\ast \mathrm{w},\,\mathrm{u}\big)_{\mathcal{U}}.
$$
For $\mathrm{w} \in \mathrm{W}$, the adjoint operator $(\nabla^2)^\ast\mathrm{w}|_\tau$ is given by: 
\begin{equation}
\begin{aligned}
(\nabla^2)^\ast\mathrm{w}|_\tau=&\frac{1}{s_\tau} \sum_{\substack{l \in B_2(\tau), \\ e^{+}\, \text{and}\, e^{-} \not \subset \partial M}} \left.\mathrm{w} \right|_l \operatorname{len}(l) \\
- &\frac{1}{s_\tau}\sum_{\substack{l \in B_1(\tau), \\ e^{+}\, \text{and} \, e^{-} \not \subset \partial M}} 2 \left.\mathrm{w} \right|_l \operatorname{len}(l) \; \forall \tau,
\end{aligned}
\label{eq35}
\end{equation}
where $B_1({\tau_i})$ and $B_2({\tau_i})$ are the sets illustrated in~\cref{fig_mesh_local_structures} (b) and (c), respectively. More discussion can be found in~\cite{liu2019triangulated}.

To deal with the vectorial data, we also extend the spaces into vectorial cases. Specifically, we denote by  $\mathbf{U}=\underbrace{\mathcal{U} \times \cdots \times \mathcal{U}}_{n}, \mathbf{V}=\underbrace{\mathcal{V} \times \cdots \times \mathcal{V}}_{n}$ and $\mathbf{W}=\underbrace{\mathcal{W} \times \cdots \times \mathcal{W}}_{n}$, then the inner products and norms in $\mathbf{U}, \mathbf{V}$ and $\mathbf{W}$ are defined as follows:
\begin{equation}
\begin{aligned}
&\left(\mathbf{u}^1, \mathbf{u}^2\right)_\mathbf{U}=\sum_{1 \leq i \leq n}\left(\mathbf{u}_i^1, \mathbf{u}_i^2\right)_\mathcal{U}, \quad\|\mathbf{u}\|_{\mathbf{U}}=\sqrt{(\mathbf{u}, \mathbf{u})_\mathbf{U}}. \\
&\left(\mathbf{v}^1, \mathbf{v}^2\right)_\mathbf{V}=\sum_{1 \leq i \leq n}\left(\mathbf{v}_i^1, \mathbf{v}_i^2\right)_\mathcal{V}, \quad\|\mathbf{v}\|_{\mathbf{V}}=\sqrt{(\mathbf{v}, \mathbf{v})_\mathbf{V}}. \\
&\left(\mathbf{w}^1, \mathbf{w}^2\right)_\mathbf{W}=\sum_{1 \leq i \leq n}\left(\mathbf{w}_i^1, \mathbf{u}_i^2\right)_\mathcal{W}, \quad\|\mathbf{w}\|_{\mathbf{W}}=\sqrt{(\mathbf{w}, \mathbf{w})_\mathbf{W}}.
\end{aligned}
\label{eq37}
\end{equation}

\section{Semi-sparsity Regularization for Mesh Denoising}
\label{semi_sparsity_model}

Semi-sparsity priors have recently been explored in various image processing tasks, such as image denoising~\cite{huang2023semi} and image decomposition~\cite{huang2023semi1}. As demonstrated in these studies, such regularization models exhibit a powerful simultaneous-fitting capability, effectively capturing both singular structures and polynomially smooth regions. Motivated by the fact that many variational and filtering models for geometric mesh denoising~\cite{he2013mesh, zhang2015variational, liu2019novel, liu2021mesh} are direct extensions of their image-processing counterparts, it is natural to extend the semi-sparse model~\cite{huang2023semi} to 3D geometry when treating mesh surfaces as piecewise constant or smoothing signals. In what follows, we demonstrate the benefits of semi-sparsity priors for triangulated mesh denoising and illustrate the cutting-edge performance in preserving sharp corners, edges and creases while producing smoothly varying surfaces without introducing noticeable staircase artifacts.

To clarify the problem, we first propose a generalized semi-sparsity regularization model based on the idea in~\cite{huang2023semi}. Given a noisy signal $\mathrm{f}$, we consider a higher-order $L_0$ regularization model in the following optimization-based framework,
\begin{equation}
\begin{aligned}
    \mathop{\min}_{\mathrm{u}} {\frac{\lambda}{2}{\left\Vert {\mathrm{u}}-{\mathrm{f}}\right\Vert }_2^2}+\alpha_i \sum_{k=1}^{n-1}  {\left\Vert {\nabla}^{k}\mathrm{u}\right\Vert }_p^p+ \beta {\left\Vert {\nabla}^{n} \mathrm{u} \right\Vert }_0,
\end{aligned}
\label{eq_semi_sparsity_model} 
\end{equation}
where $\mathrm{u}$ is the target output, and $\lambda$,  $\alpha_i$ and $\beta$ positive weights. Again, the data fidelity term imposes the output $\mathrm{u}$ to be close to $\mathrm{f}$ in the sense of least square minimization. The second term measures the $L_p(p \ge 1)$- norm similarity of higher-order gradients ${\nabla}^{k} \mathrm{u}$. The third term ${\left\Vert {\nabla}^{n} \mathrm{u} \right\Vert }_0$ favors a strict sparsity of the highest-order gradient ${ {\nabla}^{n} \mathrm{u}}$. The idea is straightforward, that is, a sparse-induced $L_0$-norm constraint is only imposed on the highest $n$-th order gradient domain, as the ones with orders less than $n$ are not fully sparse but also have a small error $L_p$ space.

In many existing mesh denoising methods~\cite{sun2007fast, zheng2010bilateral, zhang2015variational, zhang2015guided, zhang2018static, liu2019novel, liu2021mesh}, a two-stage strategy is commonly employed to achieve more accurate smoothing results. We here adopt a similar procedure, in which the face normal field is first estimated based on a semi-sparsity regularization model, and the vertices of triangle surfaces are then restored based on the estimated face normal vectors.  

\subsection{Normal Estimation}

Given a mesh $\mathcal{M}$, the normal field and its differential operators, as interpreted in~\cref{spaces_and_operators}, can be defined over triangle faces, vertices or edges. To simplify, we consider the noisy normal field $\mathbf{N}_0$ on the triangle faces. In this setting, we propose the semi-sparse regularization model in the form 
\begin{equation}
\begin{aligned}
    \mathop{\min}_{\mathbf{N} \in C_\mathbf{N}}\ \frac{\lambda}{2}{\left\Vert \mathbf{N}-\mathbf{N}_0\right\Vert}_{\mathbf{U}}^2
    +&\alpha \sum_e w_e \left\Vert  {D}_{\mathcal{M}} \mathbf{N} |_e \right\Vert_1  \operatorname{len}(e)\\+&\beta \sum_l w_l \left\Vert  {D}_{\mathcal{M}}^2 \mathbf{N} |_l \right\Vert_0  \operatorname{len}(l),
\end{aligned}
\label{eq_semi_sparsity_normal_smoothing_model}
\end{equation}
where $C_{\mathbf{N}}=\left\{\mathbf{N} \in \mathbf{V}_M:\left\|\mathbf{N}_\tau\right\|=1,\, \forall \tau\right\}$. As suggested in~\cite{huang2023semi}, we consider the vectorized model and set the highest order $n=2$ and $p=1$ in~\cref{eq_semi_sparsity_normal_smoothing_model} for the sake of simplicity and computational efficiency. We additionally follow~\cite{zhang2015variational, liu2019novel,liu2021mesh} and introduce the dynamic weight $w_e$ over each edge $e$ of the triangle as
\begin{equation}
\begin{aligned}
    w_e=\exp \left(-\frac{\left\|\mathbf{N}_{\tau^+}-\mathbf{N}_{\tau^-}\right\|^2}{2 \sigma_e^2}\right),
\end{aligned}
\label{eq_dynamic weight_e} 
\end{equation}
where $\mathbf{N}_{\tau^+}$ and $\mathbf{N}_{\tau^-}$ are normals of the triangles sharing edge $e$, and $\sigma_e$ is a user-specified parameter. Similarly, we also have the dynamic weight $w_l$ over each $l$ of the triangle $\tau$ defined by 
\begin{equation}
\begin{aligned}
    w_l=\exp \left(-\frac{\left\|\mathbf{N}_{\tau^{+}}-2 \mathbf{N}_\tau +\mathbf{N}_{\tau^{-}}\right\|^4}{2 \sigma_l^4}\right),
\end{aligned}
\label{eq_dynamic weight_l} 
\end{equation}
where $\tau^+, \tau^-, \tau$ and $l$ are illustrated in~\cref{fig_mesh_second_order_differential_operator}. Both
$w_e$ and $w_l$ are expected to be small for sharp features, and large for smooth regions,  therefore allowing the proposed method to smooth non-feature regions while preserving sharp features.

\subsection{The ADMM Solution}

Due to the non-convex and non-smooth nature of~\cref{eq_semi_sparsity_normal_smoothing_model}, we here employ a multi-block ADMM algorithm for an efficient solution. As discussed in~\cite{huang2023semi1}, the (multi-block) ADMM algorithm introduces auxiliary variables to decouple the objective function into several subproblems, and each can be solved independently by fixing the others, followed by dual updates. For the proposed semi-sparse model, this splitting yields sparse linear systems and simple component-wise thresholding/projection operators, leading to an effective iterative algorithm. Moreover, the resulting ADMM scheme scales well and remains applicable even to very large-scale problems in practice.

According to the multi-block ADMM framework, we rewrite~\cref{eq_semi_sparsity_normal_smoothing_model} as the following constrained optimization problem by introducing the auxiliary variables $\mathbf{P}$, and $\mathbf{Q}$, 
\begin{equation}
\begin{aligned}
    \min_{\mathbf{N}, \mathbf{P}, \mathbf{Q}} \ &
    \frac{\lambda}{2} \left\|\mathbf{N}-\mathbf{N}_0\right\|_{\mathbf{U}}^2
    +\alpha \sum_e w_e \left\|\mathbf{P}_e\right\|_1 \operatorname{len}(e)\\
    & +  \beta \sum_l w_l \left\|\mathbf{Q}_l\right\|_0 \operatorname{len}(l) + \eta (\mathbf{N}) , \\
    \text {s.t.} & \quad \mathbf{P}={D}_{\mathcal{M}}\mathbf{N}, \quad \mathbf{Q}={D}_{\mathcal{M}}^{2}\mathbf{N}.
\end{aligned}
\label{eq_admm_regularization_form} 
\end{equation}
Here,
\begin{equation*}
\eta(\mathbf{N})=\left\{
\begin{aligned}
    0, &\qquad\left\|\mathbf{N}_{\tau} \right\| = 1, \quad \forall \tau,\\
    +\infty, &\qquad \quad \text{otherwise}.
\end{aligned}
\right.
\label{admm_contraint}
\end{equation*}
Based on the ADMM model, the corresponding augmented Lagrangian function has the form,
\begin{equation}
\begin{aligned}
    &\mathcal{L}\left(\mathbf{N}, \mathbf{P}, \mathbf{Q}, \boldsymbol{z}_{\mathbf{P}}, \boldsymbol{z}_{\mathbf{Q}}\right) =
    \frac{\lambda}{2}\left\|\mathbf{N}-\mathbf{N}_0\right\|_{\mathbf{U}}^2 + \eta (\mathbf{N})\\
    &+\alpha\sum_e w_e \left\|\mathbf{P}_e\right\|_1 \, \operatorname{len} (e)
    +\beta \sum_l w_l \left\|\mathbf{Q}_l\right\|_{0} \, \operatorname{len} (l)  \\
    & +\langle \boldsymbol{z}_{\mathbf{P}}, {D}_{\mathcal{M}} \mathbf{N}-\mathbf{P}\rangle_{\mathbf{V}}+\frac{\rho_1}{2}\left\| {D}_{\mathcal{M}}\mathbf{N}-\mathbf{P}\right\|_{\mathbf{V}}^2 \\
    & +\langle\boldsymbol{z}_{\mathbf{Q}}, {D}_{\mathcal{M}}^{2}\mathbf{N}-\mathbf{Q}\rangle_{\mathbf{W}}+\frac{\rho_2}{2}\left\|{D}_{\mathcal{M}}^{2}\mathbf{N}-\mathbf{Q}\right\|_{\mathbf{W}}^2 ,
\end{aligned}
\label{eq_augmented_admm_form} 
\end{equation}
where $\boldsymbol{z}_{\mathbf{P}}$,and $\boldsymbol{z}_{\mathbf{Q}}$ are Lagrange multipliers, $\rho_1$ and $\rho_2$ are positive penalty weights. This formulation leads to a three-block ADMM form, which can be decomposed into three subproblems and each can be solved independently while keeping the other variables fixed. Below, we briefly describe the individual subproblems and highlight the structural properties for efficient solutions.

\textbf{The $\mathbf{N}$-subproblem:} By fixing all other variables in~\cref{eq_augmented_admm_form}, the subproblem with respect to $\mathbf{N}$ reduces to a quadratic minimization problem of the form,
\begin{equation}
\begin{aligned}
    \min_{\mathbf{N}} \ &\frac{\lambda}{2}\left\|\mathbf{N}-\mathbf{N}_0\right\|_{\mathbf{U}}^2 
    +\frac{\rho_1}{2}\left\|{D}_{\mathcal{M}}\mathbf{N}-\mathbf{P}+\frac{\boldsymbol{z}_{\mathbf{P}}}{\rho_1}\right\|_{\mathbf{V}}^2\\
    +&\frac{\rho_2}{2}\left\| {D}_{\mathcal{M}}^2\mathbf{N}-\mathbf{Q}+\frac{\boldsymbol{z}_{\mathbf{Q}}}{\rho_2}\right\|_{\mathbf{W}}^0+\eta(\mathbf{N}).
\end{aligned}
\label{admm_subproblem_n}
\end{equation}
It is evident that~\cref{admm_subproblem_n} defines a quadratic optimization problem with the unit normal constraints. As suggested in~\cite{liu2021mesh}, such a problem can be first solved by an approximation strategy without the unit normal constraints, then the solution $\mathbf{N}$ is projected onto a unit sphere. By the first-order optimal condition, the quadratic function in~\cref{admm_subproblem_n} leads to the following Euler-Lagrange equation,
\begin{equation}
\begin{aligned}
    \lambda M_s & \left(\mathbf{N} -\mathbf{N}_0\right) + \rho_1{D}_{\mathcal{M}}^T M_e \left({D}_{\mathcal{M}} \mathbf{N}-\mathbf{P}+\frac{\boldsymbol{z}_{\mathbf{P}}}{\rho_1}\right)\\
    +&\rho_2\left({D}_{\mathcal{M}}^2\right)^T M_l\left( {D}_{\mathcal{M}}^2\mathbf{N}-\mathbf{Q}+\frac{\boldsymbol{z}_{\mathbf{Q}}}{\rho_2}\right)=0,
\end{aligned}
\label{admm_subproblem_n1}
\end{equation}
where $M_s$, $M_e$, and $M_l$ are the diagonal matrices, whose diagonal elements are $s_{\tau}$ and $\operatorname{len}(e)$, $\operatorname{len}(l)$, i.e., the area of $\tau$, the length of edge $e$ and $l$, respectively. Here, ${D}_{\mathcal{M}}^T$ and $\left({D}_{\mathcal{M}}^2\right)^T$ are the adjoint operators of the first-order and second-order differential operators, respectively. Rearranging the terms in~\cref{admm_subproblem_n1} leads to the following sparse linear system:
\begin{equation}
\begin{aligned}
    &\left(\lambda M_s + \rho_1 D_{\mathcal{M}}^T M_e D + \rho_0 (D_{\mathcal{M}}^2)^T M_l D_{\mathcal{M}}^2 \right) \mathbf{N}= \lambda M_s \mathbf{N}_0 \\
    &+ \rho_1 D_{\mathcal{M}}^T M_e \left(\mathbf{P}-\frac{\boldsymbol{z}_{\mathbf{P}}}{\rho_1}\right) +  \rho_2\left({D}_{\mathcal{M}}^2\right)^T M_l\left( \mathbf{Q}-\frac{\boldsymbol{z}_{\mathbf{Q}}}{\rho_2}\right).
\end{aligned}
\label{eq_admm_subproblem_n_solution}
\end{equation}
The system matrix on the left-hand side of~\cref{eq_admm_subproblem_n_solution} is symmetric positive definite and highly sparse, allowing efficient solution by standard sparse linear solvers such as the Gauss–Seidel method or preconditioned conjugate gradient (PCG) method. Once $\mathbf{N}$ is computed, we normalize it to make sure each face normal is a unit vector.

\textbf{The $\mathbf{P}$-subproblem:} The sub-problem with respect to $\mathbf{P}$ has the form
\begin{equation}
\begin{aligned}
    \min_{\mathbf{P}} \alpha \sum_e w_e \left\|\mathbf{P}_e\right\|_1 \operatorname{len} (e) +\frac{\rho_1}{2}\left\|\mathbf{P}-\left({D}_{\mathcal{M}}\mathbf{N}+\frac{\boldsymbol{z}_{\mathbf{P}}}{\rho_1}\right)\right\|_{\mathbf{V}}^2.
\end{aligned}
\label{eq_admm_subproblem_p}
\end{equation}
It is known that~\cref{eq_admm_subproblem_p} is a classical Lasso problem, and each variable $\mathbf{P}_e$ can be solved independently. Moreover, $\mathbf{P}_e$ has a closed form solution 
\begin{equation}
    \begin{aligned}
        \mathbf{P}_e=S\left({D}_{\mathcal{M}}\mathbf{N}+\frac{\boldsymbol{z}_{\mathbf{P}}}{\rho_1}, w_e\frac{\alpha}{\rho_1}\right),
    \end{aligned}
    \label{eq_admm_subproblem_p_solution}
\end{equation}
with the soft shrinkage operator $S(x,T_1)$ defined as:
$$
S(x,T_1)= \operatorname{sign}(x) \max\left(0, \left\|x\right\|_1-T_1 \right),
$$
where $T_1$ is a specified threshold.

\textbf{The $\mathbf{Q}$-subproblem:} The $\mathbf{Q}$ sub-problems has the form
\begin{equation}
    \begin{aligned}
        \min_{\mathbf{Q}} \alpha \sum_l w_l \left\|\mathbf{Q}_l\right\|_0 \operatorname{len} (l) +\frac{\rho_2}{2}\left\|\mathbf{Q}+ \left({D}_{\mathcal{M}}^2\mathbf{N}-\frac{\boldsymbol{z}_{\mathbf{Q}}}{\rho_2}\right)\right\|_{\mathbf{W}}^2.
    \end{aligned}
    \label{eq_admm_subproblem_q}
\end{equation}

\begin{algorithm}[!t]
\small
\caption{ADMM for Semi-sparse Normal Denoising}\label{alg:alg1}
\begin{algorithmic}[1]
    \Require Noisy normal vector $\mathbf{N}_0$, the parameters $\{\lambda,\alpha,\beta\}$, and the weights $\rho_1,\rho_2$;
    \State \textbf{Initialization:} $\mathbf{N}\gets \boldsymbol{0}$; $\mathbf{P}^0\gets \boldsymbol{0}$; $\mathbf{Q}^0\gets \boldsymbol{0}$; $\boldsymbol{z}_{\mathbf{P}}\gets \boldsymbol{0}$; $\boldsymbol{z}_{\mathbf{Q}}\gets \boldsymbol{0}$; $k\gets 0$
    \Repeat
        \State Solve~\cref{eq_admm_subproblem_n_solution} for $\mathbf{N}^{k+1}$ with a linear solver and normalization;
        \State Solve~\cref{eq_admm_subproblem_p_solution} for $\mathbf{P}^{k+1}$ by the soft-shrinkage operator;
        \State Solve~\cref{eq_admm_subproblem_q_solution} for $\mathbf{Q}^{k+1}$ by the hard-shrinkage operator;
        \State Update the Lagrange multipliers $\boldsymbol{z}_{\mathbf{P}}^{k+1}$ and $\boldsymbol{z}_{\mathbf{Q}}^{k+1}$:

        $\boldsymbol{z}_{\mathbf{P}}^{k+1}=\boldsymbol{z}_{\mathbf{P}}^{k} + \rho_1\!\left(D_{\mathcal{M}} \mathbf{N}^{k} - \mathbf{P}^{k}\right)$;
        
        $\boldsymbol{z}_{\mathbf{Q}}^{k+1}=\boldsymbol{z}_{\mathbf{Q}}^{k} + \rho_2\!\left(D_{\mathcal{M}}^2 \mathbf{N}^{k} - \mathbf{Q}^{k}\right)$;
         \State  update weights $w_e$ and $w_l$ based on~\cref{eq_dynamic weight_e} and~\cref{eq_dynamic weight_l}; 
        \State Increment $k$: $k \gets k+1$;
    \Until{ $\bigl\lVert \mathbf{N}^{k}-\mathbf{N}^{k-1}\bigr\rVert_2^2 \leq \varepsilon$ \textbf{or} $k > K$};
    \Ensure Filtered normal vector $\mathbf{N} = \mathbf{N}^{k}$.
\end{algorithmic}
\end{algorithm}

The $\mathbf{Q}$-subproblem~\cref{eq_admm_subproblem_q} is a $L_0$ norm minimization problem, which has a similar separable property as~\cref{eq_admm_subproblem_p} and each variable is given by the formula
\begin{equation}
    \begin{aligned}
        \mathbf{Q}_l=H\left({D}_{\mathcal{M}}^2\mathbf{N}+\frac{\boldsymbol{z}_{\mathbf{Q}}}{\rho_2}, w_l\frac{\beta}{\rho_2}\right),
    \end{aligned}
    \label{eq_admm_subproblem_q_solution}
\end{equation}
with the hard-threshold operator defined as:
$$
H(x,T_2)=\left\{
\begin{aligned}
    0, &\qquad\left\|x\right\|\le T_2,\\
    x, &\qquad\text{otherwise},
\end{aligned}
\right.
$$
where $T_2$ is a specified threshold. Notice that both~\cref{eq_admm_subproblem_p_solution} and~\cref{eq_admm_subproblem_q_solution} are fully separable, in which each variable can be computed independently. This separability makes them well-suited for parallel implementation in practice.

Finally, the Lagrange multipliers $\boldsymbol{z}_{\mathbf{P}}$ and $\boldsymbol{z}_{\mathbf{Q}}$ are updated in the form,
\begin{equation}
    \begin{aligned}
        \boldsymbol{z}_{\mathbf{P}} = &\boldsymbol{z}_{\mathbf{P}} + \rho_1\left( {D}_{\mathcal{M}}\mathbf{N}-\mathbf{P} \right),\\
        \boldsymbol{z}_{\mathbf{Q}} = &\boldsymbol{z}_{\mathbf{Q}} + \rho_2\left( {D}_{\mathcal{M}}^2\mathbf{N}-\mathbf{Q} \right).
    \end{aligned}
    \label{eq_admm_subproblem_z_solution}
\end{equation}

The three-block ADMM algorithm alternatively solves the sub-problems and the Lagrange multipliers until the given stop criteria are met, which leads to an iterative procedure for the proposed semi-sparsity normal filtering model. Since all sub-problems have closed-form solutions in low computational complexity, the problem is empirically solvable even with a large number of variables. The effectiveness and efficiency of the multi-block ADMM algorithm will be further demonstrated by various experimental results in the next section. The reader is referred to~\cite{huang2023semi1} for more details of the convergence analysis in different settings. 

\subsection{Vertex Updating}

Once the face normal field is restored by~\cref{alg:alg1}, one needs to reconstruct the vertex positions to match the updated normals. A simple vertex updating model~\cite{sun2007fast}, which has been used in many existing methods~\cite{yagou2002mesh, zhang2015variational, zhang2015guided, zheng2010bilateral}, minimizes the orthogonality residual of edges to their incident face normals:
$$
\min _v \sum_\tau \sum_{\left(v_i, v_j\right) \in \tau} s_\tau\left(\mathbf{N}_\tau \cdot\left(v_i-v_j\right)\right)^2,
$$
where $s_\tau$ is the area and $\mathbf{N}_\tau$ is the filtered normal of $\tau$. This model minimizes the orthogonal error between the filtered face normal and the three edges at each face over the surface. Simple it is, this method may introduce folding faces when a surface is corrupted by noise in random directions, even with the exact recovered face normals. In the case of large-scale noise, this phenomenon is even worse. The reason is twofold: (i) the squared dot product is invariant to the sign of $\mathbf{N}_\tau$, so the cost does not distinguish between a target normal and its opposite orientation; and (ii) the objective does not explicitly regularize local triangle shape, allowing inverted configurations under large perturbations. To address this limitation, we follow the orientation-consistent, shape-preserving update
strategy~\cite{zhang2018static}, which enforces consistency with target normals while preserving local triangle shapes. In our experiments, we fix the number of vertex-update iterations to 35, which consistently yields satisfactory reconstructions. We refer the reader to~\cite{zhang2018static} for further discussion and implementation details.

\subsection{Differing from TV, HO, and TGV}

From a variational perspective, several mesh-denoising models are closely related to the proposed semi-sparsity formulation. Below, we clarify the key differences between our approach and three representative regularization-based models: the TV model~\cite{zhang2015variational}, the higher-order total variation (HO) model~\cite{liu2019novel}, and the second-order TGV model~\cite{liu2021mesh}.

Let $\mathrm{u}\in\mathcal{U}$ be a discrete (normal) vector restricted on a triangulated mesh $\mathcal{M}$. The (anisotropic) discrete TV, for example, as derived in~\cite{zhang2015variational}(see also in~\cref{discrete_operators}), is typically formulated as:
$$
\mathrm{TV}(\mathrm{u})=\left\|D_{\mathcal{M}} \mathrm{u}\right\|_\mathcal{V},
$$
which penalizes first-order jumps across mesh edges. This TV regularizer captures first-order variations over mesh edges and is well-known for its ability to preserve sharp features. However, it is also prone to producing staircase artifacts in smoothly varying regions.

To mitigate the staircase effect, a second-order regularization was proposed in~\cite{liu2019novel}, based on discrete curvature-like differences:
$$
\operatorname{HO}(\mathrm{u})=\sum_{l\in B_1(\tau)}\left\|2 \mathrm{u}_\tau-\mathrm{u}_{\tau^{+}}-\mathrm{u}_{\tau^{-}}\right\| \, \operatorname{len}(l) .
$$
This HO regularizer is essentially a second-order total variation measure over mesh edges, which is effective at restoring smooth surfaces, yet it tends to blur sharp features, particularly in high-noise scenarios.

\begin{figure*}[!t]
	\centering
	\begin{subfigure}{0.18\linewidth}
		\includegraphics[width=\textwidth]{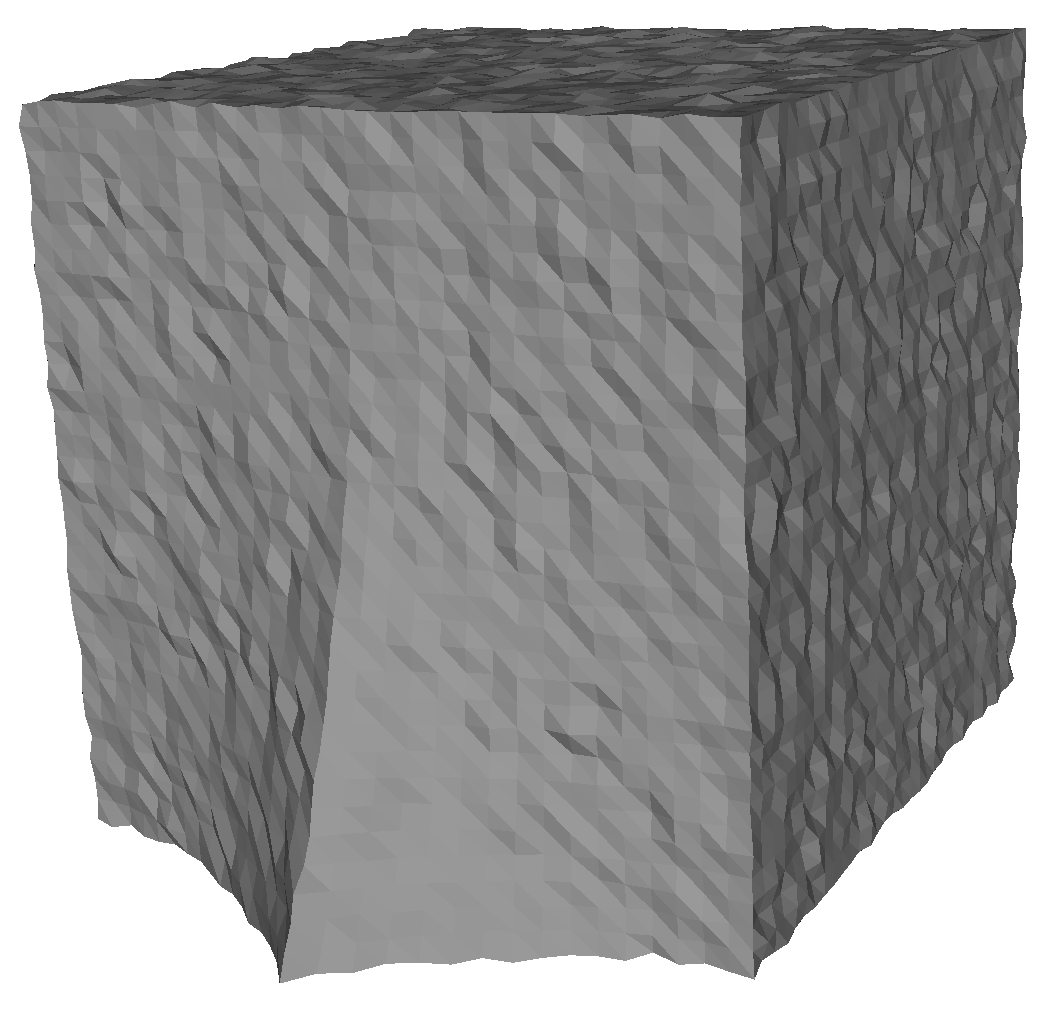}
		\caption{Noisy}
		\label{fig:varying_params_alpha-a}
	\end{subfigure}
	\begin{subfigure}{0.18\linewidth}
		\includegraphics[width=\textwidth]{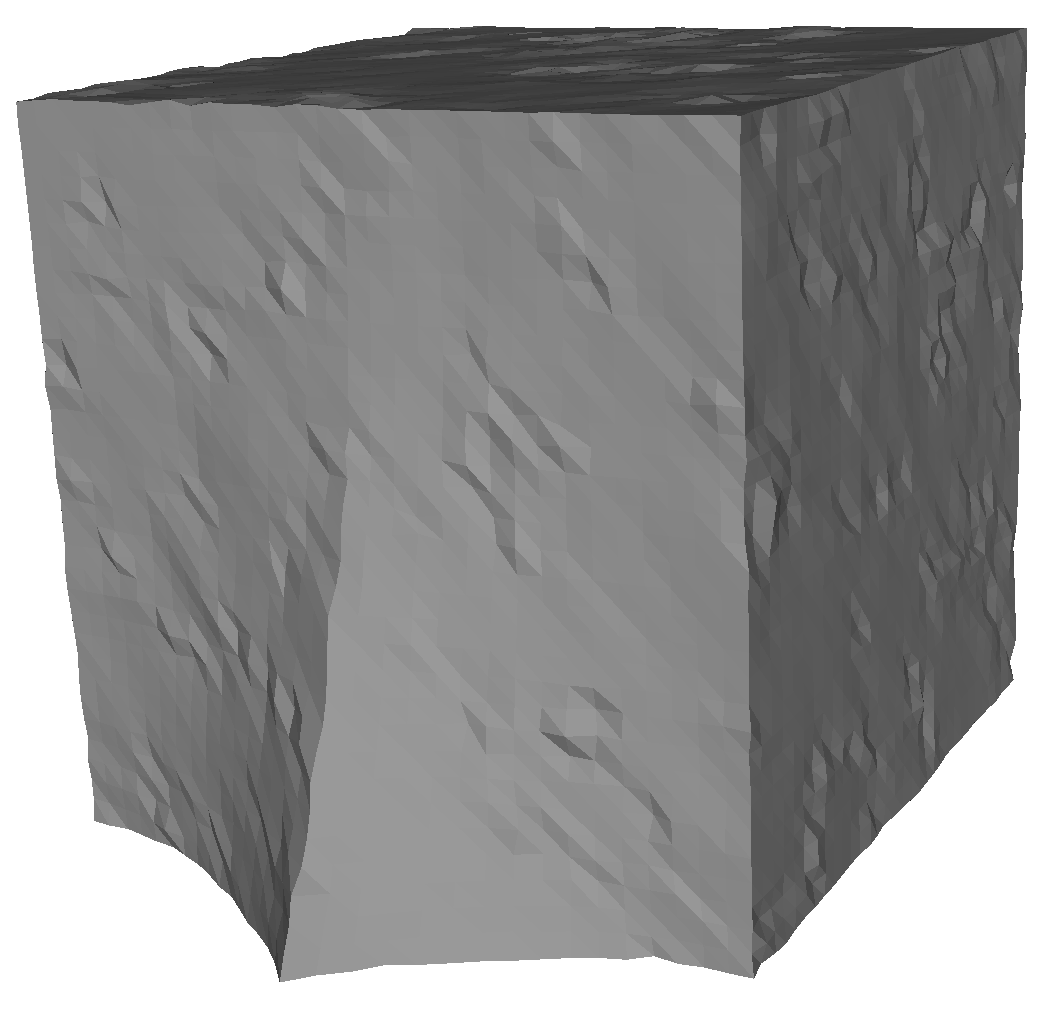}
		\caption{$\alpha = 0.1$}
		\label{fig:varying_params_alpha-b}
	\end{subfigure}
	\begin{subfigure}{0.18\linewidth}
		\includegraphics[width=\textwidth]{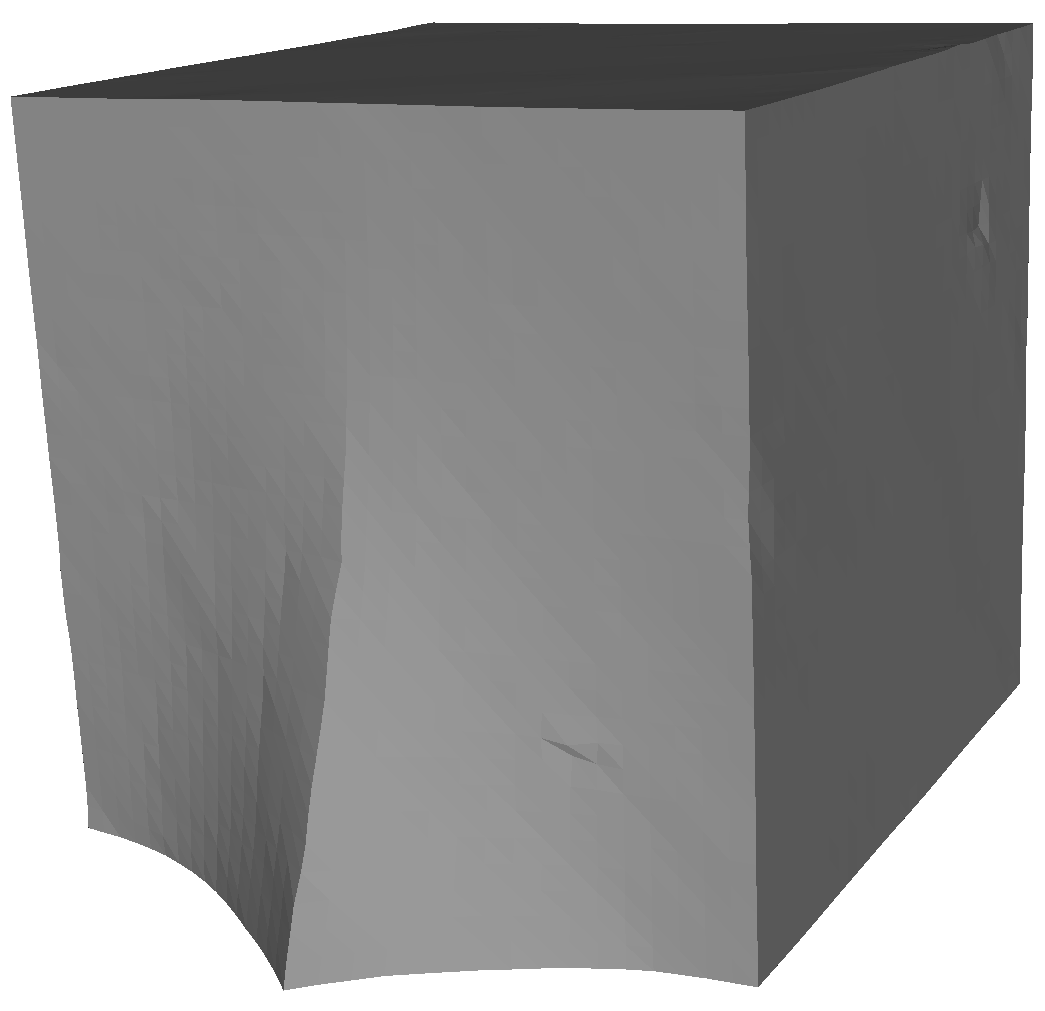}
		\caption{$\alpha = 0.5$}
		\label{fig:varying_params_alpha-c}
	\end{subfigure}
	\begin{subfigure}{0.18\linewidth}
		\includegraphics[width=\textwidth]{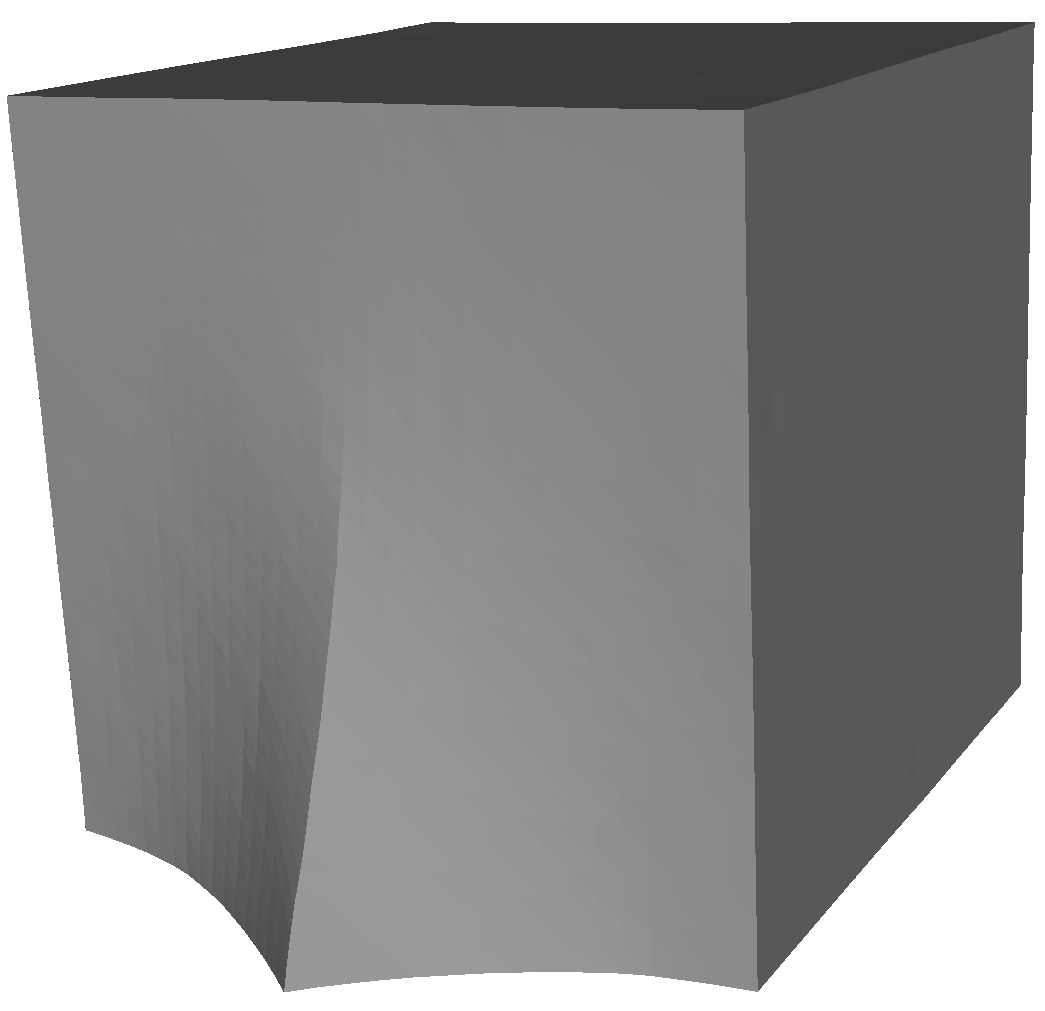}
		\caption{$\alpha = 1.2$}
		\label{fig:varying_params_alpha-d}
	\end{subfigure}
	\begin{subfigure}{0.18\linewidth}
		\includegraphics[width=\textwidth]{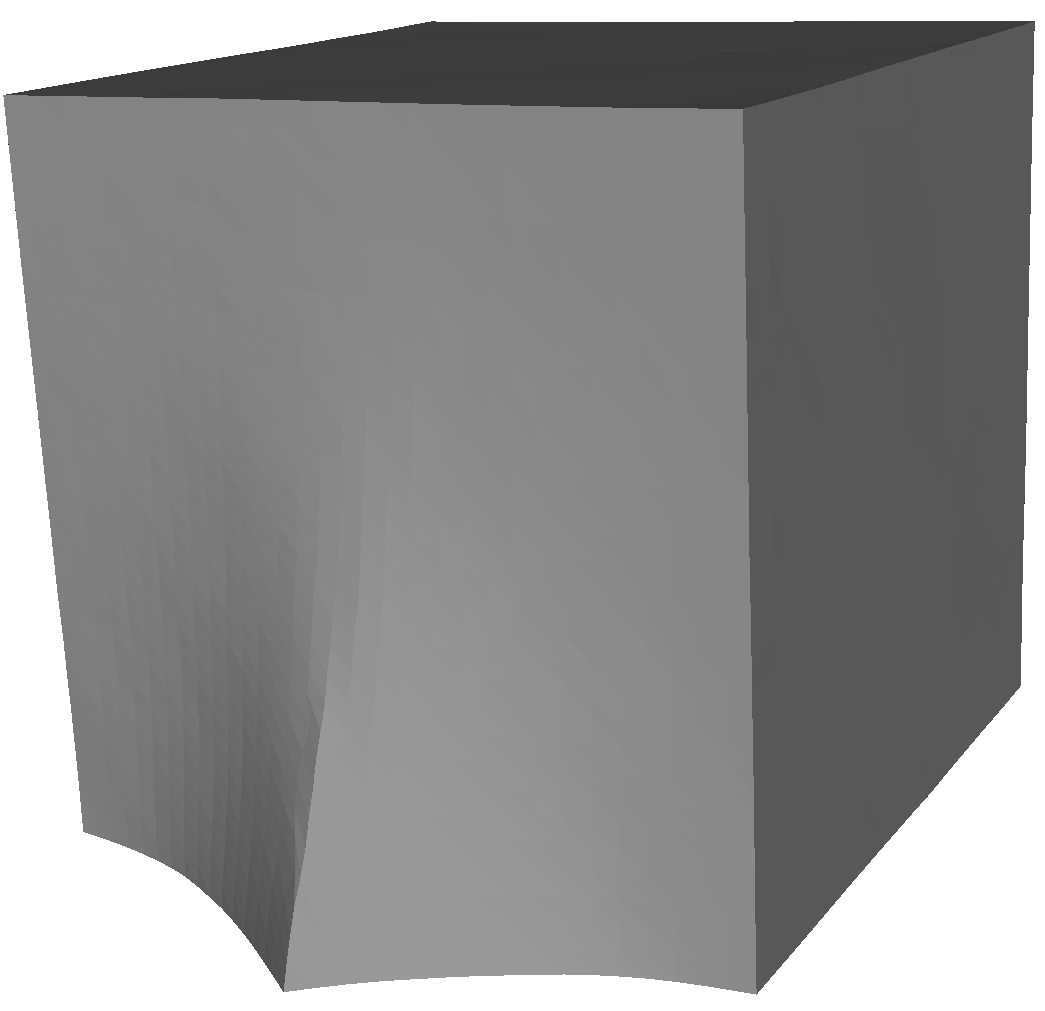}
		\caption{$\alpha = 2.8$}
		\label{fig:varying_params_alpha-e}
	\end{subfigure}
	\caption{Visual denoising comparison of varying $\alpha$ with the fixed $\beta$ and $\lambda$. The noisy mesh is corrupted by Gaussian noise ($\sigma=0.1\,\bar{l}_e$).}
	 \label{fig:varying_params_alpha}
\end{figure*}

\begin{figure*}[!t]
	\centering
	\begin{subfigure}{0.18\linewidth}
		\includegraphics[width=\textwidth]{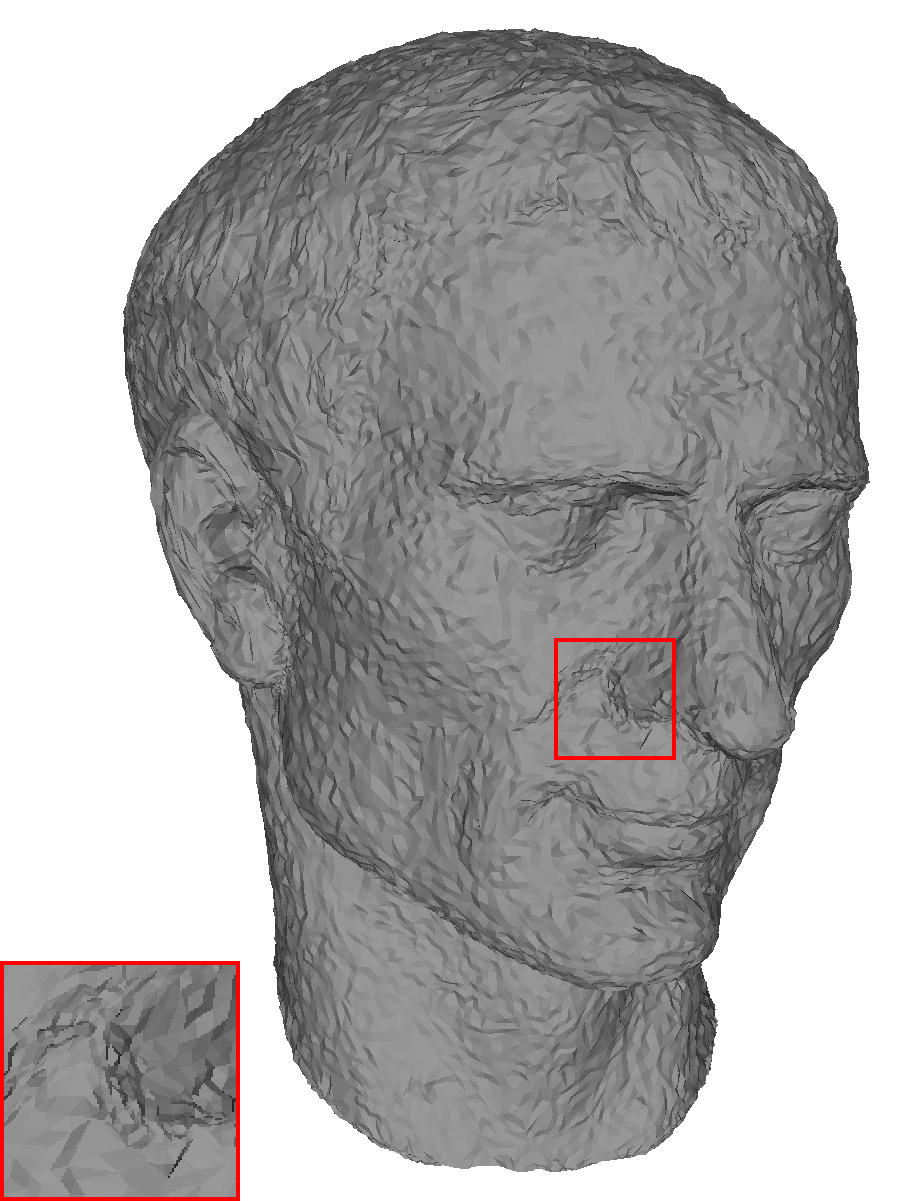}
		\caption{Noisy}
		\label{fig:varying_params_beta-a}
	\end{subfigure}
	\begin{subfigure}{0.18\linewidth}
		\includegraphics[width=\textwidth]{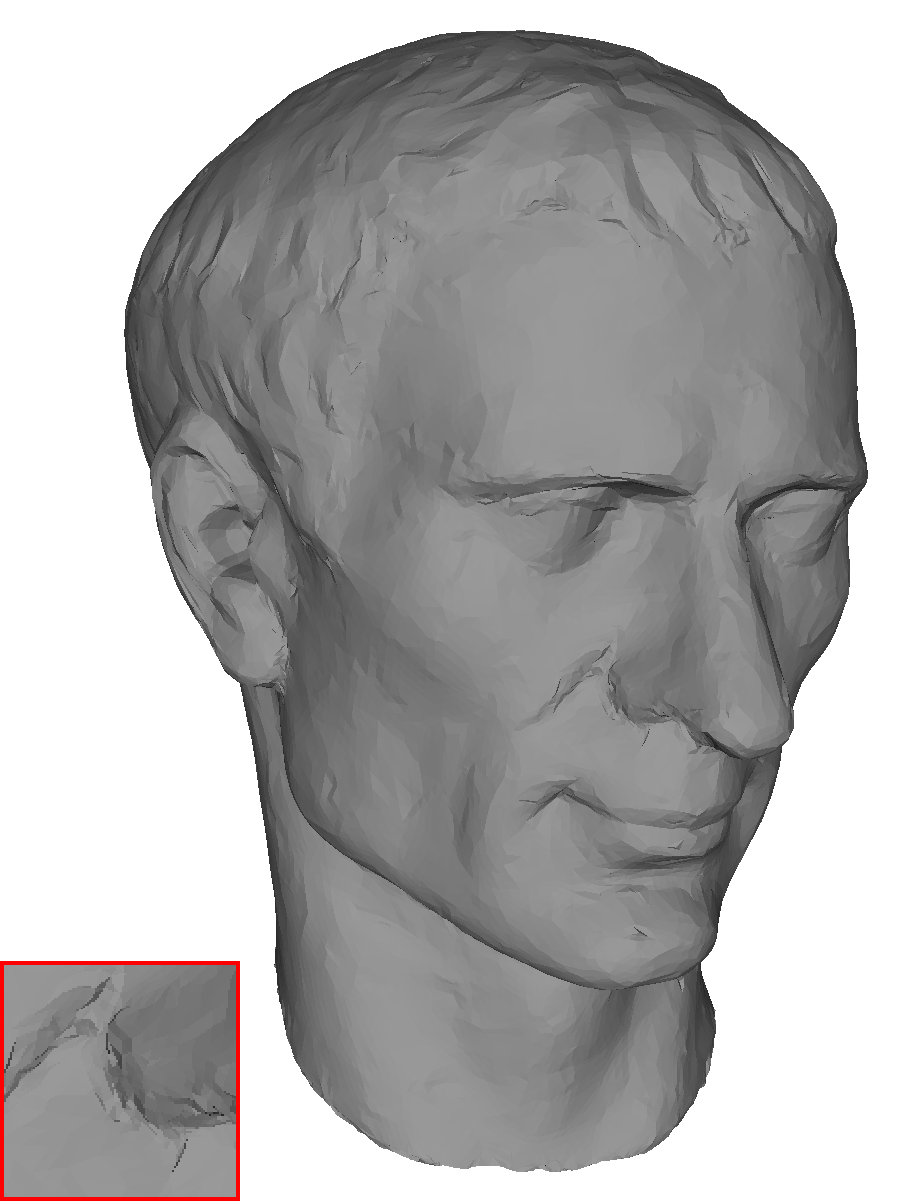}
		\caption{$\beta = 0.05$}
		\label{fig:varying_params_beta-b}
	\end{subfigure}
	\begin{subfigure}{0.18\linewidth}
		\includegraphics[width=\textwidth]{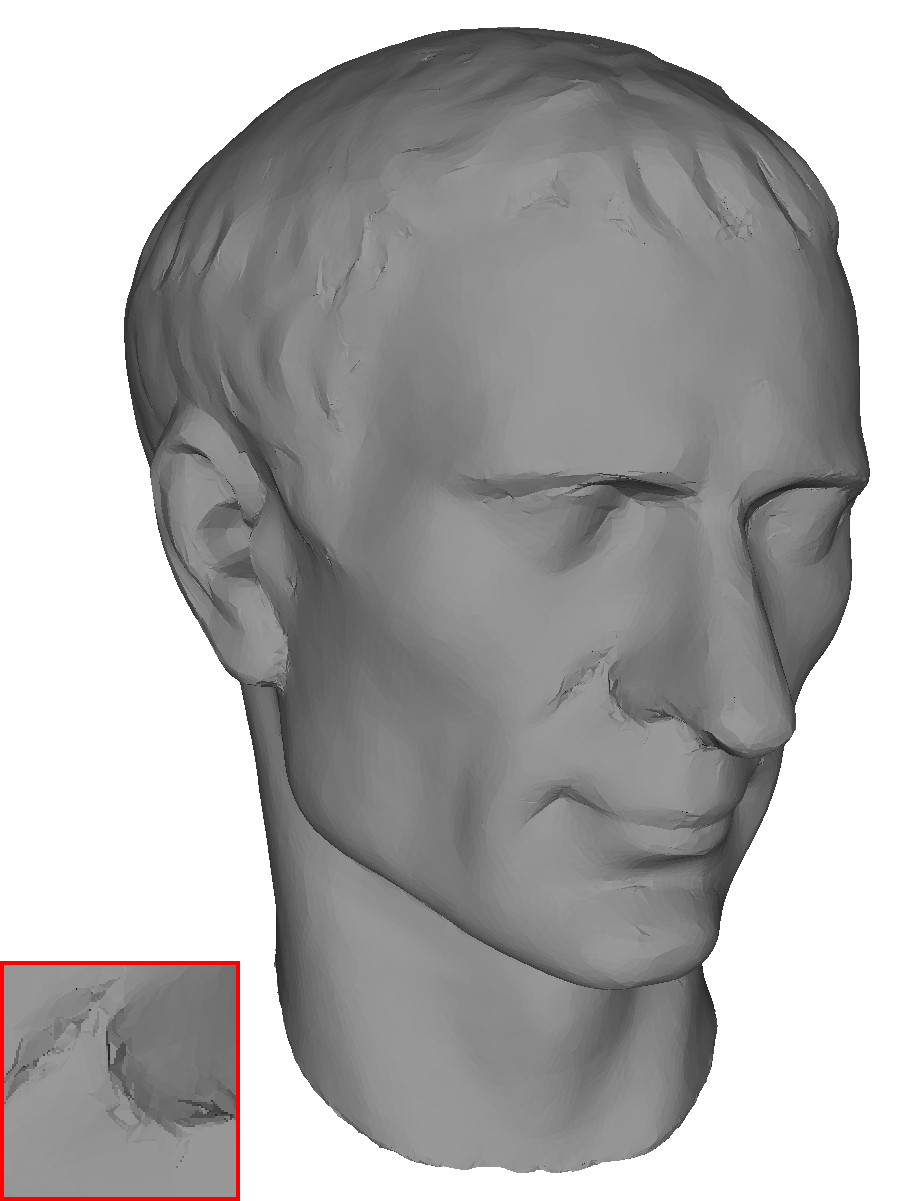}
		\caption{$\beta = 0.2$}
		\label{fig:varying_params_beta-c}
	\end{subfigure}
	\begin{subfigure}{0.18\linewidth}
		\includegraphics[width=\textwidth]{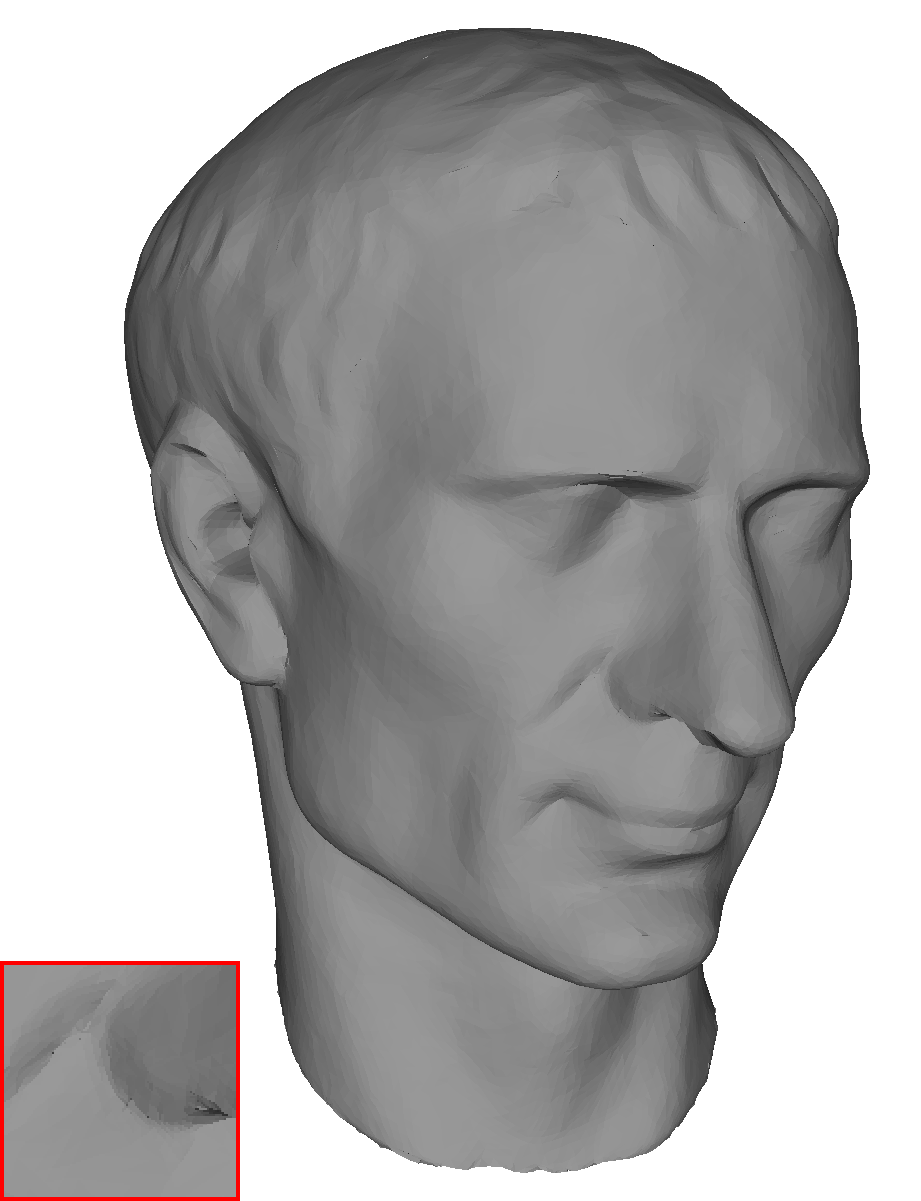}
		\caption{$\beta = 0.5$}
		\label{fig:varying_params_beta-d}
	\end{subfigure}
	\begin{subfigure}{0.18\linewidth}
		\includegraphics[width=\textwidth]{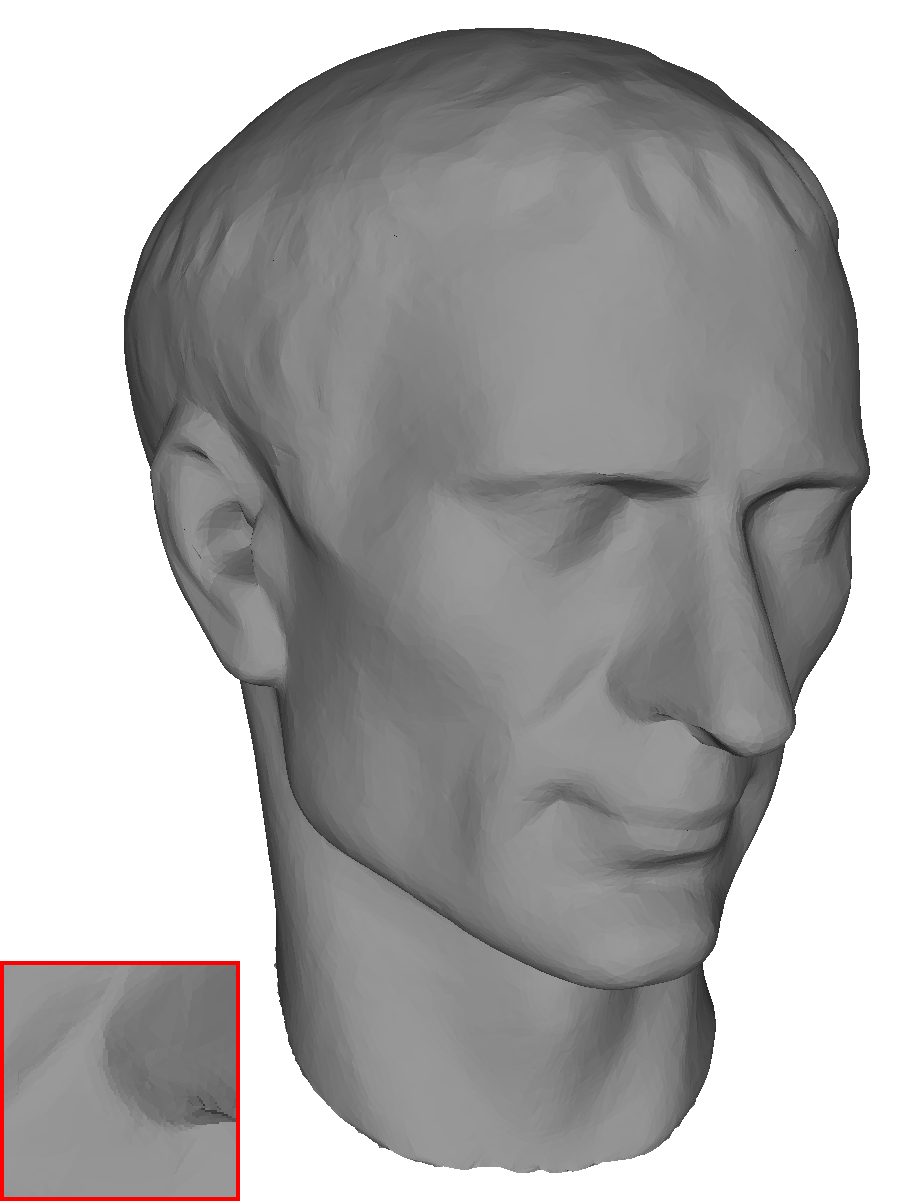}
		\caption{$\beta = 0.8$}
		\label{fig:varying_params_beta-e}
	\end{subfigure}
	\caption{Visual comparison of varying $\beta$ with the fixed $\alpha$ and $\lambda$. The noisy mesh is corrupted by Gaussian noise ($\sigma=0.15\,\bar{l}_e$). Zoom in for better view.} 
    \label{fig:varying_params_beta}
\end{figure*}

A more flexible second-order regularizer is introduced in the TGV-based method~\cite{liu2021mesh}, defined as:
$$
\mathrm{TGV}(\mathrm{u})=\min _{\mathrm{v} \in \mathcal{V}}\left\{\alpha\left\|D_{\mathcal{M}} \mathrm{u} -\mathrm{v}\right\|_{\mathcal{V}}+\beta\|\mathcal{E}\mathrm{v}\|_{\mathcal{W}}\right\},
$$
where $\mathrm{v}$ is an auxiliary variable that decouples the gradient of $\mathrm{u}$, $\mathcal{E}\mathrm{v}$ represents a higher-order differential operator, and $\alpha,\beta>0$ balance the first- and second-order terms. This formulation combines the first- and second-order information and typically improves the recovery of smoothly varying regions while retaining edges better than the pure HO regularizer.

By definition, the HO regularizer extends the TV-based model to the second-order derivative case in the same direction $\left(\partial_i \partial_i u\right)$, which helps alleviate staircase effects and spurious “cracks’’ in locally smooth regions. Likewise, the second-order TGV model simultaneously controls directional second derivatives $\partial_i\partial_i u$ and mixed (symmetric) derivatives $\partial_i\partial_j u+\partial_j\partial_i$. Notice that both HO and TGV regularizers can be viewed as TV-type penalties applied to second-order quantities, yielding convex formulations with favorable mathematical properties; however, the performance of their edge preservation is typically weaker than that of well-designed nonconvex formulations. 

By definition, the proposed semi-sparsity regularizer adopts a higher-order modeling perspective but replaces the conventional HO and TGV regularizers with a strictly sparse penalty on the highest-order differences and a milder (e.g., $L_1$-norm) penalty on the lower-order terms. This hybrid design inherits the advantages of both TV-based models and their higher-order extensions—preserving sharp geometric features while promoting piecewise-polynomial smoothness in regular regions. We note that the semi-sparsity model is a higher-order, nonsmooth, and nonconvex formulation, which is typically more difficult to optimize than convex HO/TGV counterparts.

\section{Experimental Results}
\label{experimental_results}


In this section, we demonstrate the performance of our semi-sparse regularization model for triangulated mesh denoising. We highlight its high-quality fitting ability in both sharpening edges and polynomial smoothing surfaces, which contributes the core of the proposed denoising model to remove small-scale details but preserve important shape features. We present both visual comparison and quantitative evaluation between our method and several state-of-the-art approaches, including total variation (TV) filtering~\cite{zhang2015variational}, higher-order (HO) total variation extension~\cite{liu2019novel}, bilateral filtering (BF)~\cite{zheng2010bilateral} $L_0$ minimization~\cite{he2013mesh}, the cascaded normal regression (CNR) method~\cite{wang2016mesh}, and TGV-based regularization model~\cite{liu2021mesh}. The experimental results are verified on a diverse collection of triangulated meshes, including CAD, non-CAD, and scanned data. For synthetic CAD meshes, we add a zero-mean Gaussian noise with mean edge length $\bar{l}_e$ and standard deviation $\sigma$.  For the fairness, the parameters in each filtering algorithm are either carefully configured with a greedy search strategy to yield visually optimal results or fine-tuned to reach a comparable level of smoothness. All experiments are conducted on a desktop PC with Intel Core i7-9800X 16 core CPU 3.80GHz and 64G RAM.

\subsection{Parameter Settings}

We first investigate how the parameters of the semi–sparsity model influence the mesh denoising behavior. Notice that the formulation~\cref{eq_admm_regularization_form} involves three global parameters, i.e., $\lambda$, $\alpha$, and $\beta$, which control the relative contributions of the data-fidelity, first-order, and second-order regularization terms, respectively. For simplicity, we here assume $\lambda$ is fixed with a suitable value in our experiments, as it primarily serves as a scaling factor to rescale the problem into an appropriate range for stable numerical implementation.

\begin{figure*}[!t]
	\centering
	\begin{subfigure}{0.12\linewidth}
		\includegraphics[width=\textwidth]{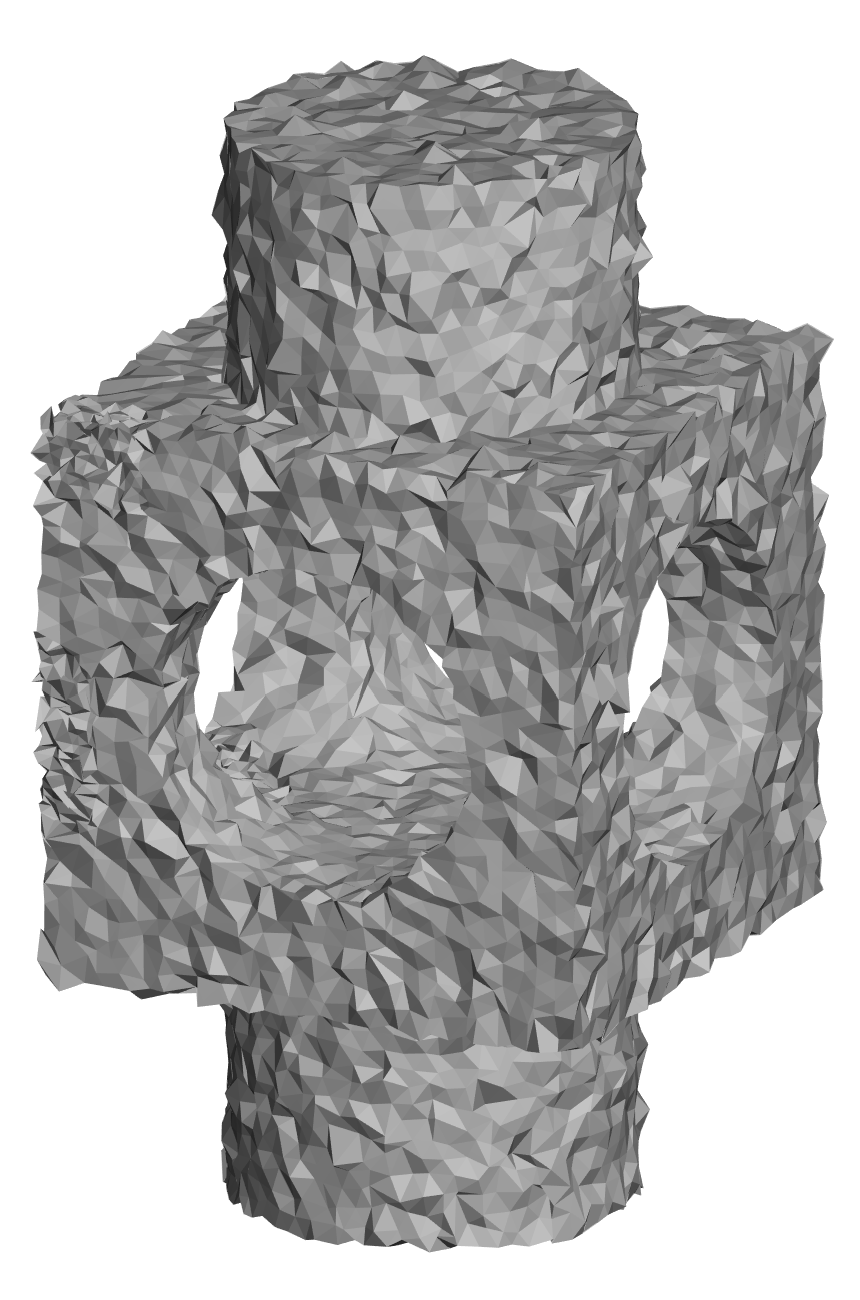}
        \includegraphics[width=\textwidth]{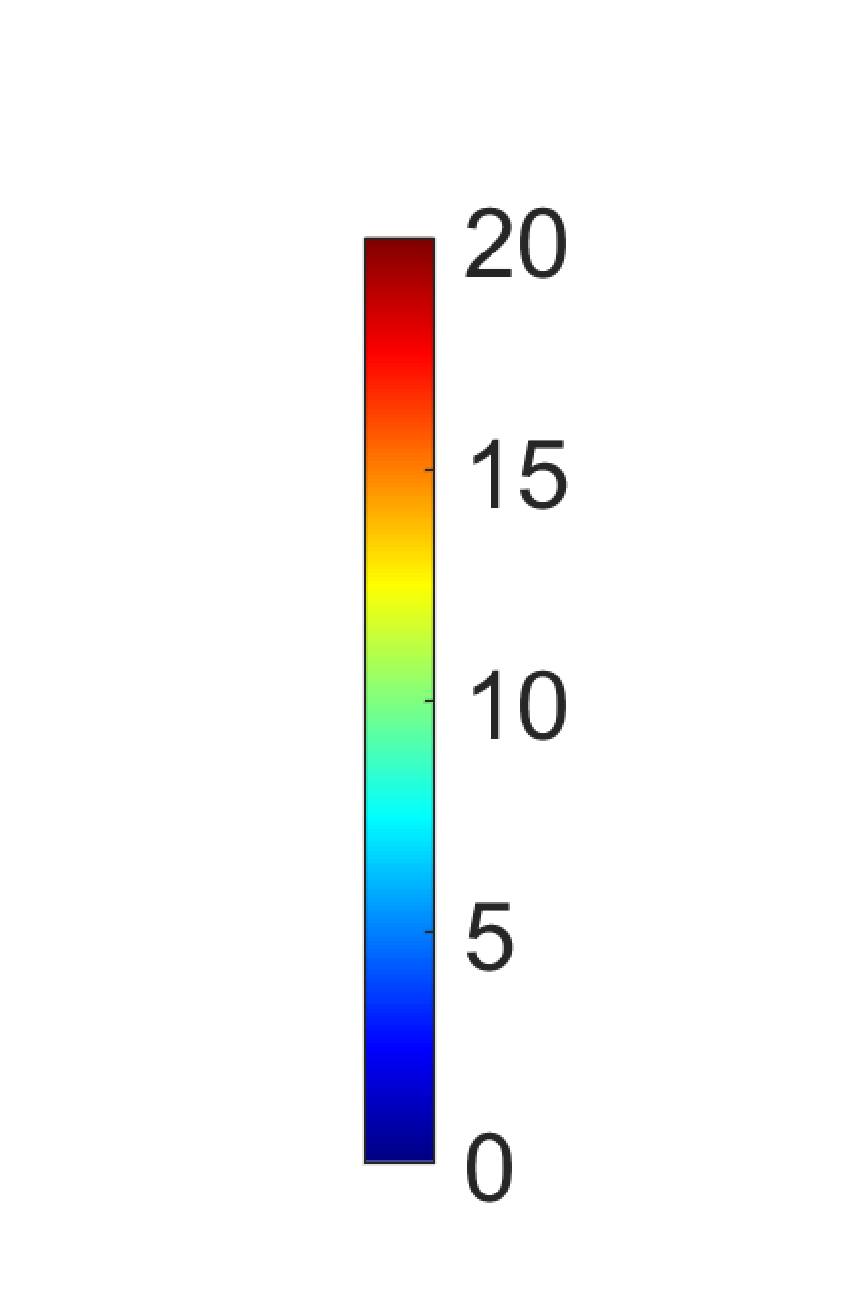} 
		\caption{Noisy}
		\label{fig_block-a}
	\end{subfigure}
	\begin{subfigure}{0.12\linewidth}
		\includegraphics[width=\textwidth]{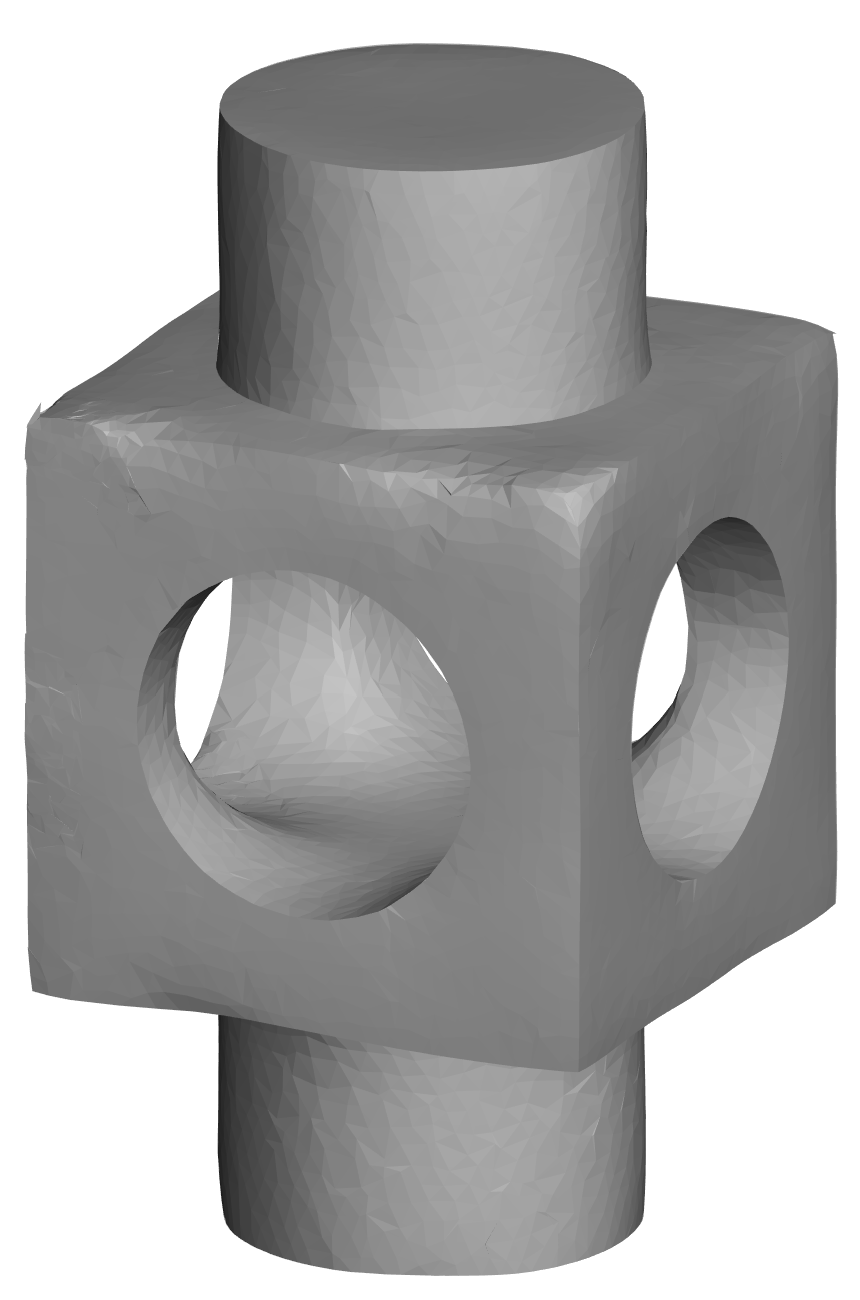}
		\includegraphics[width=\textwidth]{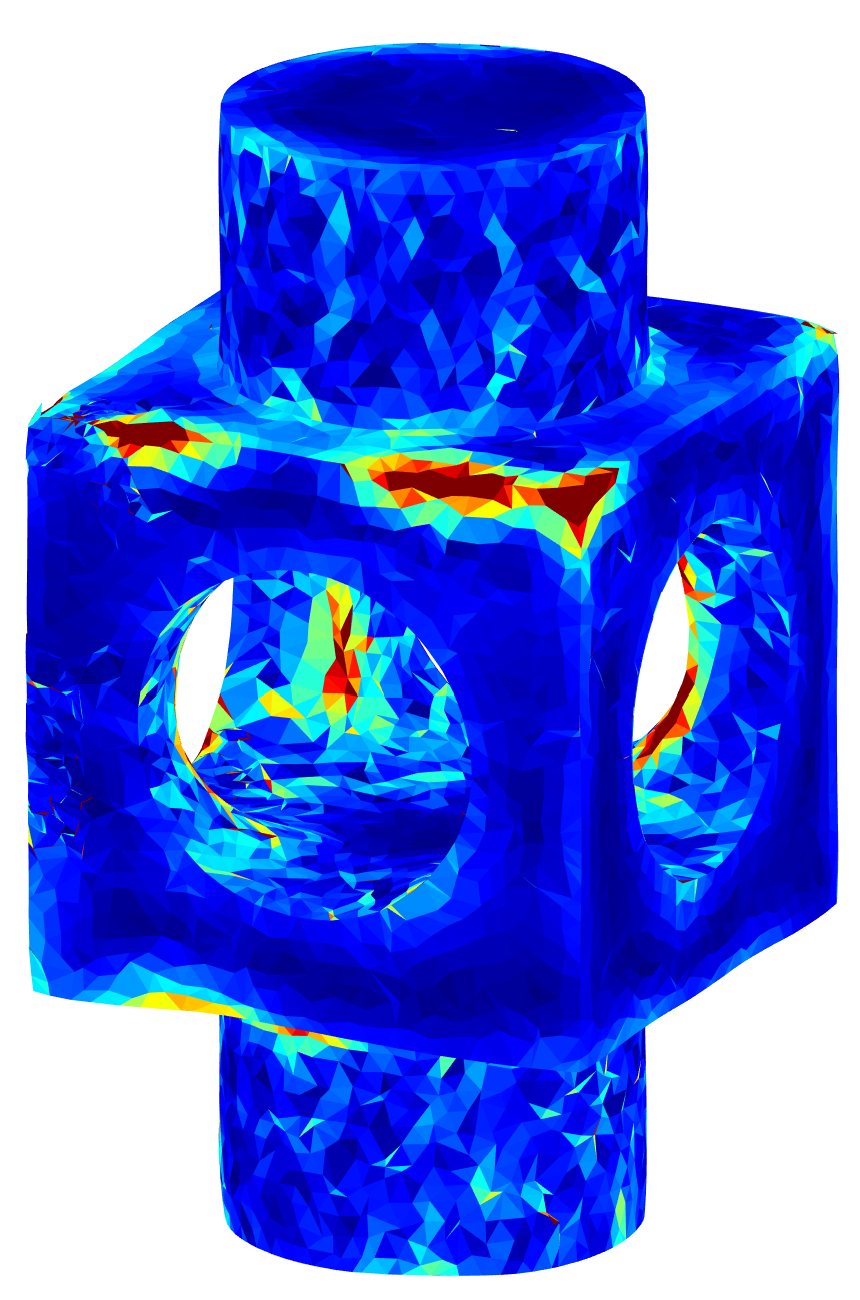}
		\caption{BF}
		\label{fig_block-b}
	\end{subfigure}
	\begin{subfigure}{0.12\linewidth}
		\includegraphics[width=\textwidth]{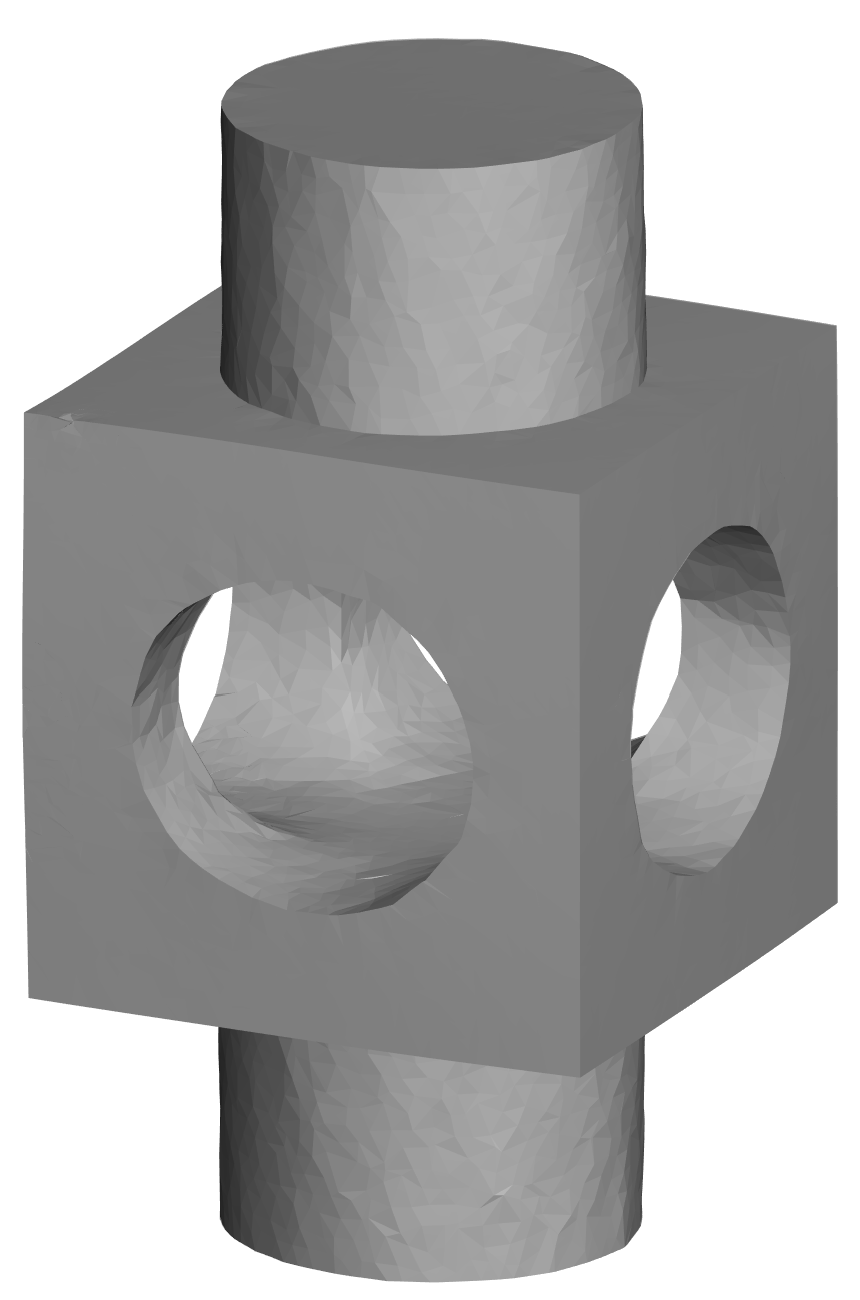}
		\includegraphics[width=\textwidth]{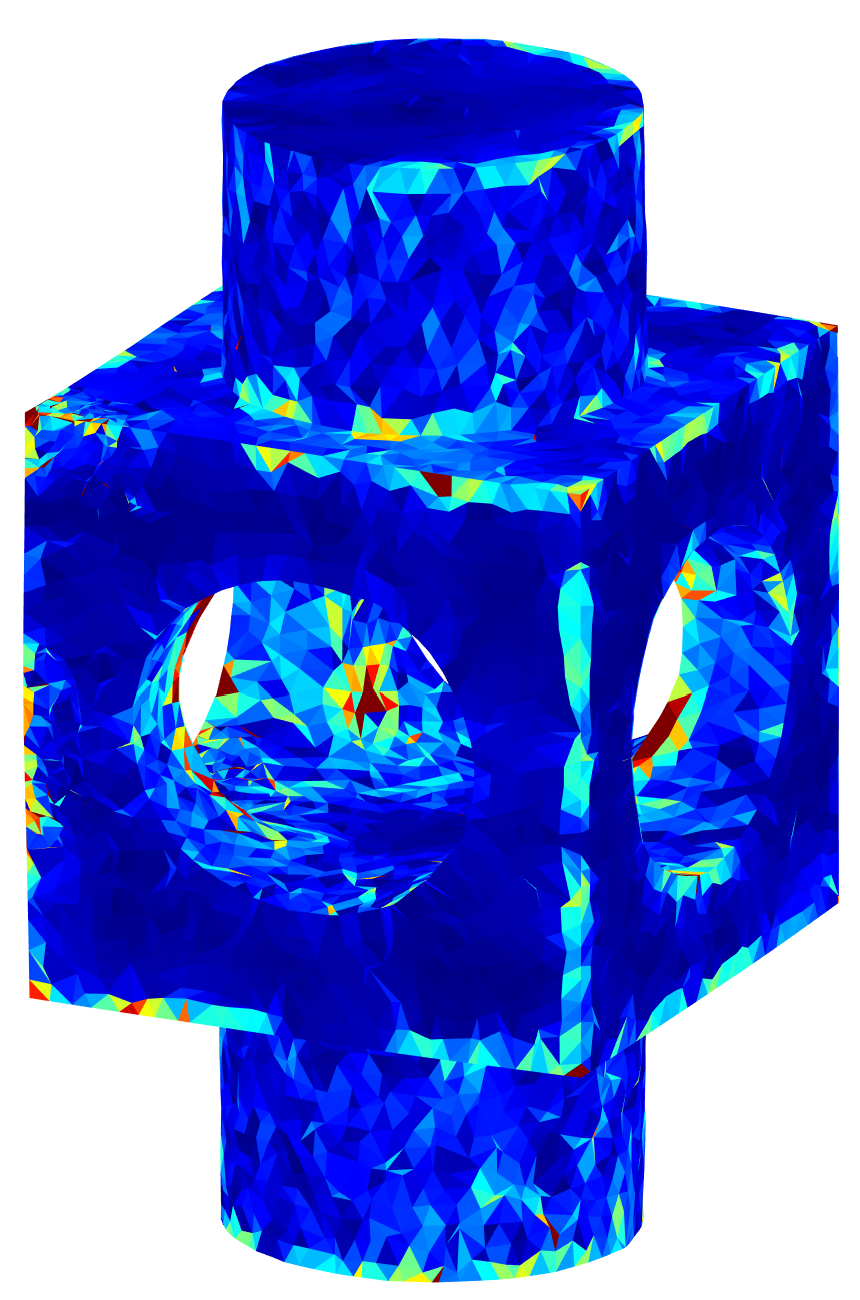}
		\caption{TV}
		\label{fig_block-c}
	\end{subfigure}
	\begin{subfigure}{0.12\linewidth}
		\includegraphics[width=\textwidth]{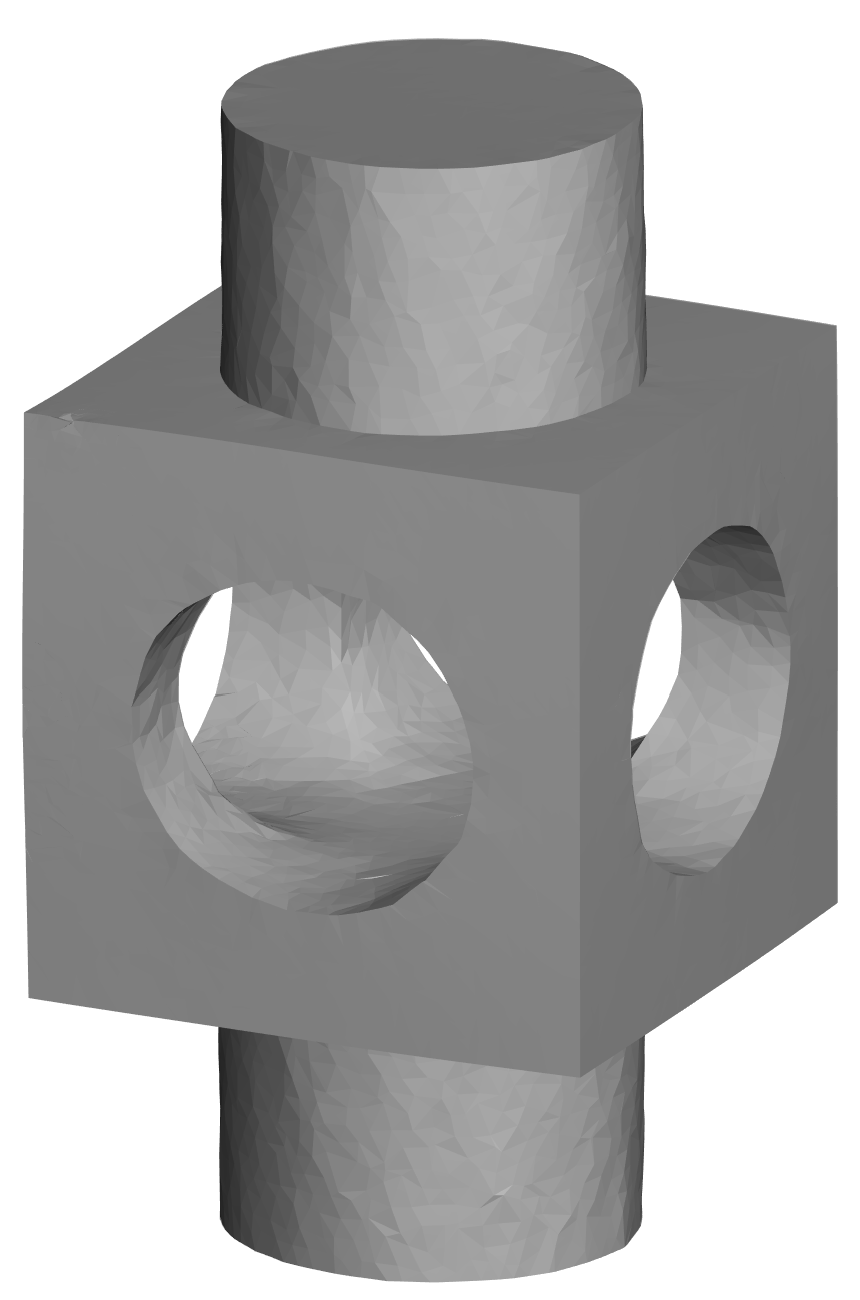}
		\includegraphics[width=\textwidth]{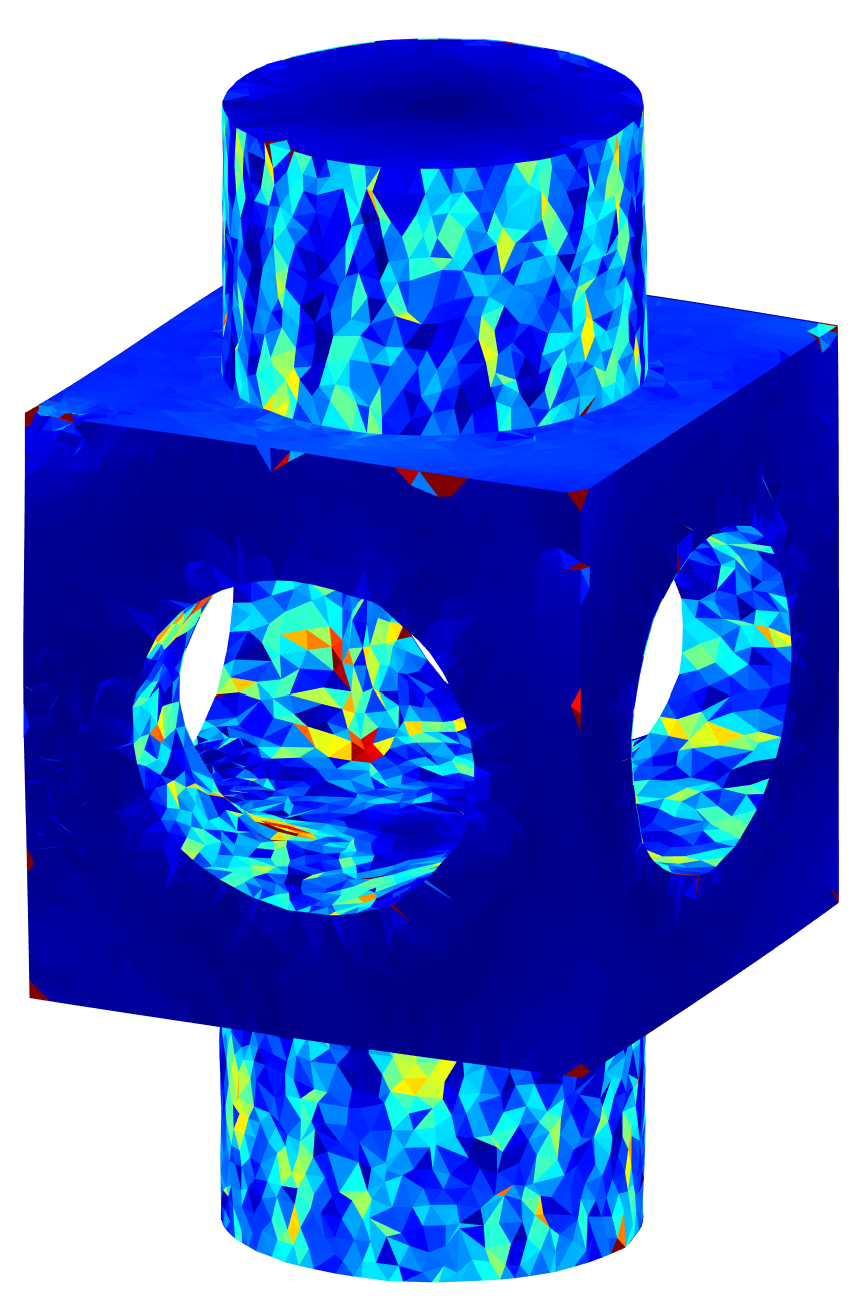}
		\caption{L0}
		\label{fig_block-d}
	\end{subfigure}
	\begin{subfigure}{0.12\linewidth}
		\includegraphics[width=\textwidth]{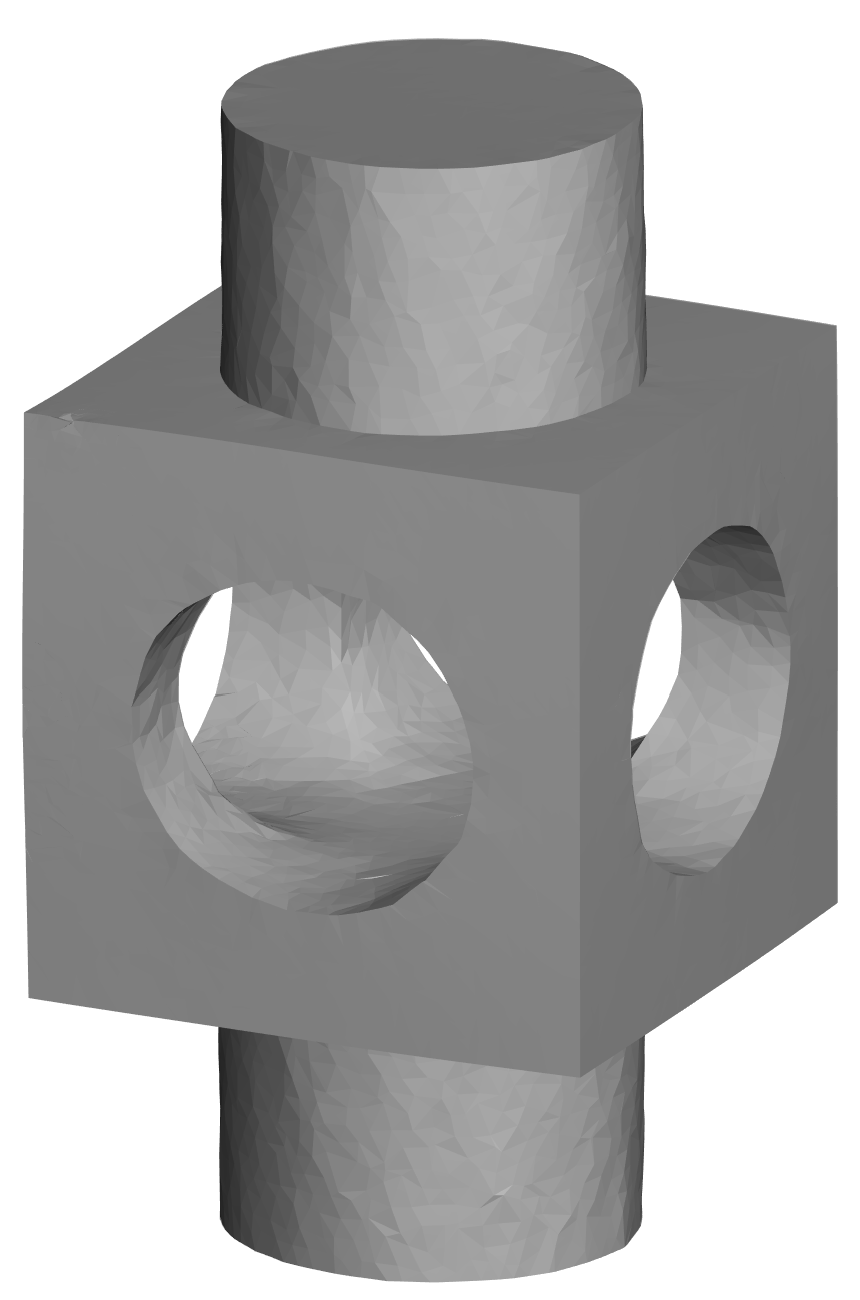}
		\includegraphics[width=\textwidth]{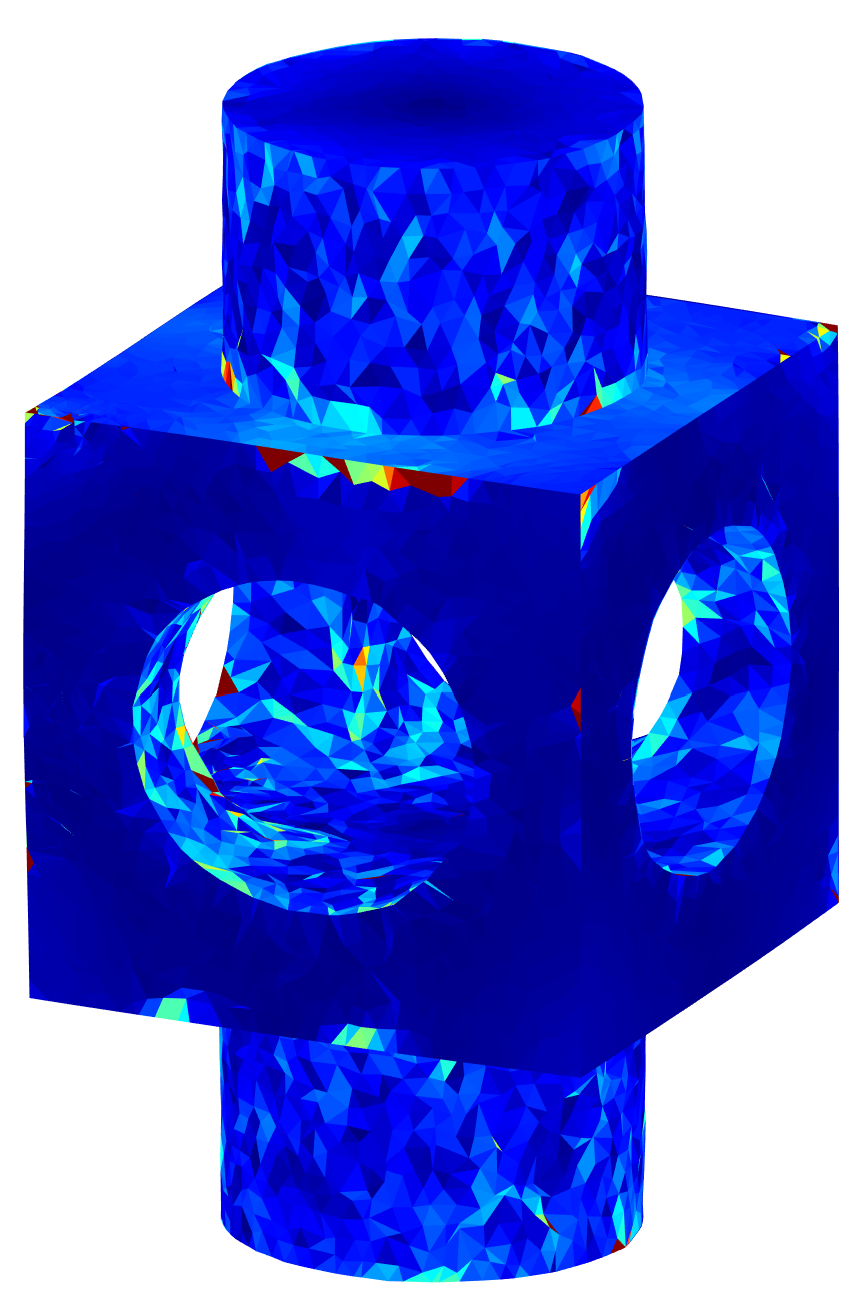}
		\caption{HO}
		\label{fig_block-e}
	\end{subfigure}
    \begin{subfigure}{0.12\linewidth}
		\includegraphics[width=\textwidth]{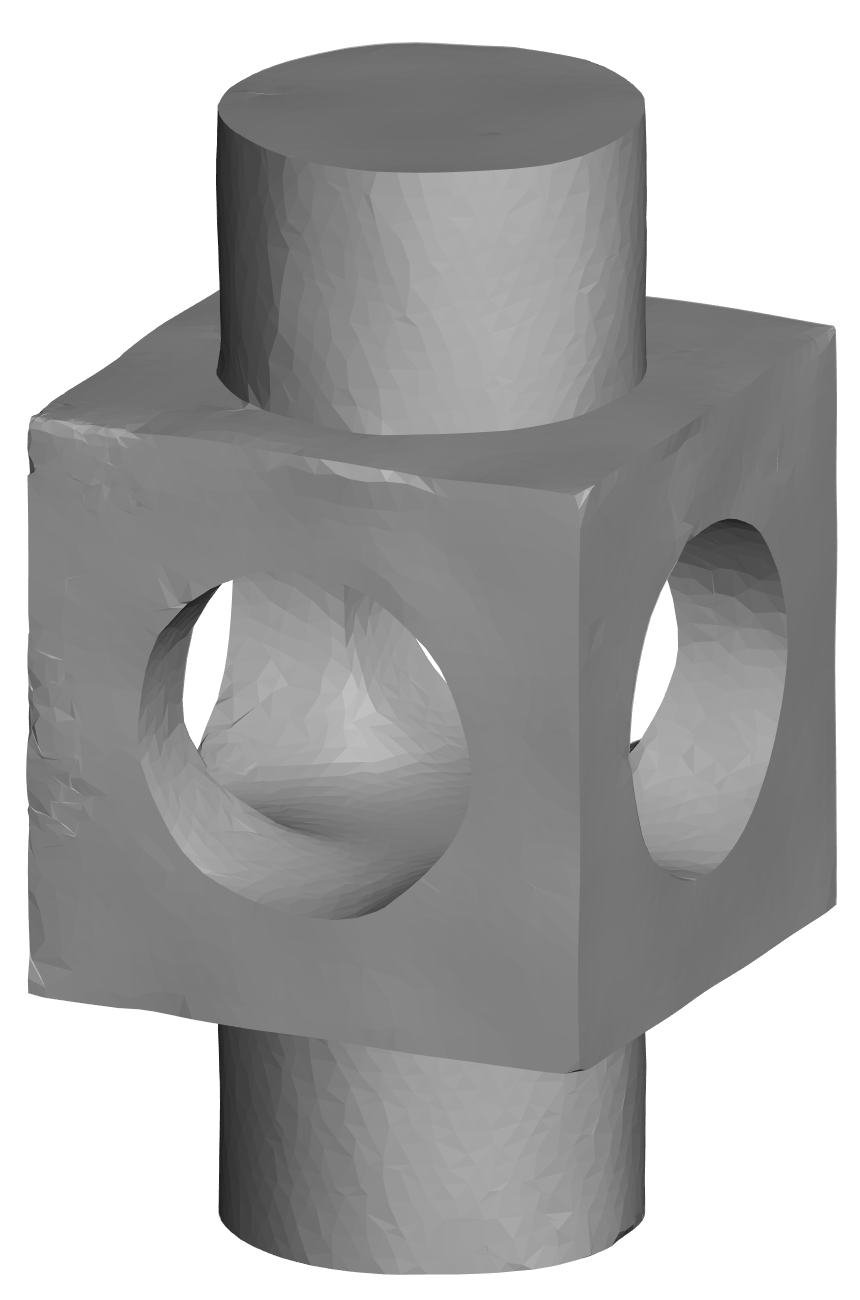}
		\includegraphics[width=\textwidth]{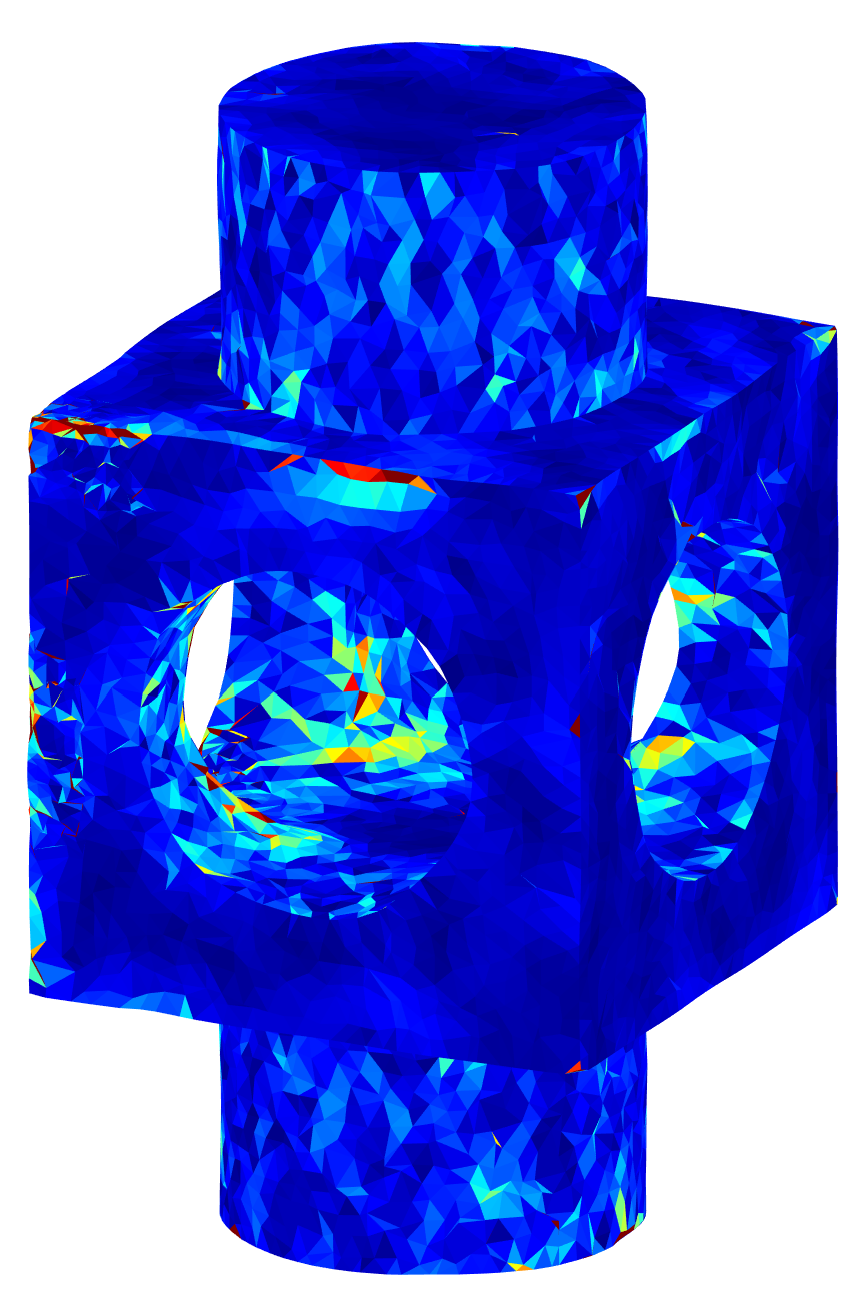}
		\caption{CNR}
		\label{fig_block-f}
	\end{subfigure}
	\begin{subfigure}{0.12\linewidth}
		\includegraphics[width=\textwidth]{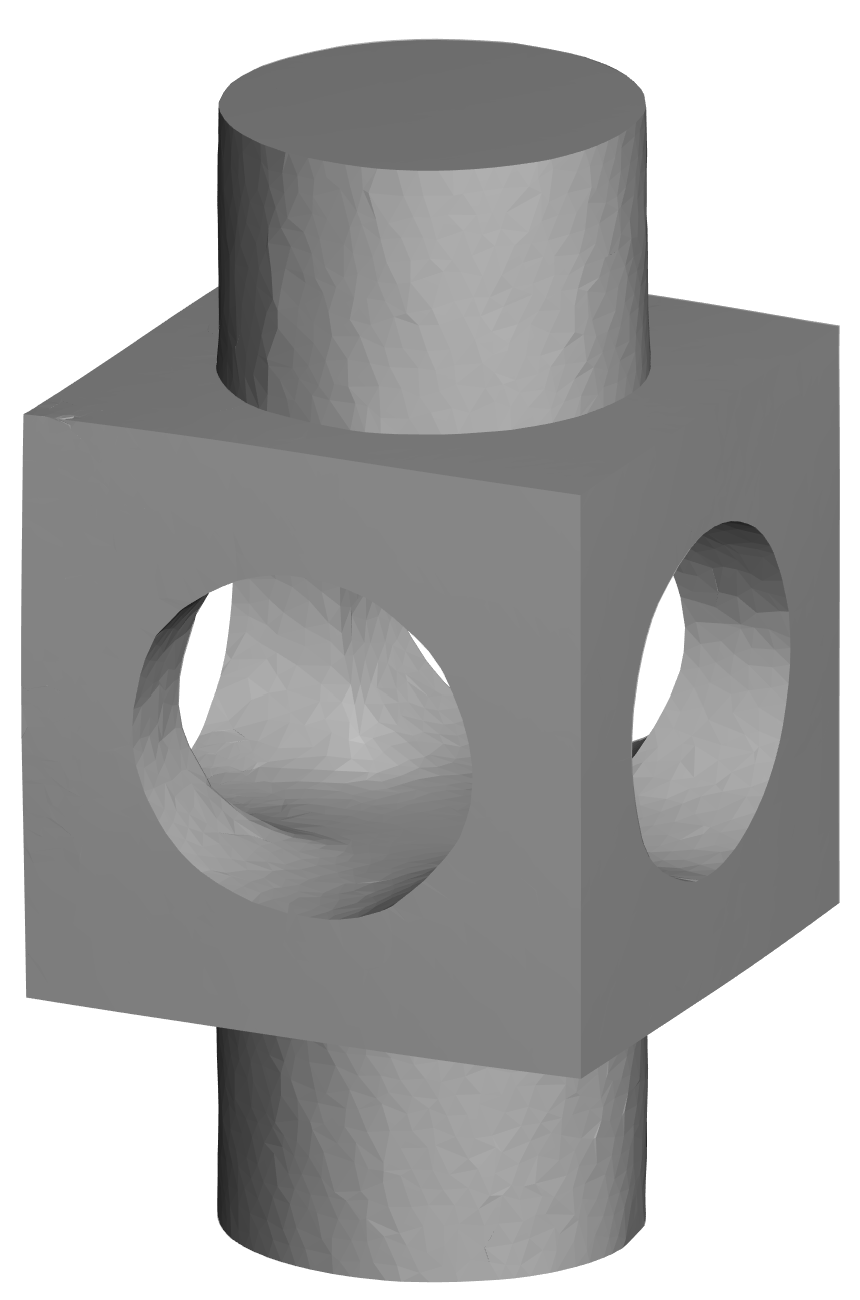}
		\includegraphics[width=\textwidth]{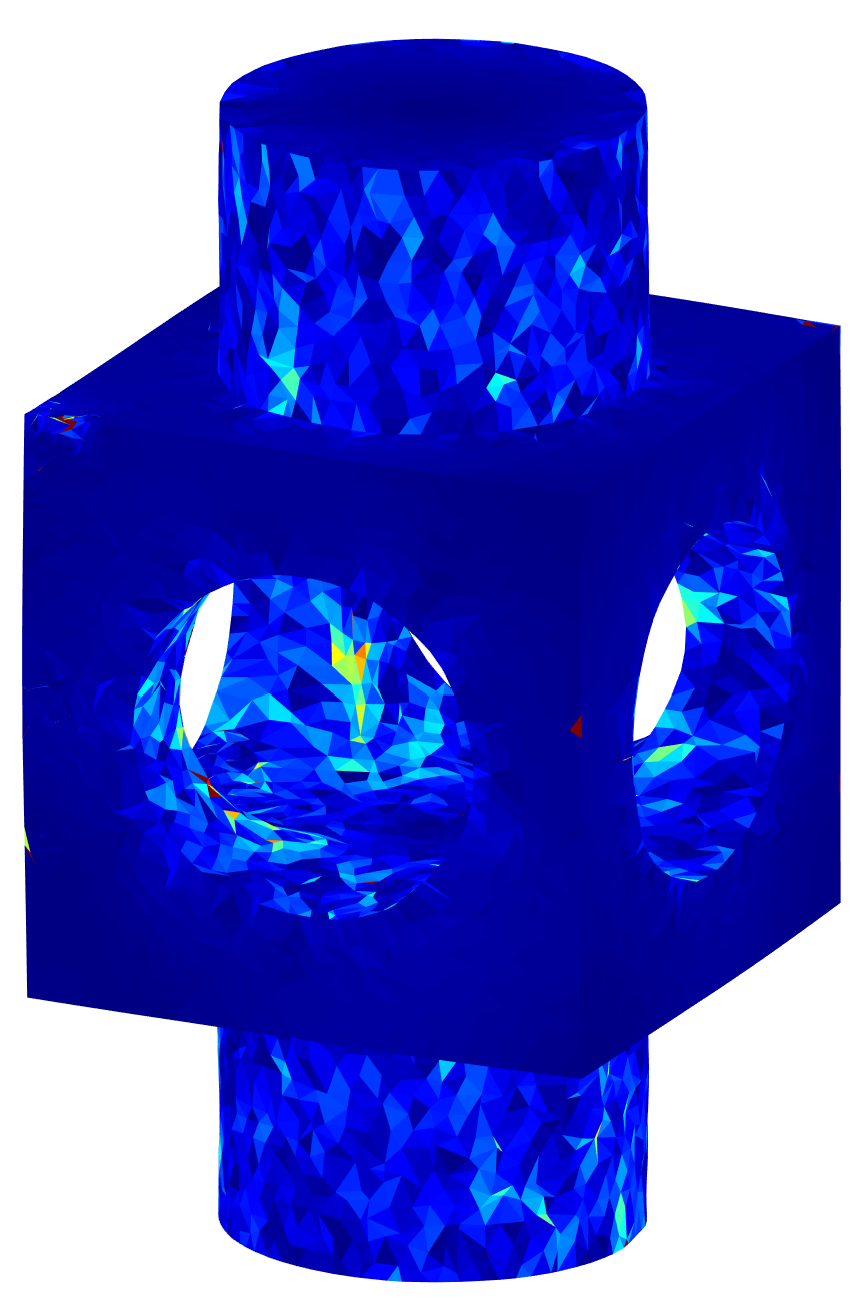}
		\caption{TGV}
		\label{fig_block-g}
	\end{subfigure}
	\begin{subfigure}{0.12\linewidth}
		\includegraphics[width=\textwidth]{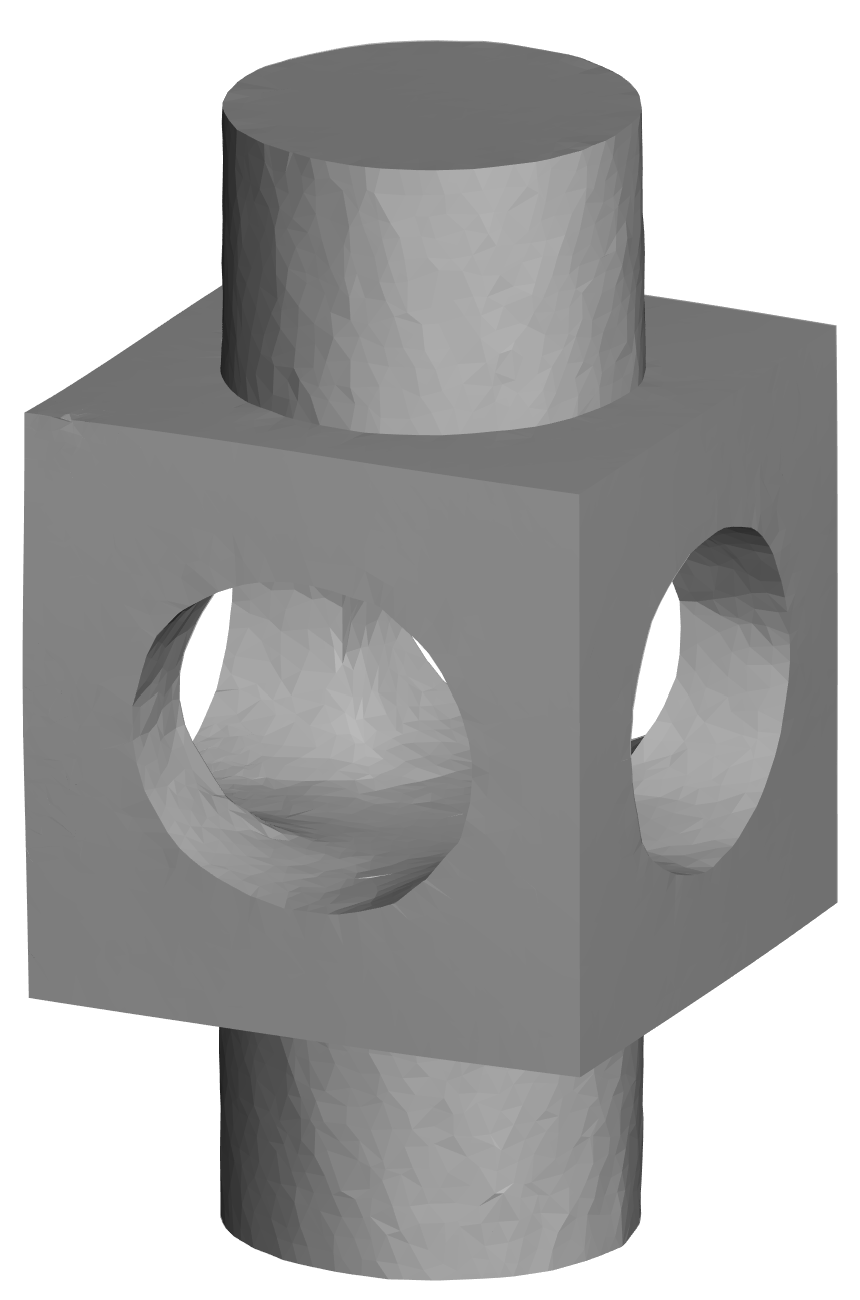}
		\includegraphics[width=\textwidth]{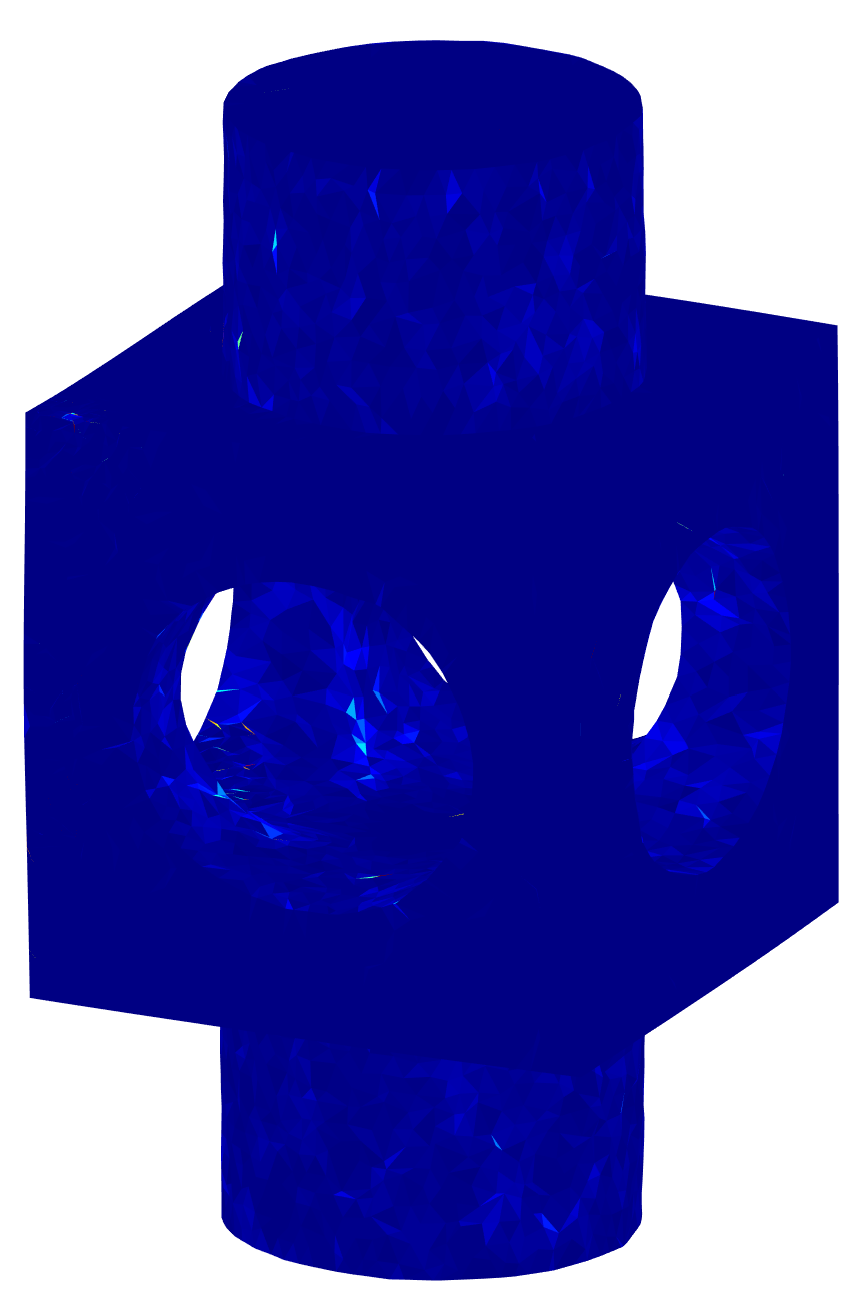}
		\caption{Ours}
		\label{fig_block-h}
	\end{subfigure}
	\caption{Visual comparison of different denoising methods under noise level $\sigma = 0.35\, \bar{l}_e$. The second row shows the corresponding error maps of the angular difference between the face normals of the denoised meshes and the ground truth. (Zoom in for a better view.)}
	\label{fig_block}
\end{figure*}


Intuitively, $\alpha$ governs the strength of the first-order regularization term. As shown in \cref{fig:varying_params_alpha}, for fixed $\lambda$ and $\beta$, Gaussian noise is progressively attenuated as $\alpha$ increases. Moreover, the proposed semi-sparse model effectively smooths mild surface fluctuations while preserving strong geometric features even for relatively large $\alpha$, in contrast to the TGV-based regularization method~\cite{liu2021mesh}. This behavior aligns with the limiting case $\alpha \!\to\! 0$, where the first-order contribution becomes negligible and the model degenerates into a purely second-order scheme. In this regime, the regularizer favors piecewise-polynomial smoothing, provided that the remaining parameters are chosen appropriately. A similar trend is observed in~\cref{fig:varying_params_beta} as $\beta$ increases: the model enhances local feature refinement while inducing comparatively less global smoothing. In the limiting case $\beta \!\to\! 0$, the second-order term vanishes, and the formulation reduces to a TV-based filtering model~\cite{zhang2015variational}, which is well known for its strong edge-preserving property but may introduce staircase artifacts and spurious discontinuities in otherwise smooth regions.

\begin{figure*}[!t]
\centering
\begin{subfigure}{0.12\linewidth}
    \includegraphics[width=\textwidth]{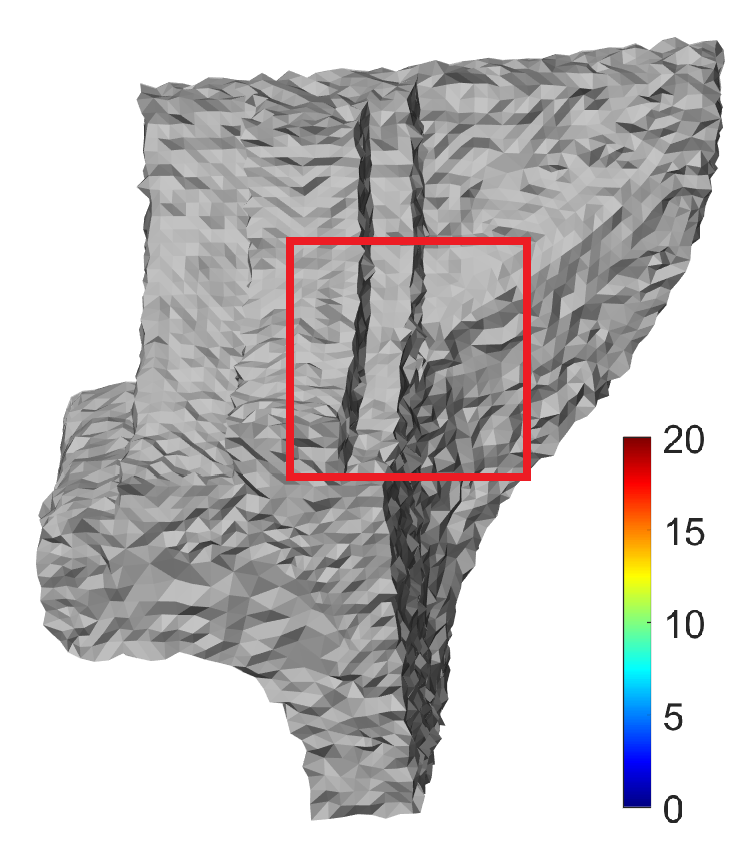}
    \caption{Noisy}
    \label{fig_fandisk-a}
\end{subfigure}
\begin{subfigure}{0.12\linewidth}
    \includegraphics[width=\textwidth]{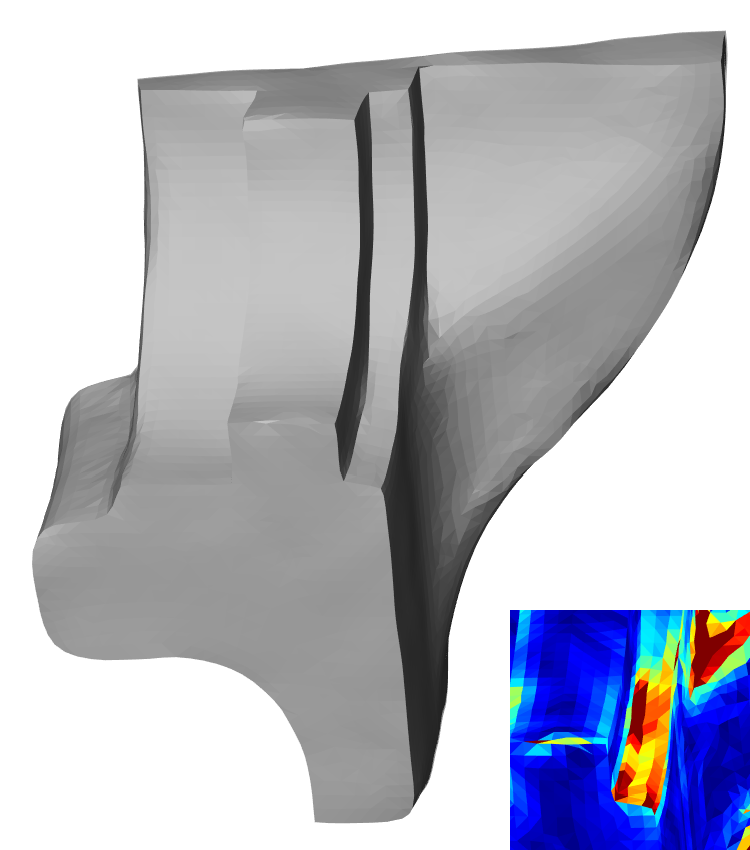} 
    \caption{BF~\cite{zheng2010bilateral}}
    \label{fig_fandisk-b}
\end{subfigure}
\begin{subfigure}{0.12\linewidth}
    \includegraphics[width=\textwidth]{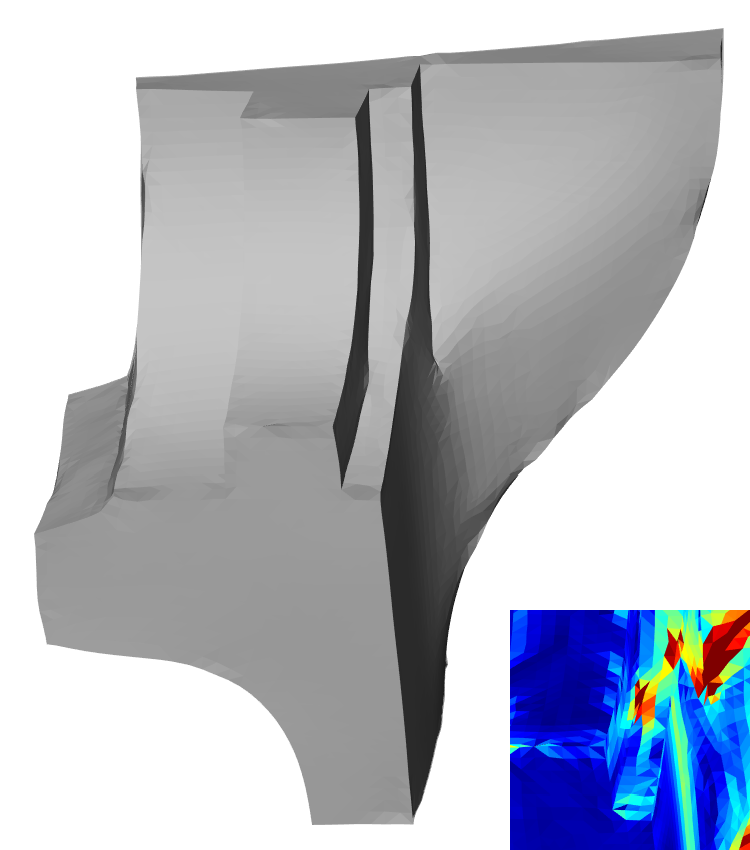}
    \caption{TV~\cite{zhang2015variational}}
    \label{fig_fandisk-c}
\end{subfigure}
\begin{subfigure}{0.12\linewidth}
    \includegraphics[width=\textwidth]{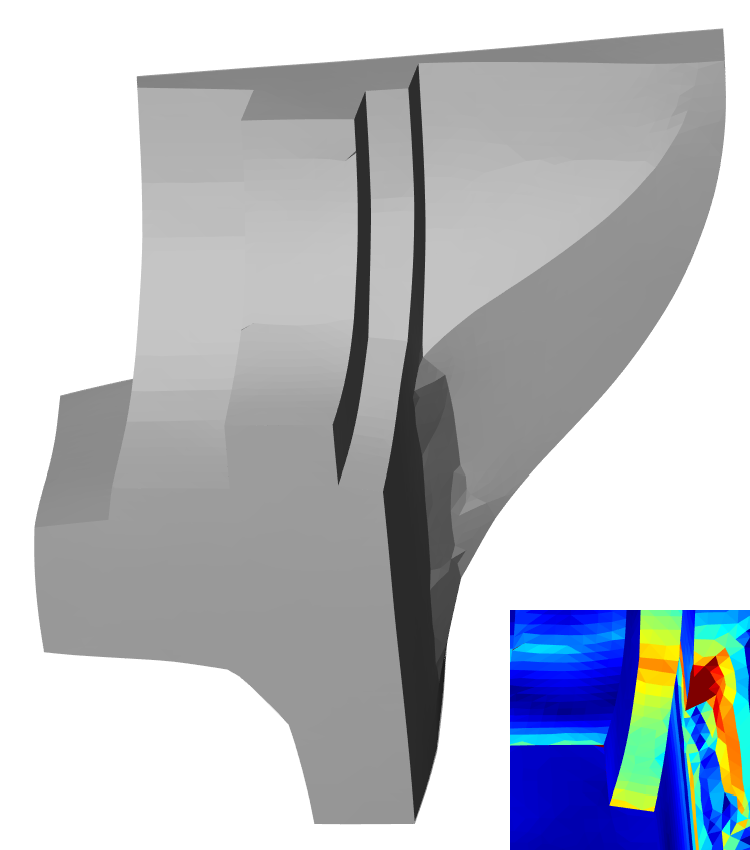}
    \caption{L0~\cite{he2013mesh}}
    \label{fig_fandisk-d}
\end{subfigure}
\begin{subfigure}{0.12\linewidth}
    \includegraphics[width=\textwidth]{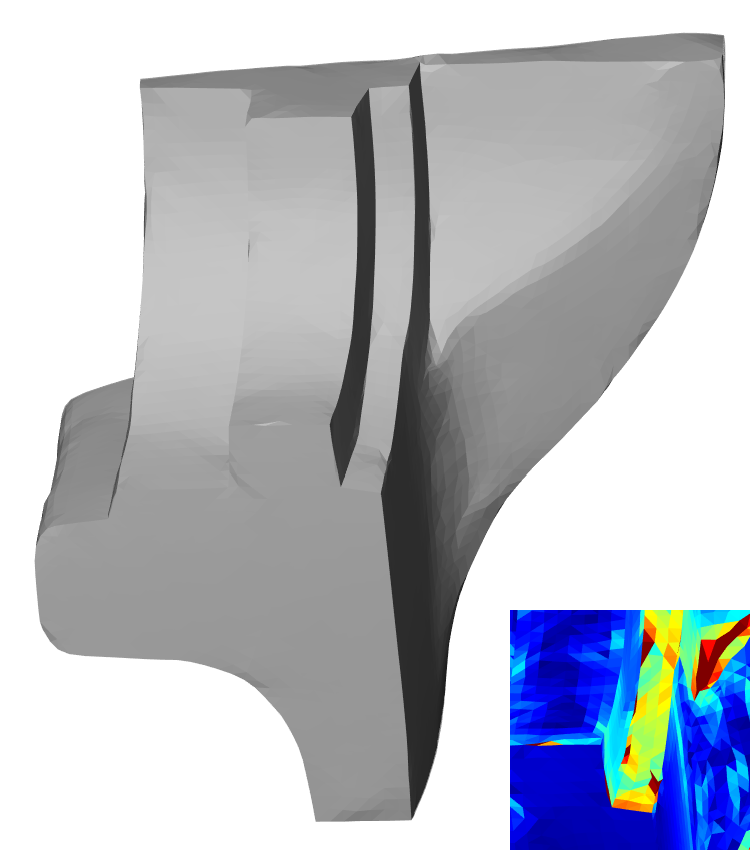}
    \caption{HO~\cite{liu2019novel}}
    \label{fig_fandisk-e}
\end{subfigure}
    \begin{subfigure}{0.12\linewidth}
    \includegraphics[width=\textwidth]{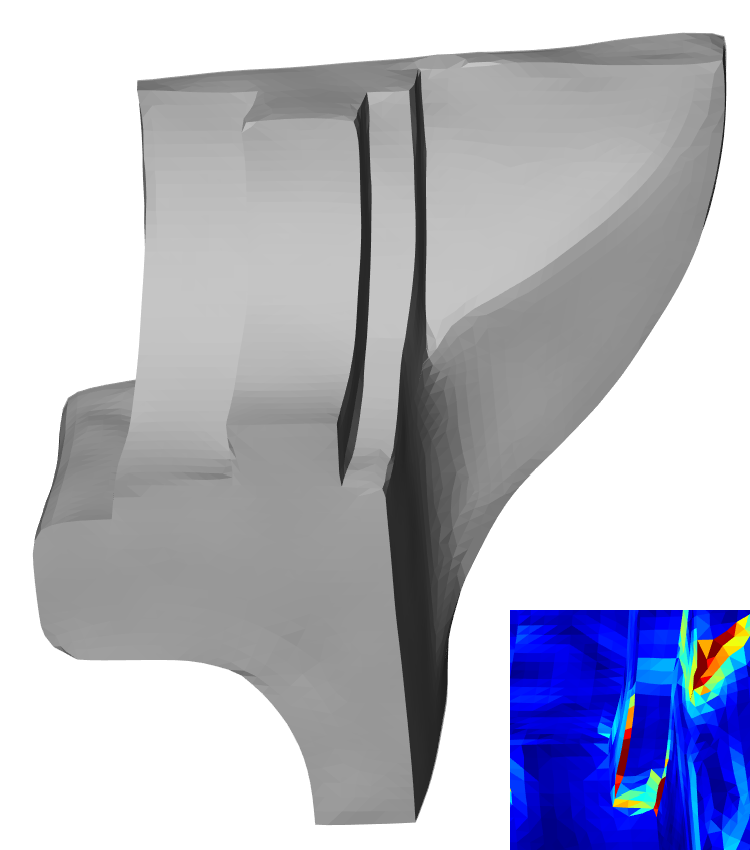}
    \caption{CNR~\cite{wang2016mesh}}
    \label{fig_fandisk-f}
\end{subfigure}
\begin{subfigure}{0.12\linewidth}
    \includegraphics[width=\textwidth]{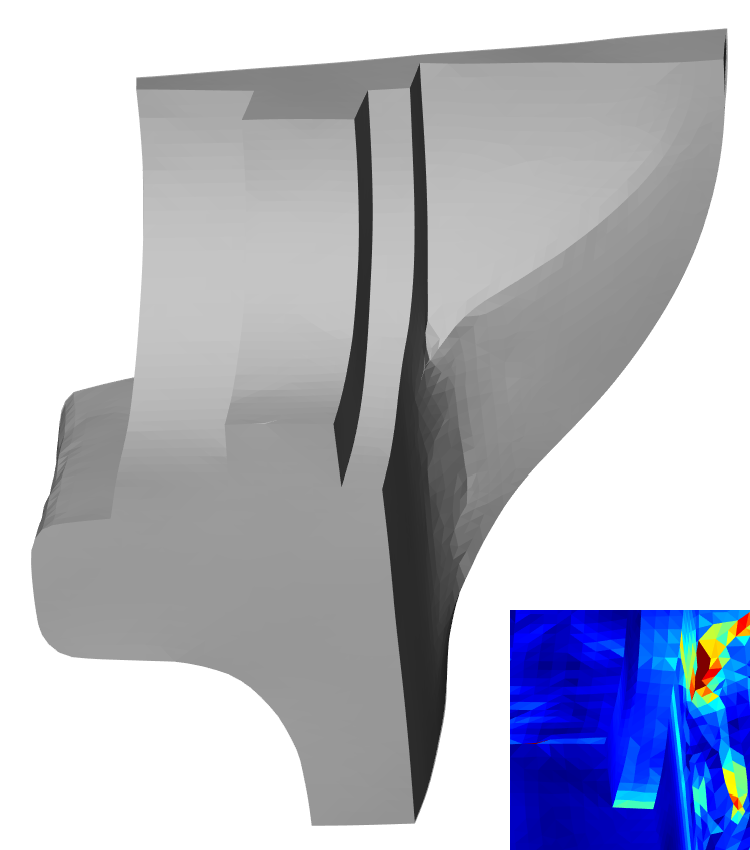}
    \caption{TGV~\cite{liu2021mesh}}
    \label{fig_fandisk-g}
\end{subfigure}
\begin{subfigure}{0.12\linewidth}
    \includegraphics[width=\textwidth]{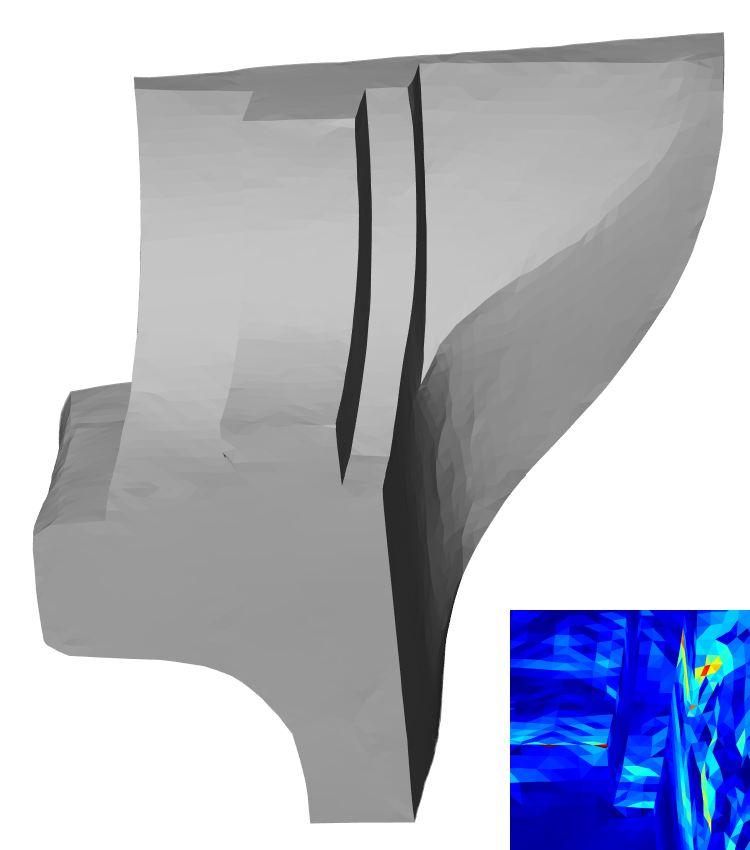}
    \caption{Ours}
    \label{fig_fandisk-h}
\end{subfigure}
\caption{Visual comparison of different denoising methods under noise level $\sigma = 0.25\, \bar{l}_e$. (Zoom in for a better view.)}
\label{fig_fandisk}
\end{figure*}

We also observe that $\alpha$ and $\beta$ act relatively independently in the proposed model in contrast to TGV-based regularization~\cite{liu2021mesh}, where the two weights are tightly coupled. This independence offers a practical tuning strategy: one may first select $\alpha$ (e.g., following the TV-based guideline according to the noise level) and subsequently fine-tune $\beta$ to achieve a desirable balance between sharp-feature preservation and smooth-surface recovery. This sequential adjustment yields robust and stable denoising performance in practice. It is also worth noting that excessively large values of $\alpha$ or $\beta$ tend to oversuppress fine-scale geometric details, leading to over-smoothing effects.

\subsection{Visual Comparison}

We now compare the proposed semi-sparsity model with several existing methods and show its advantage on a diverse collection of triangulated meshes, including CAD models, non-CAD shapes, and raw scanned data.

\textbf{CAD Surfaces:} We first evaluate the denoising performance on a CAD surface containing both sharp features and smooth regions. As shown in~\cref{fig_block}, the BF method~\cite{zheng2010bilateral} suffers from noticeable over-smoothing around strong edges and corners, whereas the other methods achieve better edge feature preservation. However, both TV-based regularization~\cite{zhang2015variational} and $L_0$ minimization~\cite{he2013mesh} tend to introduce crease edges in in smoothly varying surface regions. These artifacts are substantially reduced by the higher-order (HO) method~\cite{liu2019novel}, the TGV-based model~\cite{liu2021mesh}, and our semi-sparse model. This also demonstrate the benefit of higher-order regularization. The learn-based CNR method~\cite{wang2016mesh} also received the results comparable to the TGV-based results, which indicates the potential of learning-based methods for high-quality performance. Compared with $L_{0}$ minimization~\cite{he2013mesh}, our semi-sparse model markedly suppresses flattened false features. This attributes to the strict sparsity prior imposed on the highest-order differences. Meanwhile, our method also achieves better edge preservation along sharp features than both the HO method~\cite{liu2019novel} and TGV-based~\cite{liu2021mesh} regularization due to the strict higher-order sparse regularization. For all methods, we also present the normal error maps (angular differences between filtered and ground-truth normals). These error maps further verify that our method produces normals that are consistently closer to the ground truth than all the compared methods.

\begin{figure*}[!t]
	\centering
	\begin{subfigure}{0.12\linewidth}
        \includegraphics[width=\textwidth]{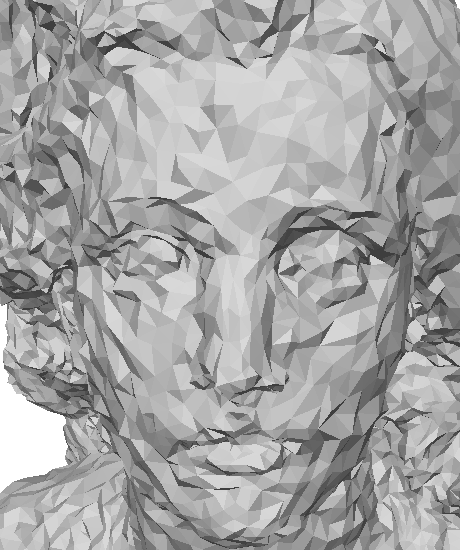} 
        \includegraphics[width=\textwidth]{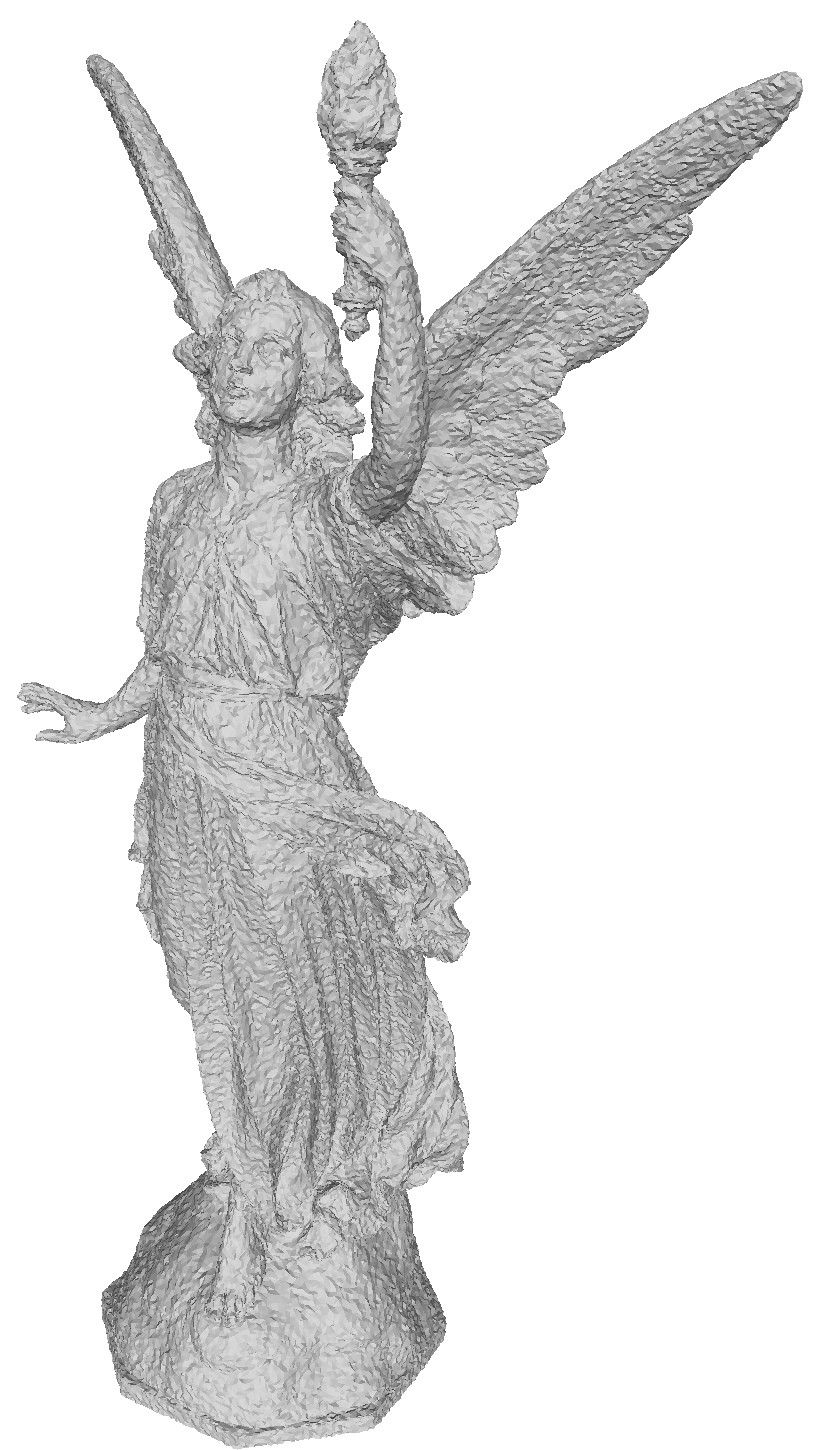}
		\caption{Noisy}
		\label{fig_lucy-a}
	\end{subfigure}
	\begin{subfigure}{0.12\linewidth}
		\includegraphics[width=\textwidth]{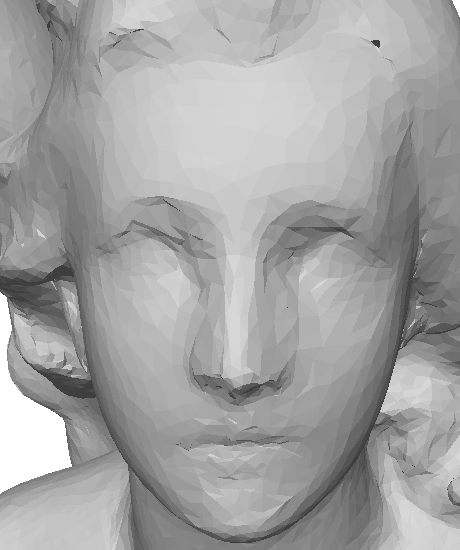}
		\includegraphics[width=\textwidth]{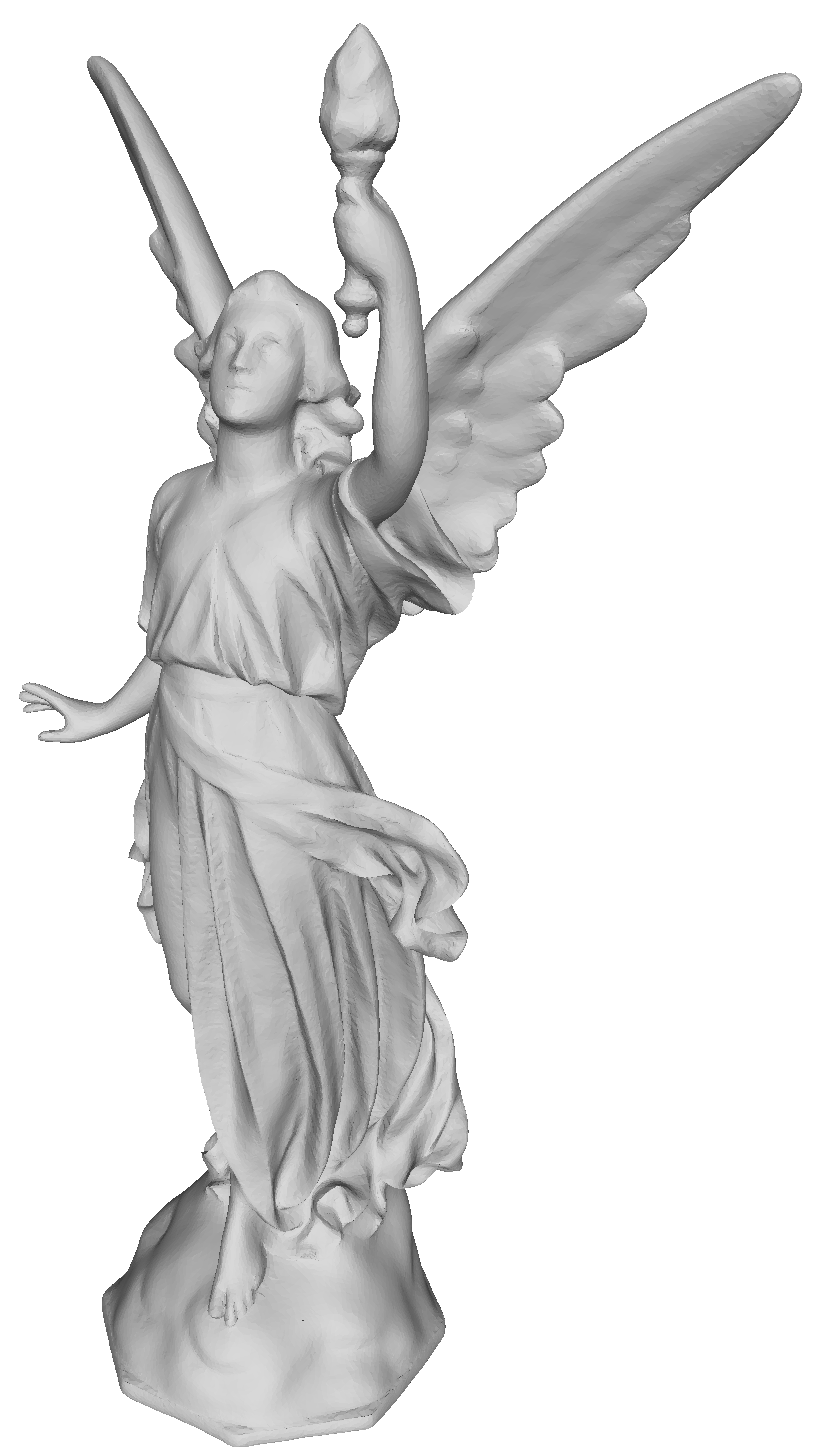} 
		\caption{BF}
		\label{fig_lucy-b}
	\end{subfigure}
	\begin{subfigure}{0.12\linewidth}
		\includegraphics[width=\textwidth]{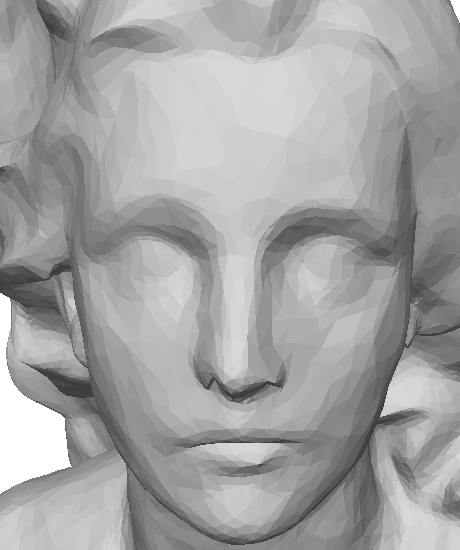}
		\includegraphics[width=\textwidth]{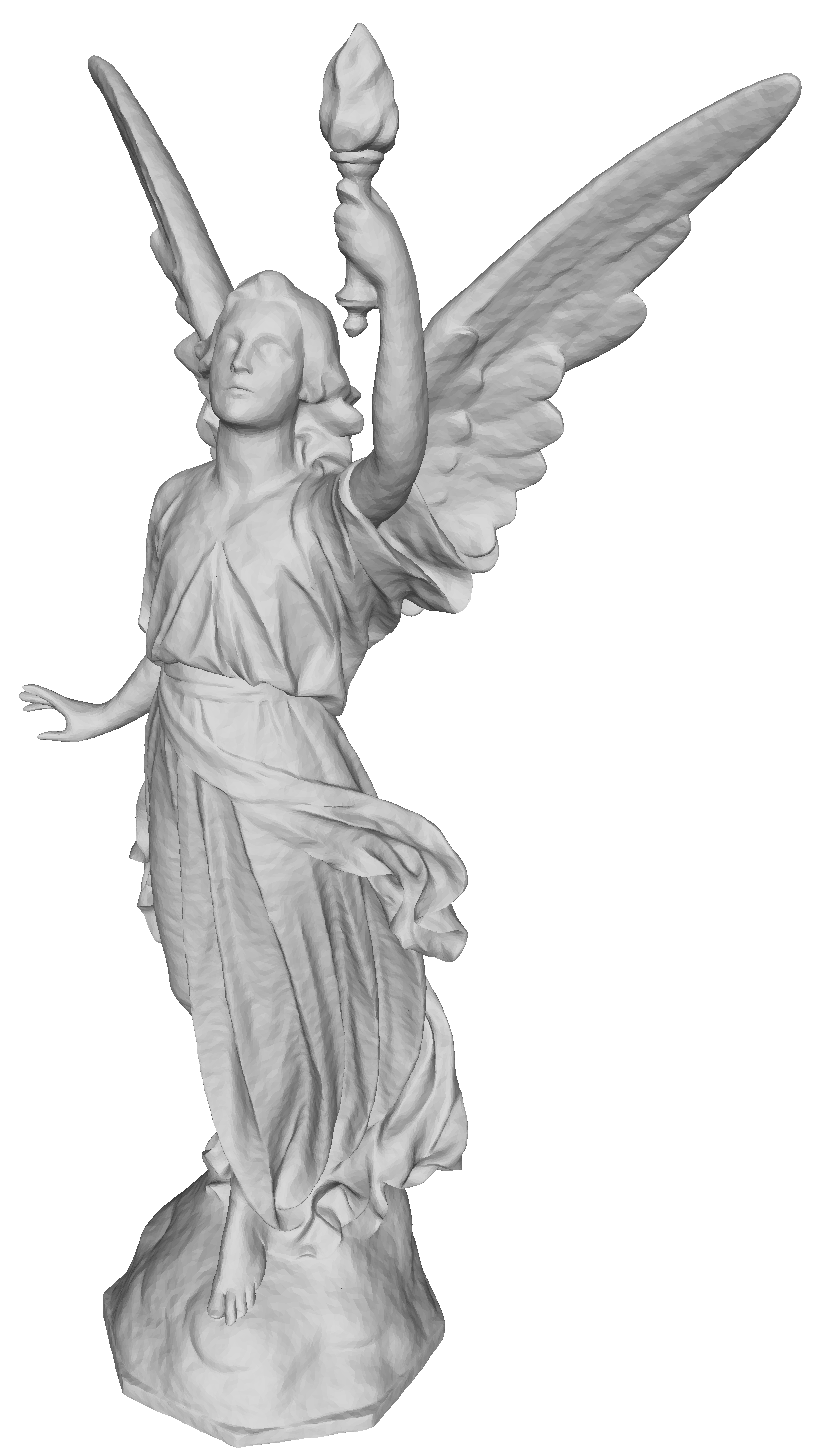}
		\caption{TV}
		\label{fig_lucy-c}
	\end{subfigure}
	\begin{subfigure}{0.12\linewidth}
		\includegraphics[width=\textwidth]{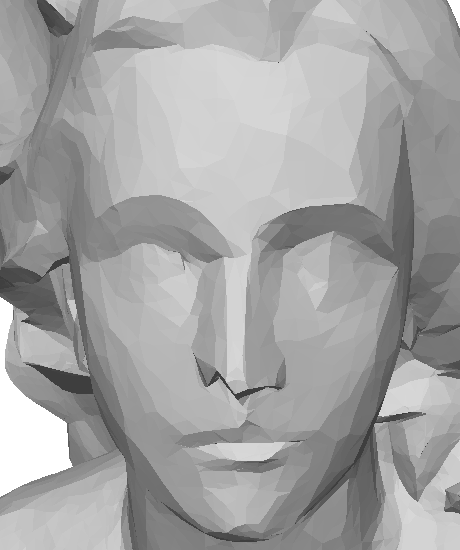}
		\includegraphics[width=\textwidth]{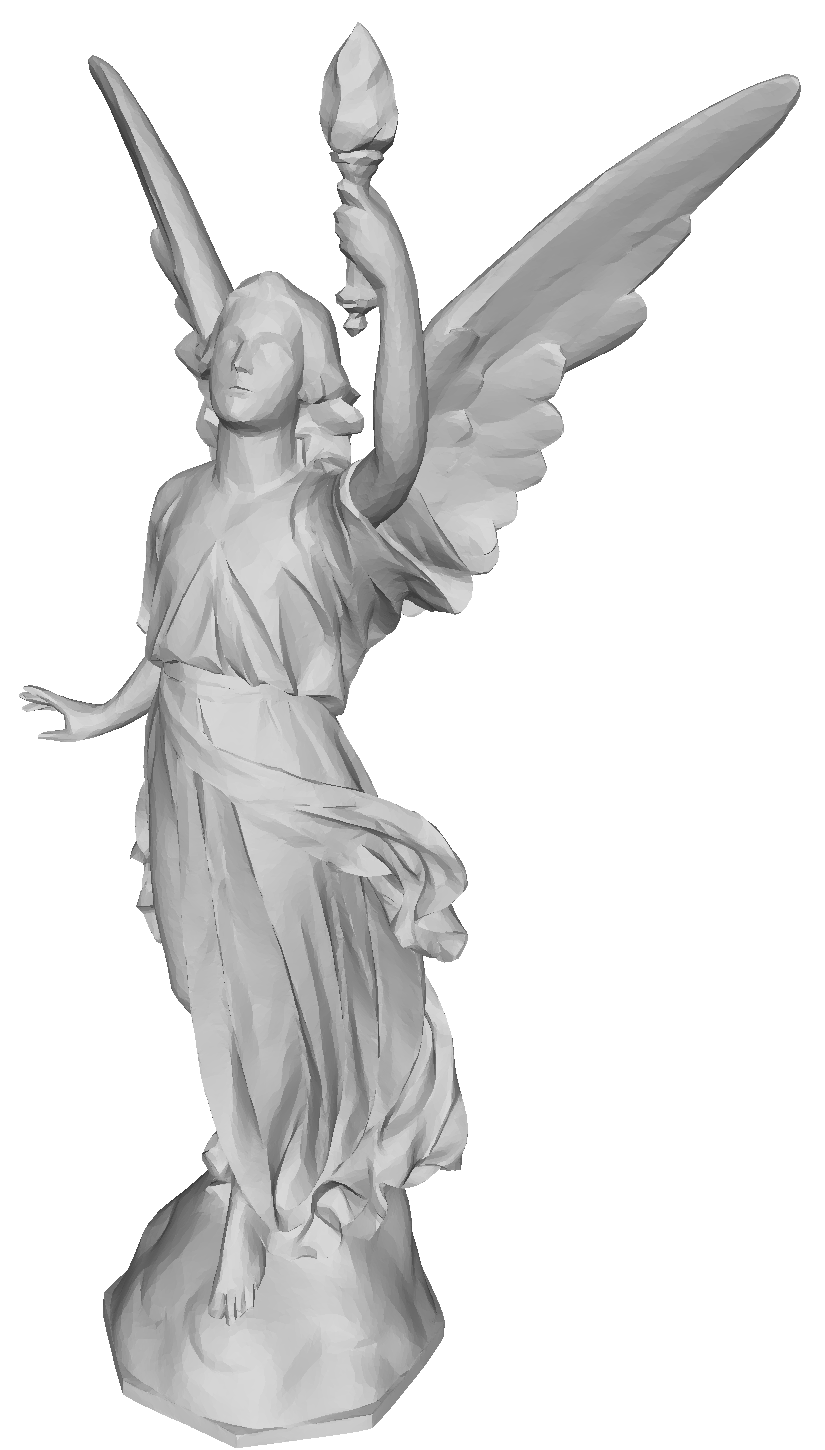}
		\caption{L0}
		\label{fig_lucy-d}
	\end{subfigure}
	\begin{subfigure}{0.12\linewidth}
		\includegraphics[width=\textwidth]{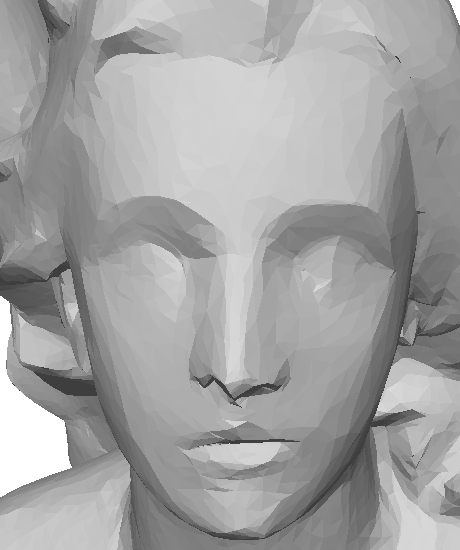}
		\includegraphics[width=\textwidth]{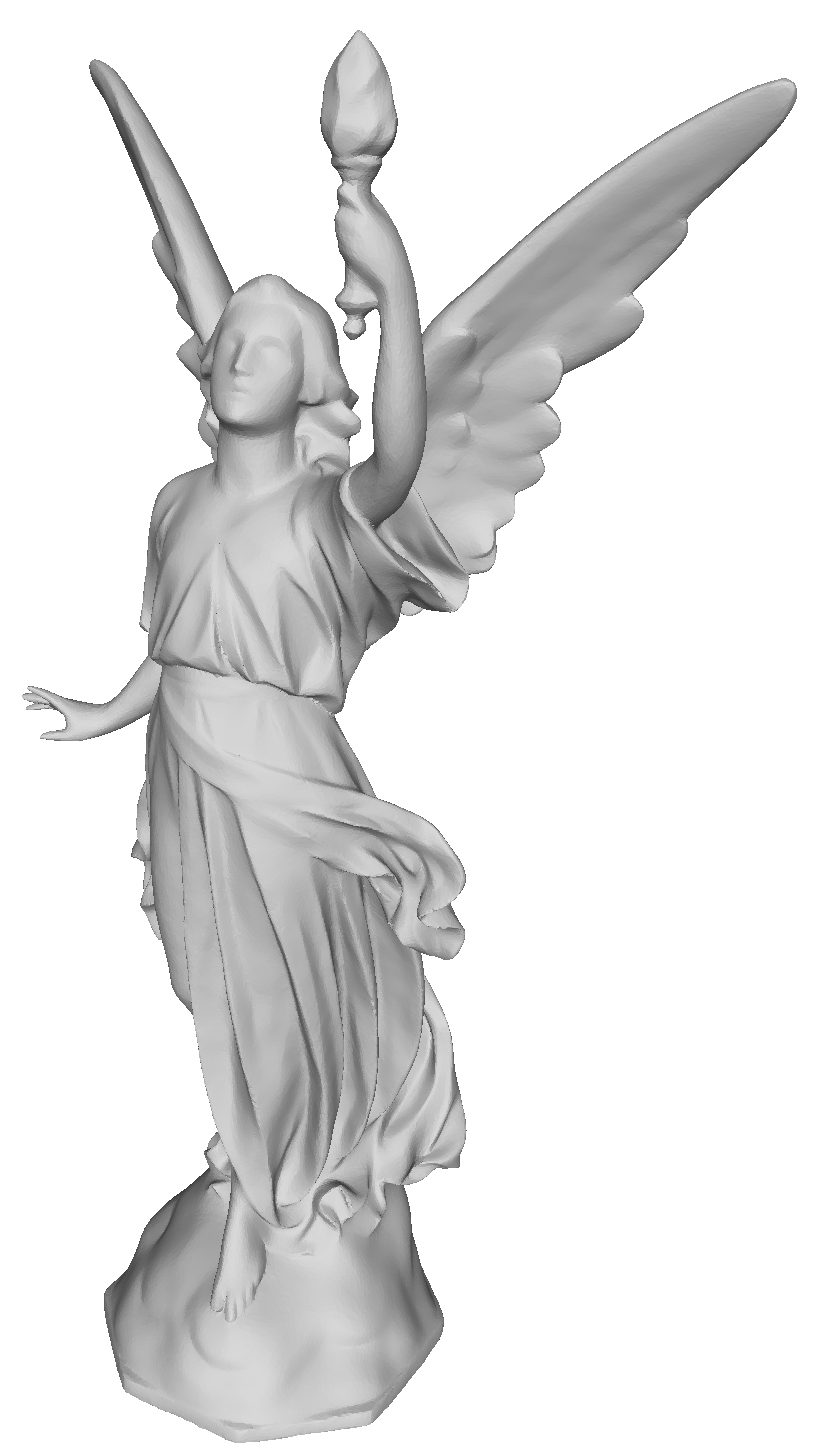}
		\caption{HO}
		\label{fig_lucy-e}
	\end{subfigure}
    \begin{subfigure}{0.12\linewidth}
		\includegraphics[width=\textwidth]{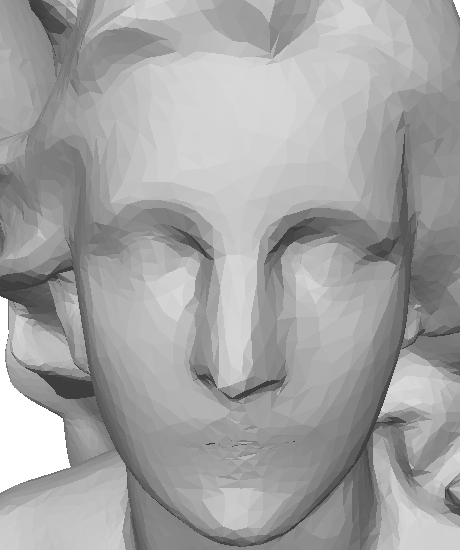}
		\includegraphics[width=\textwidth]{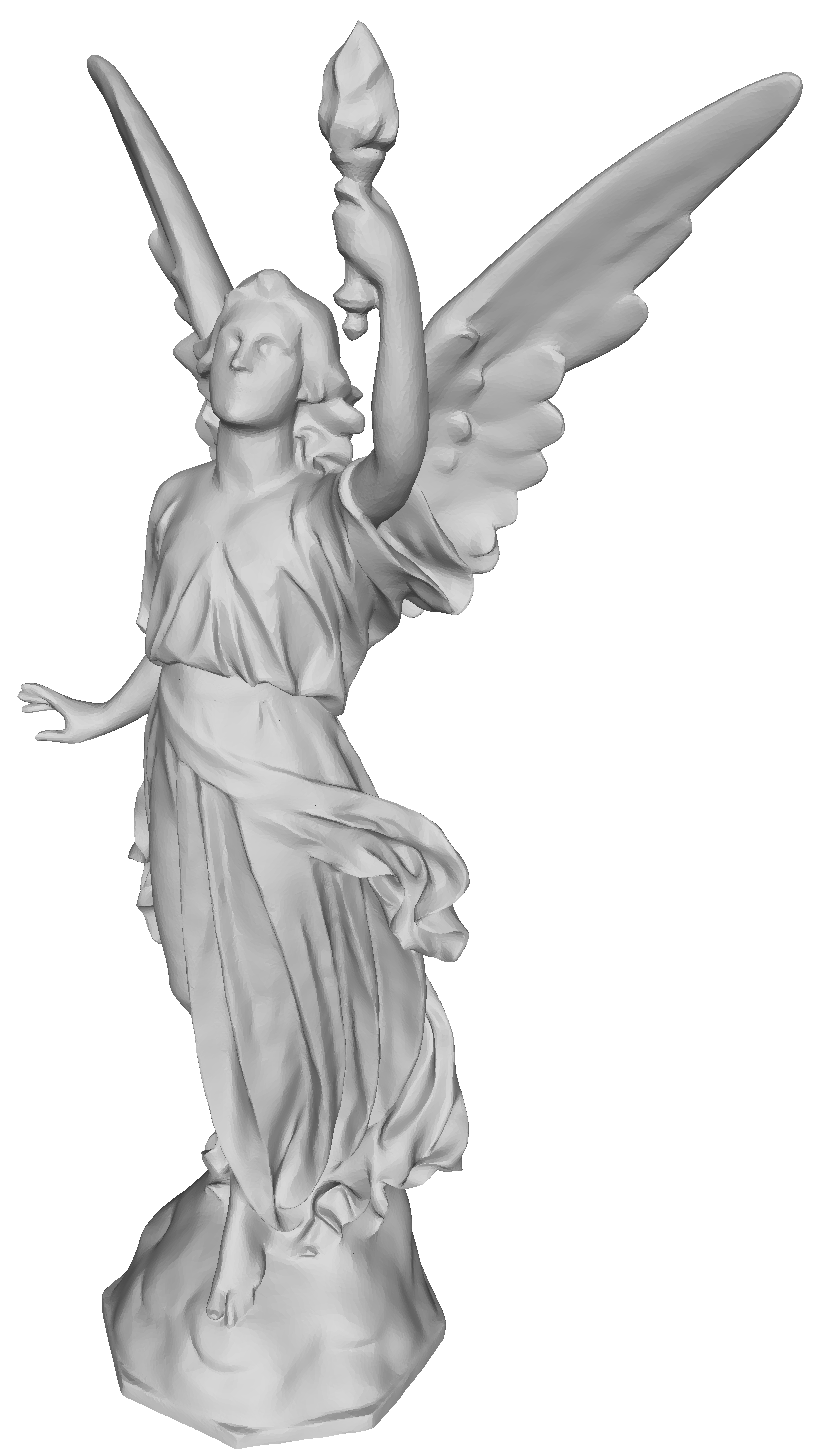}
		\caption{CNR}
		\label{fig_lucy-f}
	\end{subfigure}
	\begin{subfigure}{0.12\linewidth}
		\includegraphics[width=\textwidth]{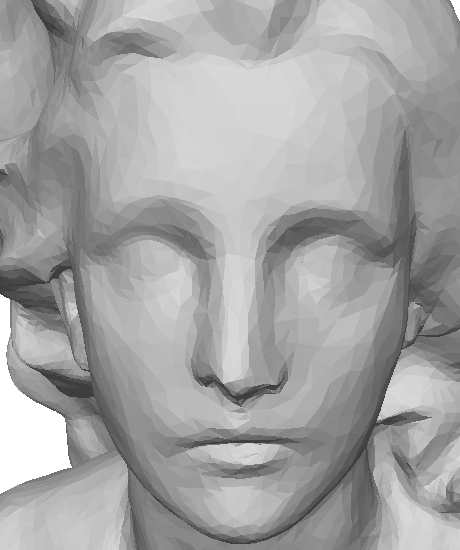}
		\includegraphics[width=\textwidth]{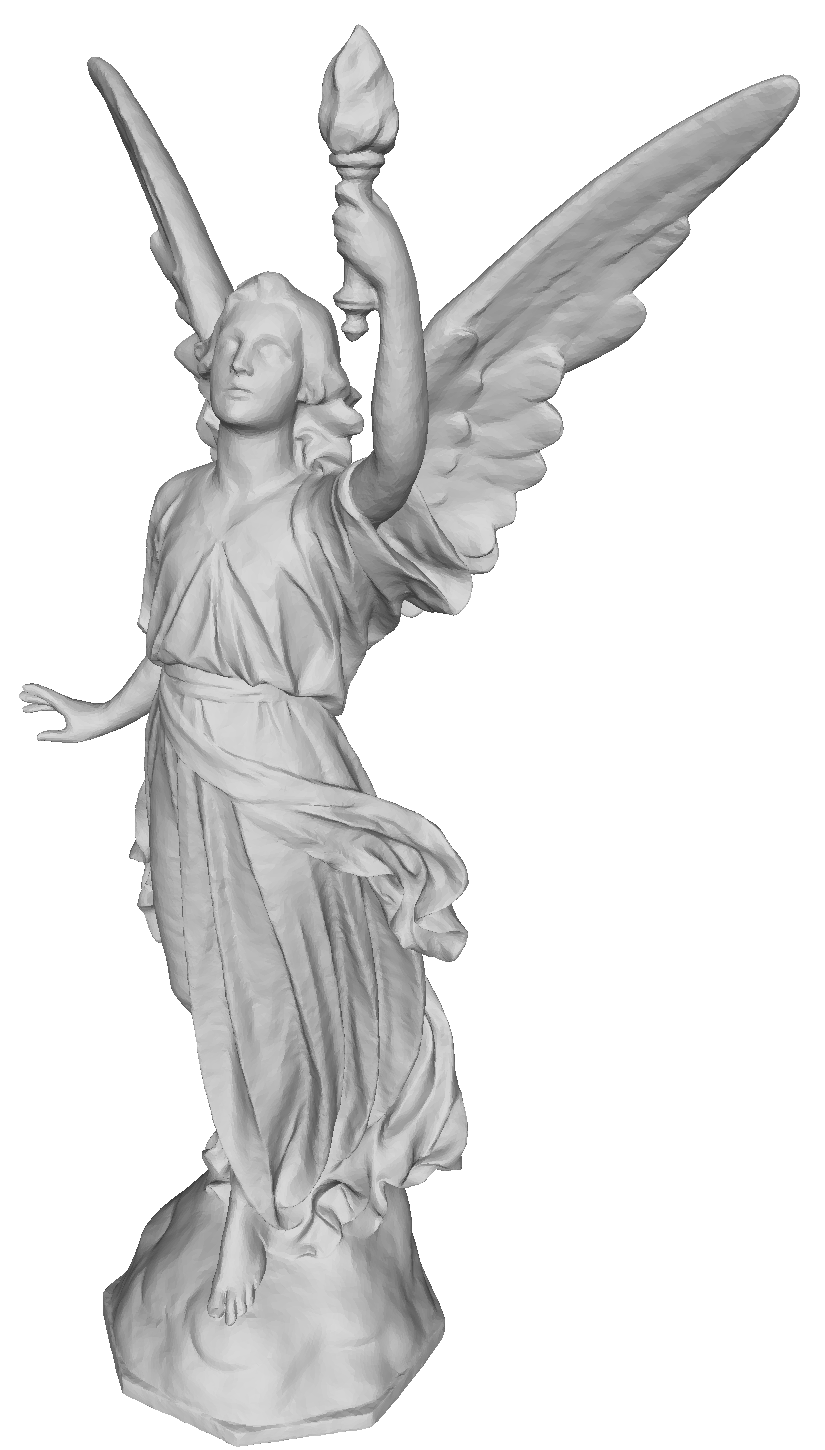}
		\caption{TGV}
		\label{fig_lucy-g}
	\end{subfigure}
	\begin{subfigure}{0.12\linewidth}
		\includegraphics[width=\textwidth]{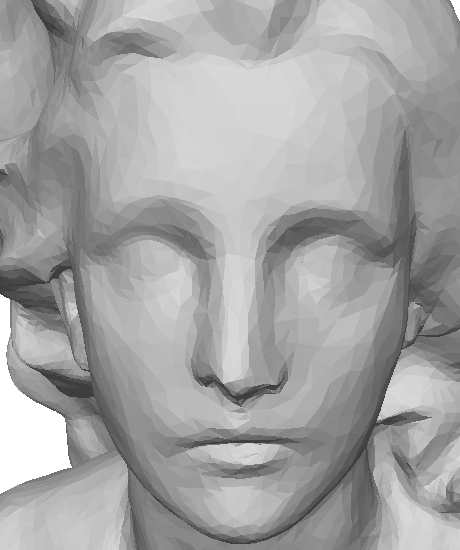}
		\includegraphics[width=\textwidth]{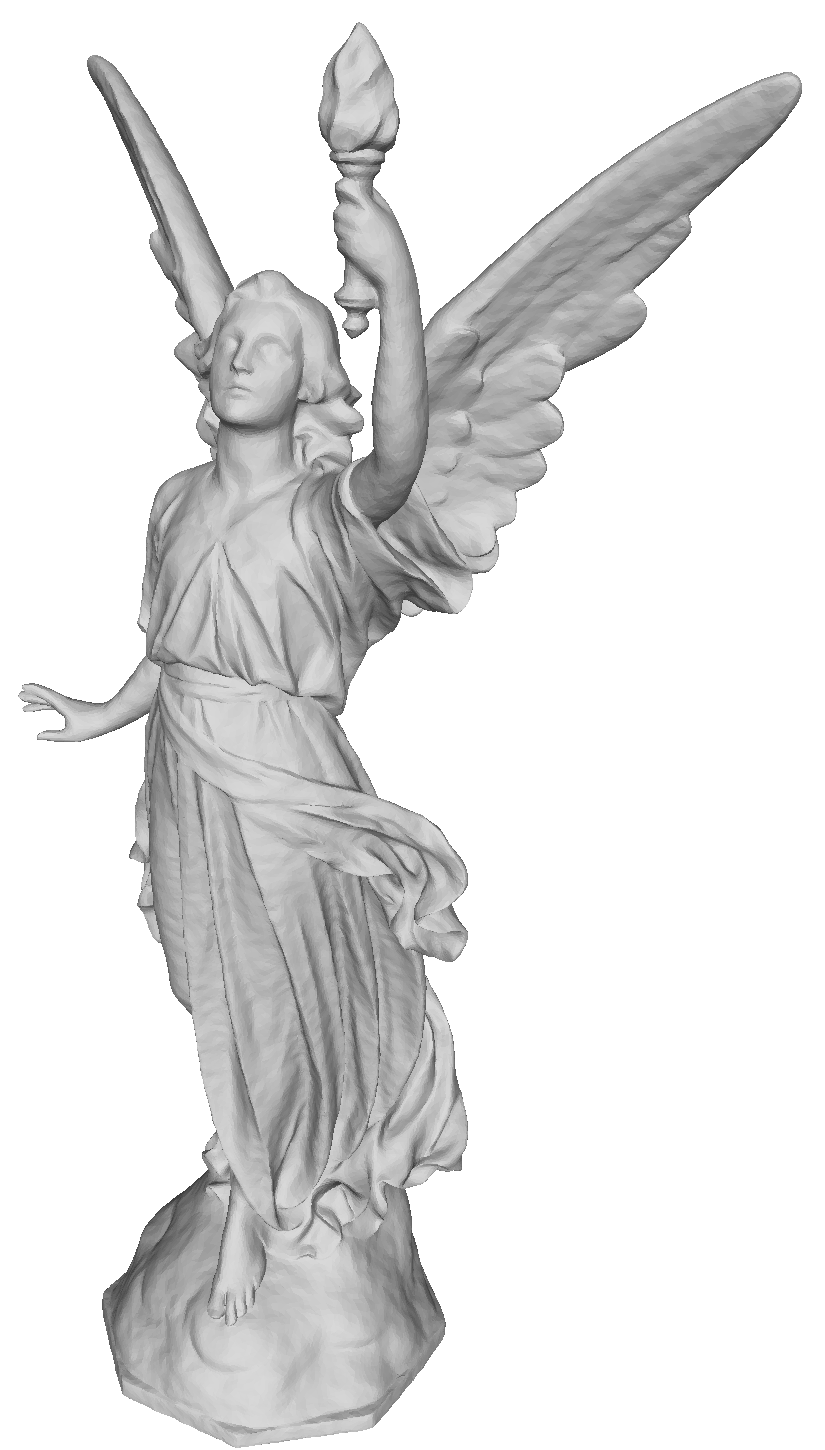}
		\caption{Ours}
		\label{fig_lucy-h}
	\end{subfigure}
	\caption{Visual comparison of different denoising methods under noise level $\sigma = 0.2\, \bar{l}_e$.}
	\label{fig_lucy}
\end{figure*}

We also show another visual comparison in~\cref{fig_fandisk}, where the CAD surface includes sharp edge features, as well as shallow edges. Again, we obtain similar results as shown in~\cref{fig_block}. Clearly, both the BF method~\cite{zheng2010bilateral} and TV regularization~\cite{zhang2015variational} significantly blur sharp features to varying extents, whereas $L_0$ minimization~\cite{he2013mesh} preserves strong edge features but also introduces crease edges in smoothing varying regions. The HO model~\cite{liu2019novel} recovers smooth regions more effectively than the TV regularization. Since this higher-order model relies only on second-order information, it tends to bend straight-line edges and blur shallow edge features. In contrast, the TGV-based~\cite{liu2021mesh} and the learn-based CNR method~\cite{wang2016mesh} alleviate the drawbacks with comparable or more favorable effects. Again, the proposed semi-sparsity minimization typically retains a similar level of TGV performance in terms of fitting/smoothing ability in both sharp features and smooth varying surfaces.

\textbf{Non-CAD Surfaces:} We now compare the performance on non-CAD surfaces. Differing from the CAD surfaces, the later typically exhibit abundant coarse-to-fine geometric details, which pose challenges for mesh denoising, particularly under high noise levels. As shown in~\cref{fig_lucy}, the BF method~\cite{zheng2010bilateral} and TV regularization~\cite{zhang2015variational} tend to over-smooth fine-scale details, and the $L_0$ model introduces slight staircase artifacts towards piecewise constant patches. The HO method~\cite{liu2019novel} performs reasonably well in preserving medium-scale features, whereas the learning-based CNR method~\cite{wang2016mesh} oversmooths small-scale details and yields comparatively poor results, likely due to limitations in the training data. In contrast, both the TGV-based regularization~\cite{liu2021mesh} and our semi-sparsity model effectively combine the strengths of first- and second-order regularization, successfully reconstructing sharp features while maintaining smooth regions. The visual comparisons highlight the superior performance of our approach in preserving geometric fidelity across diverse feature scales.

\begin{figure}[!t]
	\centering
	\begin{subfigure}{0.23\linewidth}
        \includegraphics[width=\textwidth]{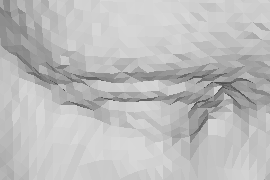} 
        \includegraphics[width=\textwidth]{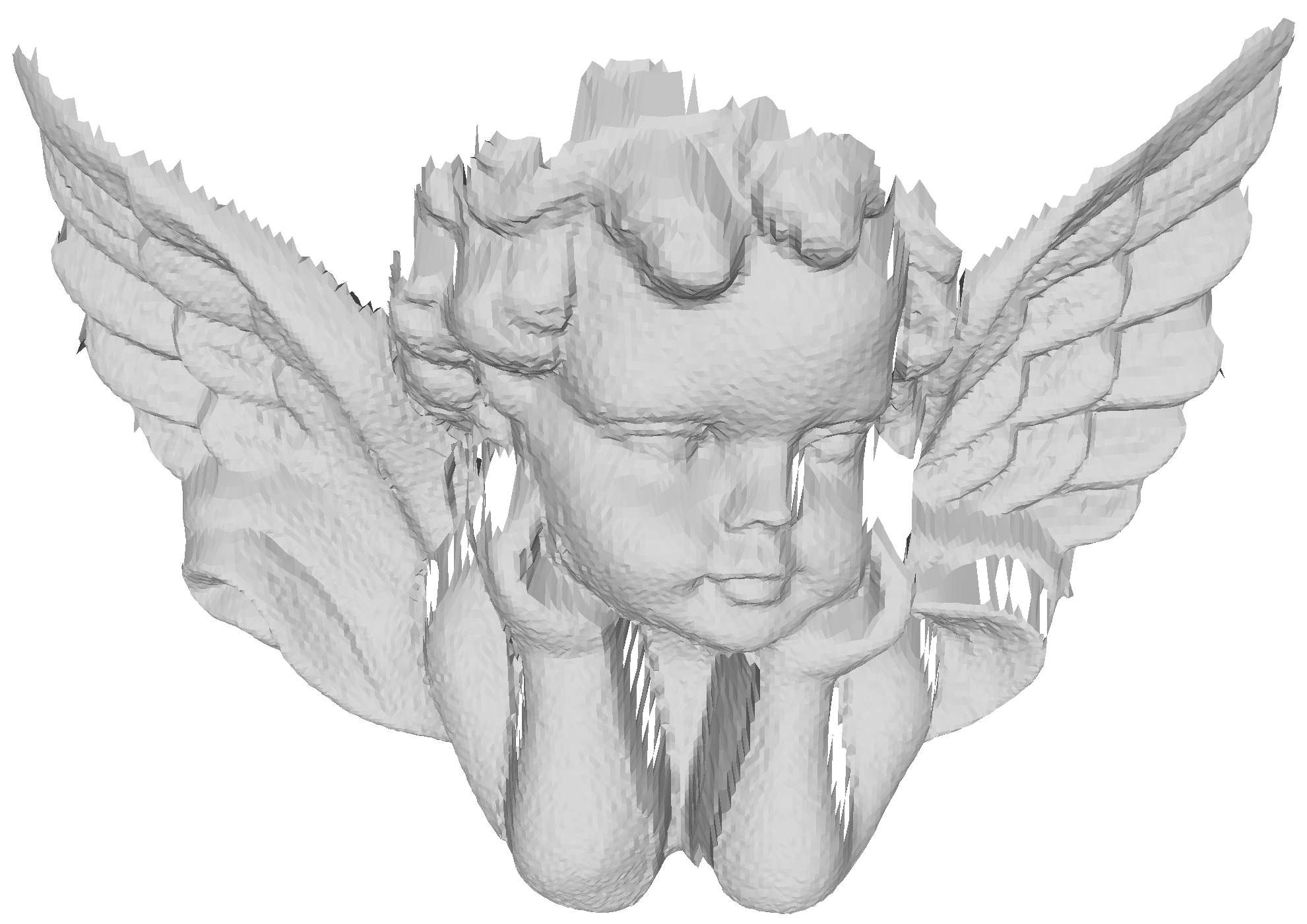}
		\caption{Noisy}
		\label{fig_scan-a}
	\end{subfigure}
	\begin{subfigure}{0.23\linewidth}
		\includegraphics[width=\textwidth]{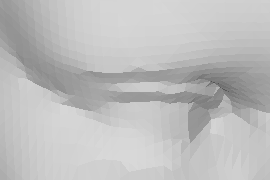}
		\includegraphics[width=\textwidth]{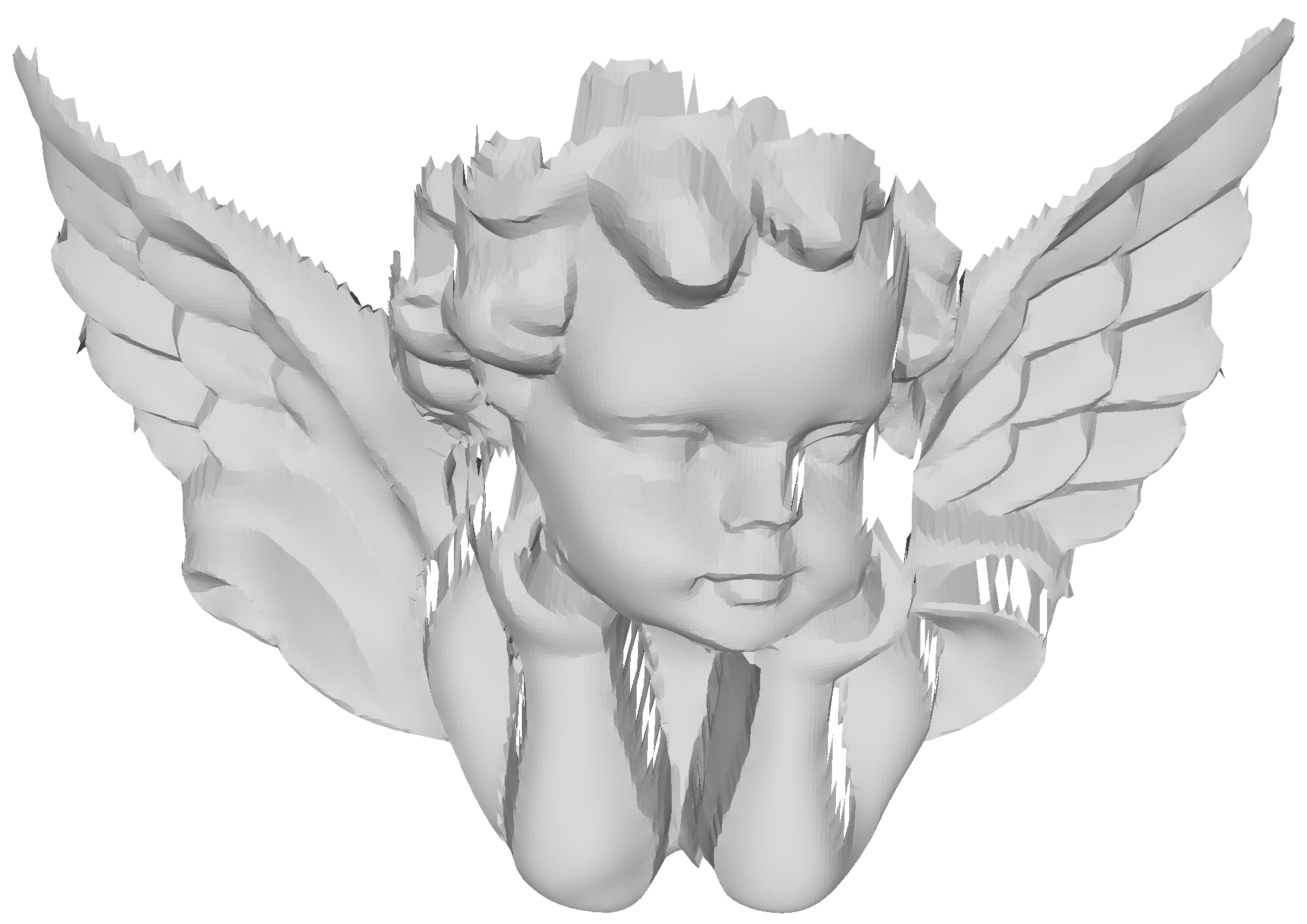} 
		\caption{BF~\cite{zheng2010bilateral}}
		\label{fig_scan-b}
	\end{subfigure}
	\begin{subfigure}{0.23\linewidth}
		\includegraphics[width=\textwidth]{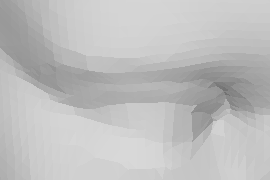}
		\includegraphics[width=\textwidth]{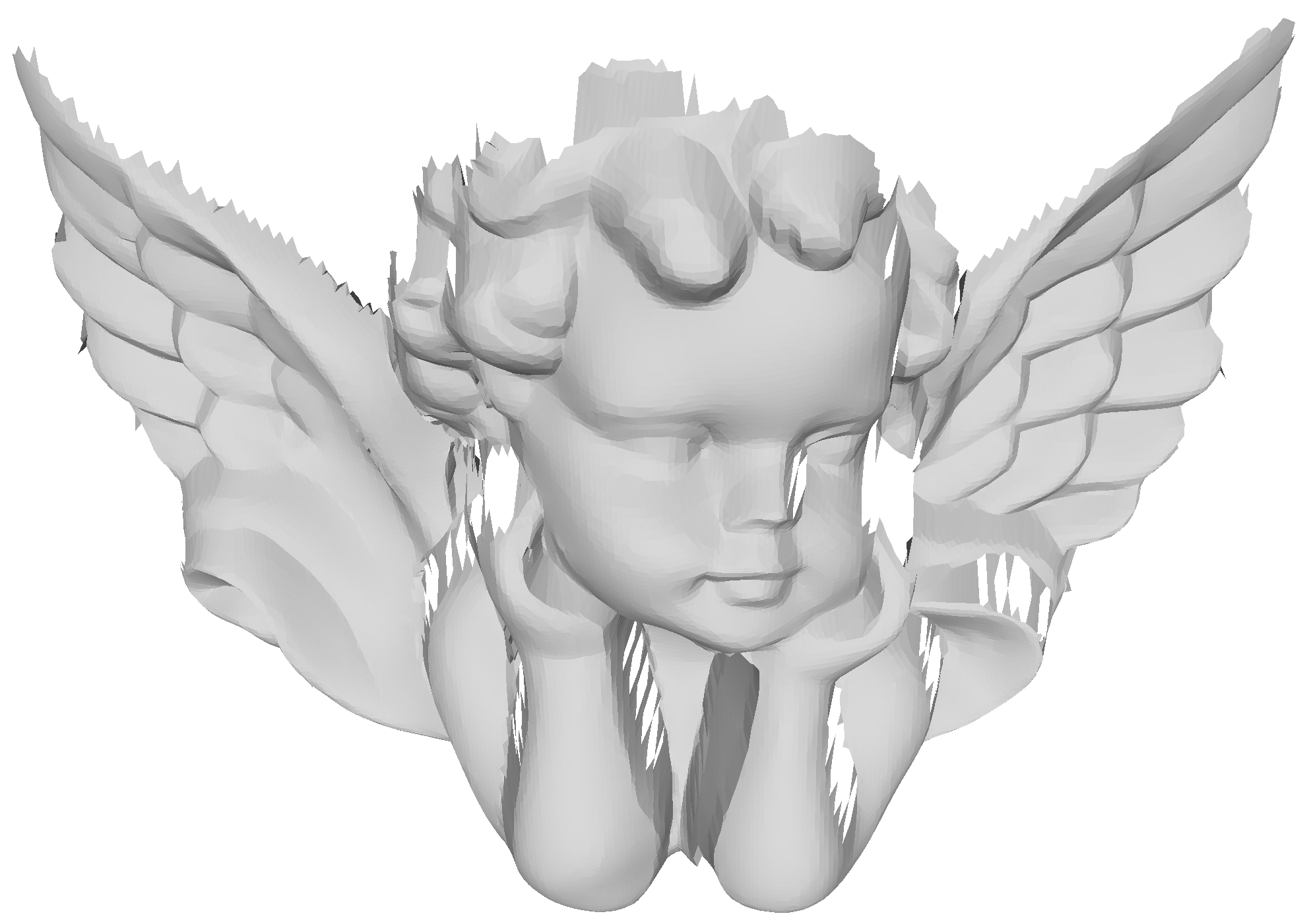}
		\caption{TV~\cite{zhang2015variational}}
		\label{fig_scan-c}
	\end{subfigure}
	\begin{subfigure}{0.23\linewidth}
		\includegraphics[width=\textwidth]{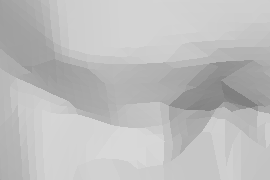}
		\includegraphics[width=\textwidth]{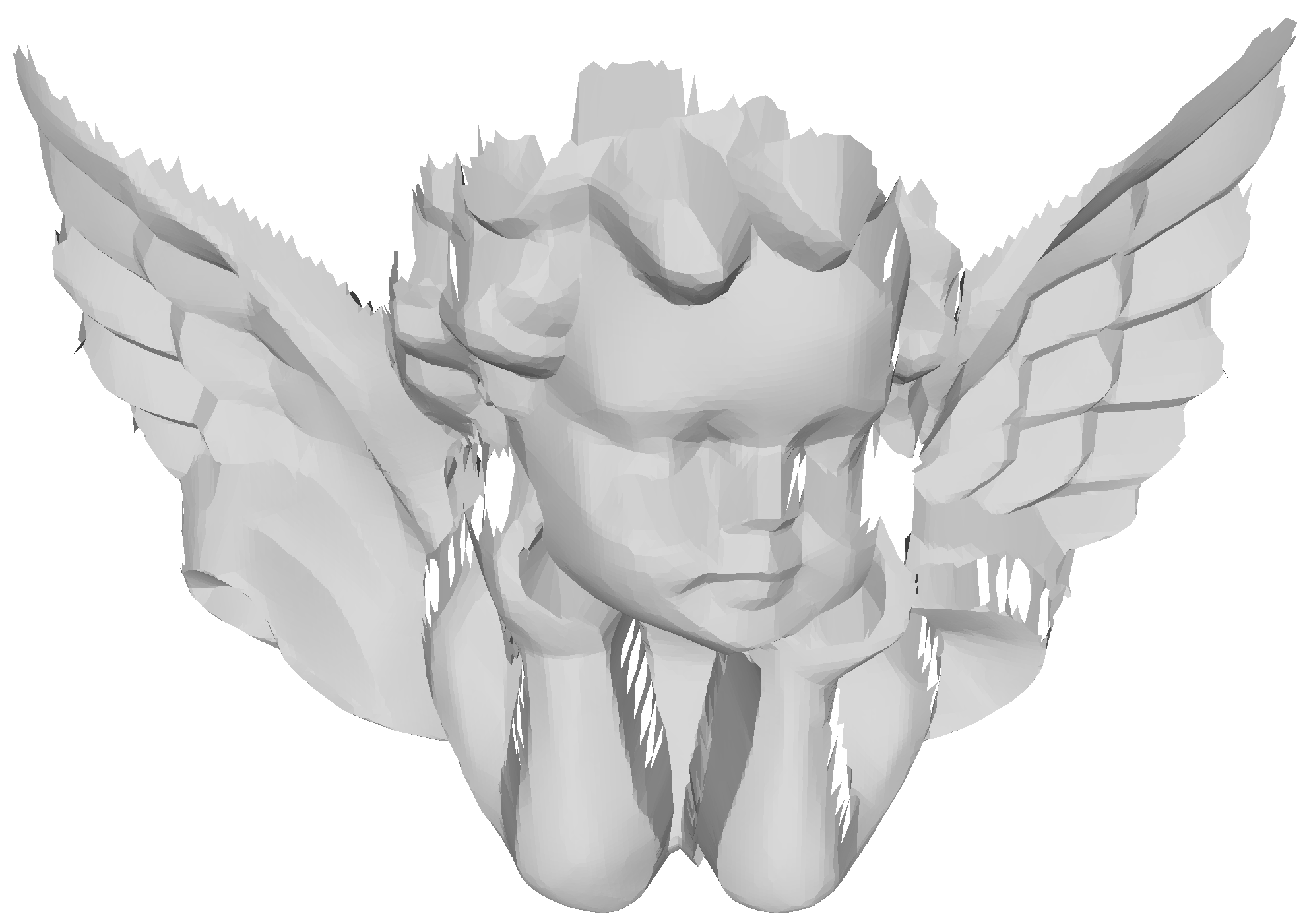}
		\caption{L0~\cite{he2013mesh}}
		\label{fig_scan-d}
	\end{subfigure}
	\begin{subfigure}{0.23\linewidth}
		\includegraphics[width=\textwidth]{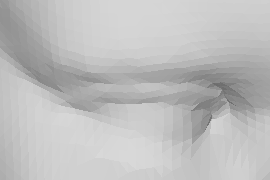}
		\includegraphics[width=\textwidth]{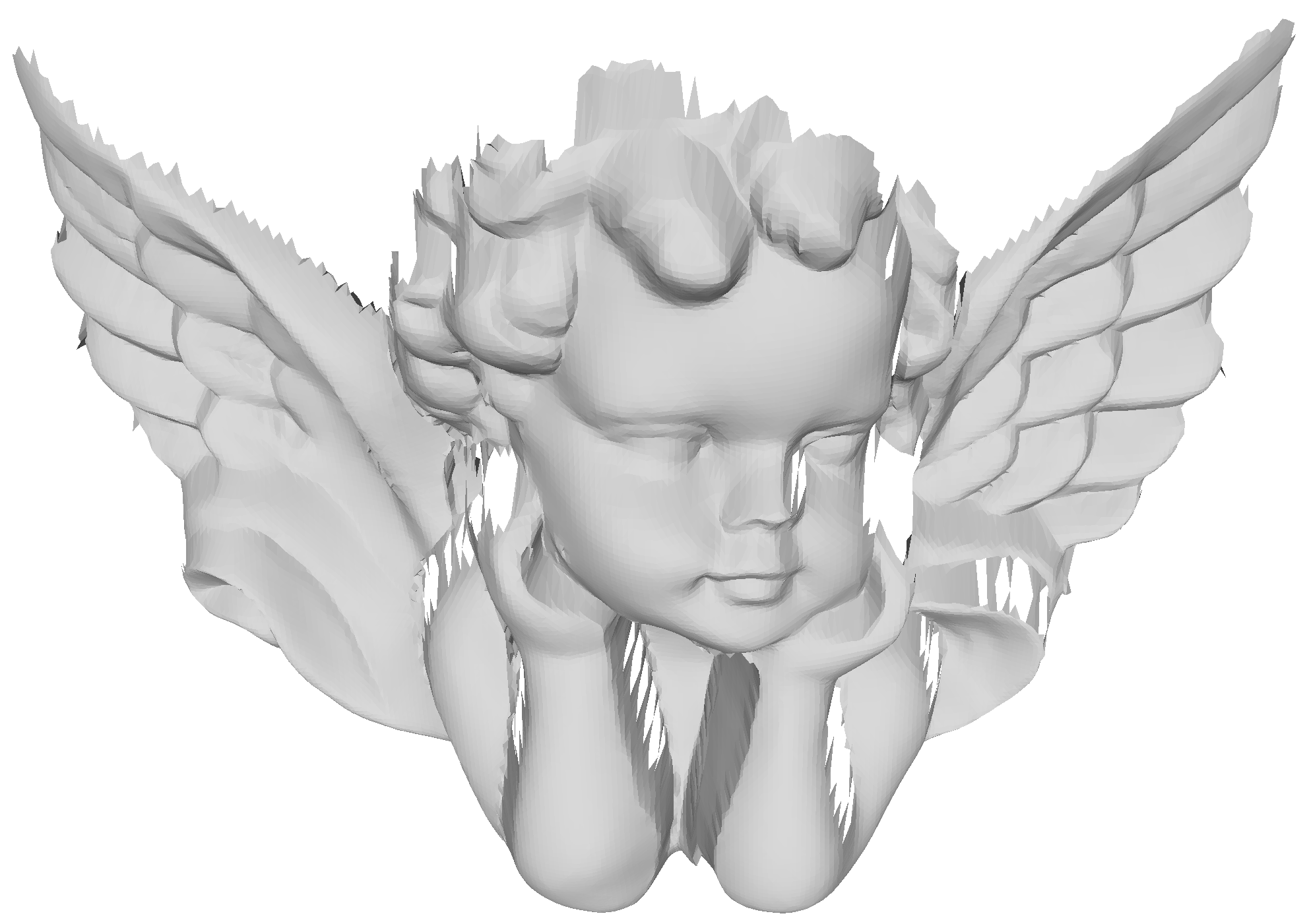}
		\caption{HO~\cite{liu2019novel}}
		\label{fig_scan-e}
	\end{subfigure}
        \begin{subfigure}{0.23\linewidth}
		\includegraphics[width=\textwidth]{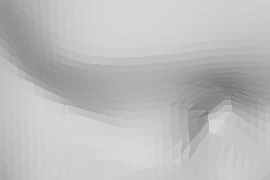}
		\includegraphics[width=\textwidth]{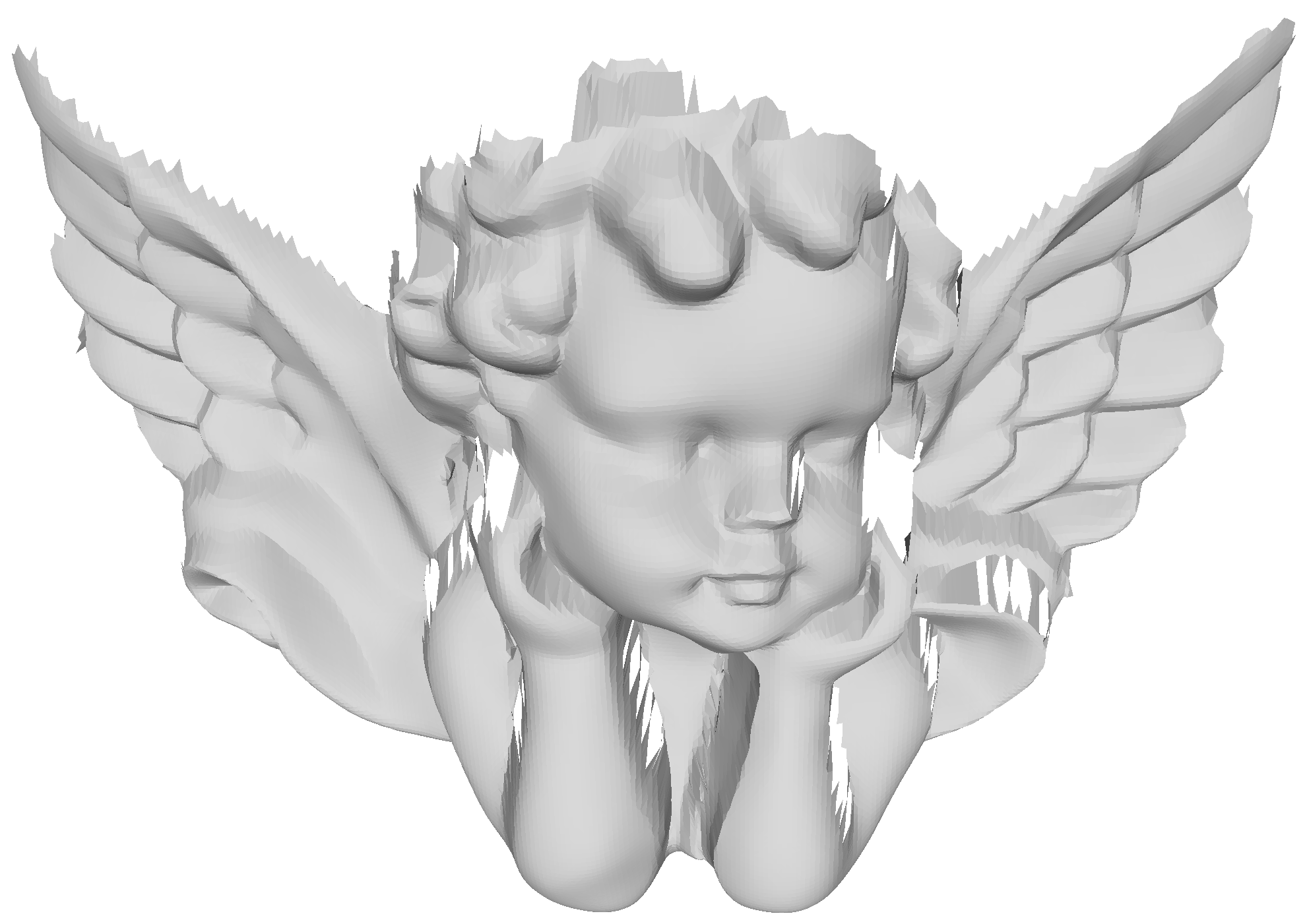}
		\caption{CNR~\cite{wang2016mesh}} 
		\label{fig_scan-f}
	\end{subfigure}
	\begin{subfigure}{0.23\linewidth}
		\includegraphics[width=\textwidth]{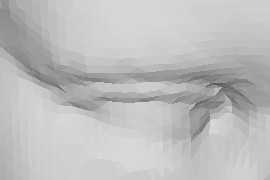}
		\includegraphics[width=\textwidth]{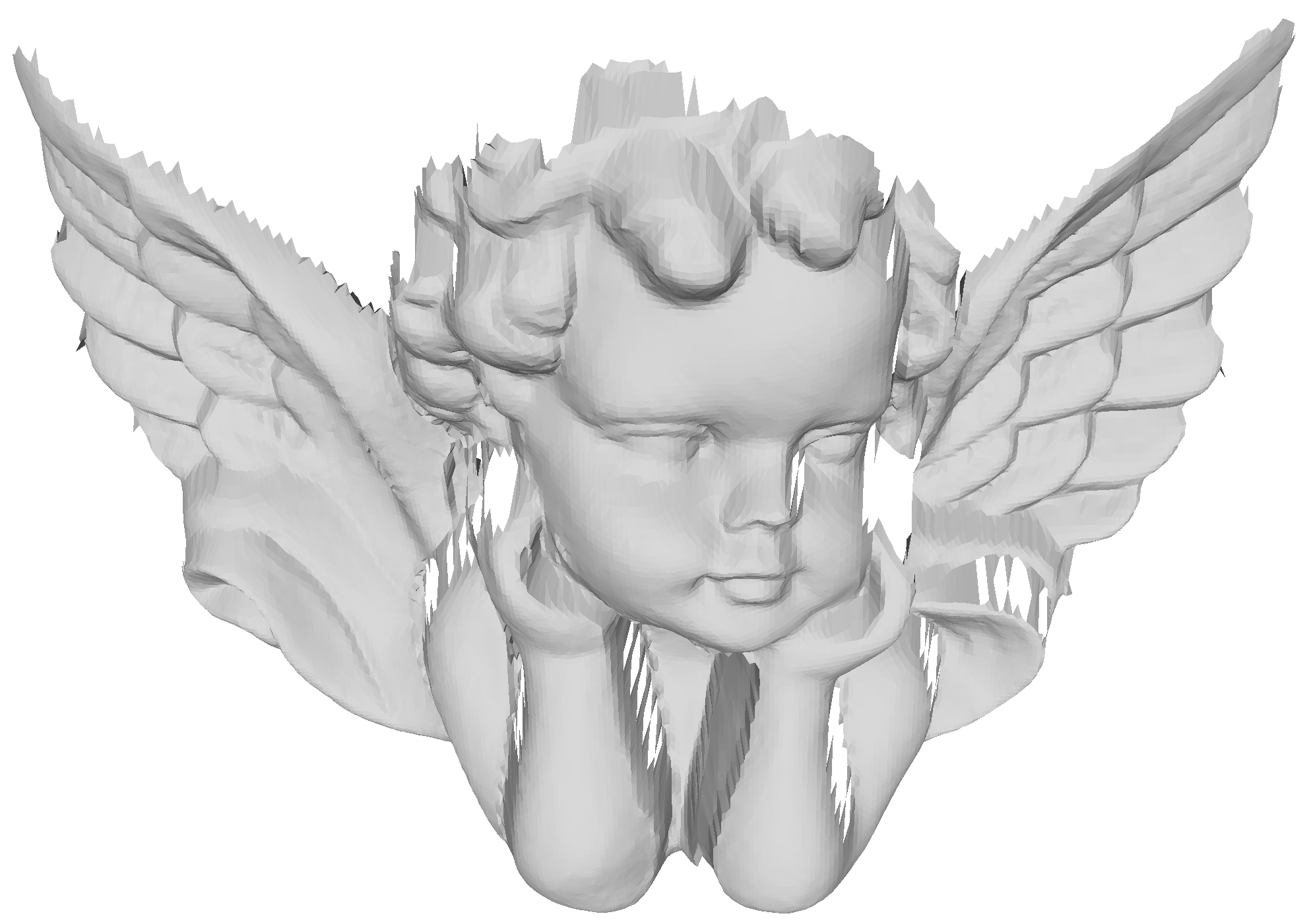}
		\caption{TGV~\cite{liu2021mesh}}
		\label{fig_scan-g}
	\end{subfigure}
	\begin{subfigure}{0.23\linewidth}
		\includegraphics[width=\textwidth]{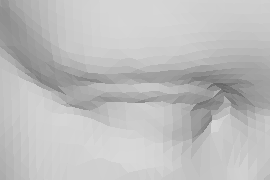}
		\includegraphics[width=\textwidth]{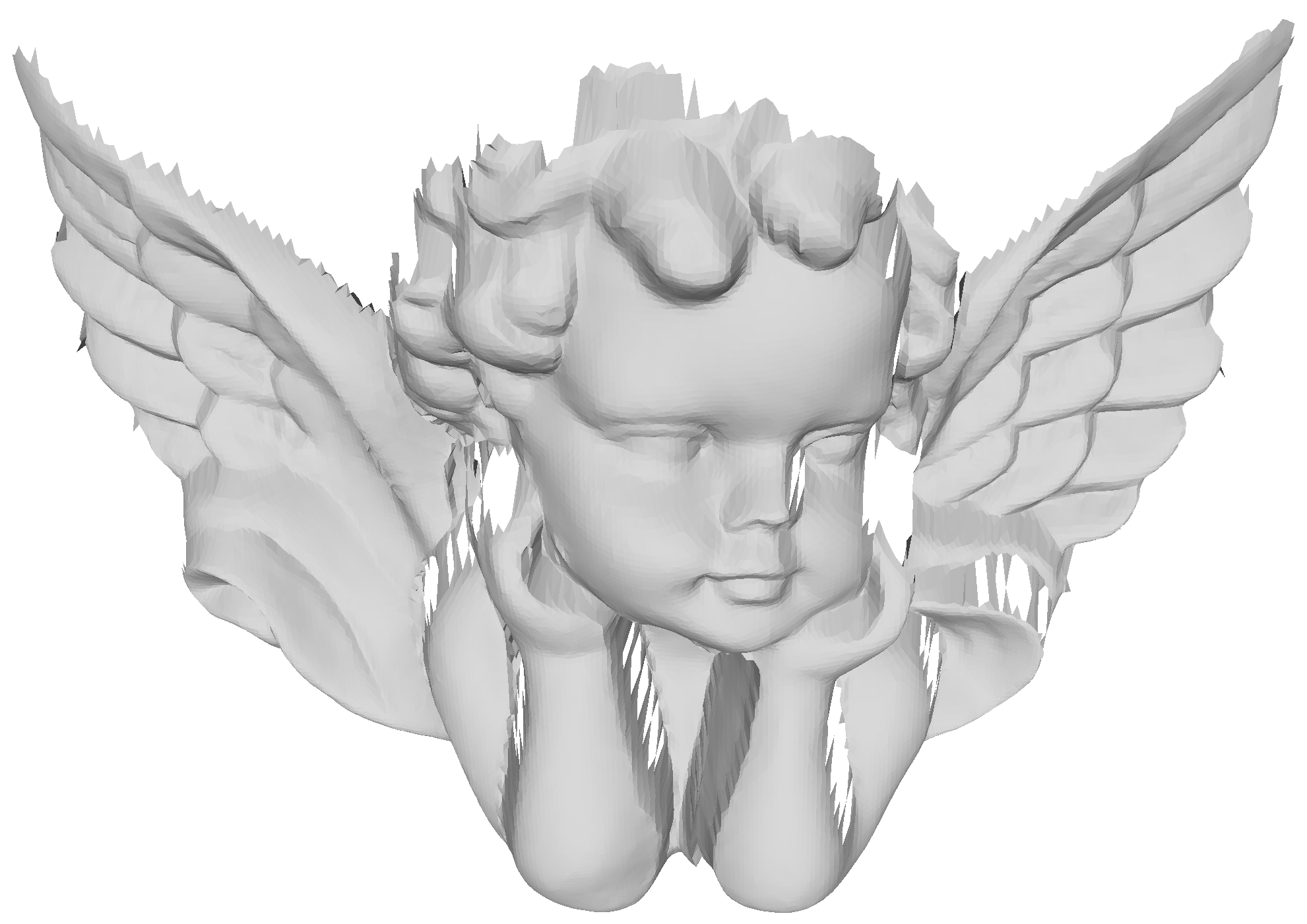}
		\caption{Ours}
		\label{fig_scan-h}
	\end{subfigure}
	\caption{Visual comparison of different denoising methods for the real scanned data.}
	\label{fig_scan}
\end{figure}

\textbf{Scanned Data:} Furthermore, we evaluate the proposed semi-sparsity denoising model on real scanned data acquired from a laser scanner. In this setting, the level and type of noise are generally unknown. To ensure a fair comparison, we employ a greedy search strategy to optimize the parameters of each method for the best performance. As shown in~\cref{fig_scan}, the BF method~\cite{zheng2010bilateral} is able to restore mid-level textural details, while TV regularization~\cite{zhang2015variational}, $L_0$ minimization~\cite{he2013mesh}, and the CNR method~\cite{wang2016mesh} tend to oversmooth fine details, thereby impairing visual naturalness. In contrast, higher-order methods, including the HO method~\cite{liu2019novel}, the TGV-based regularization~\cite{liu2021mesh}, and our semi-sparsity minimization scheme, achieve much better performance in retaining both sharp features and smooth surface regions.

\begin{figure*}[!t]
\centering
\begin{subfigure}{0.12\linewidth}
    \includegraphics[width=\textwidth]{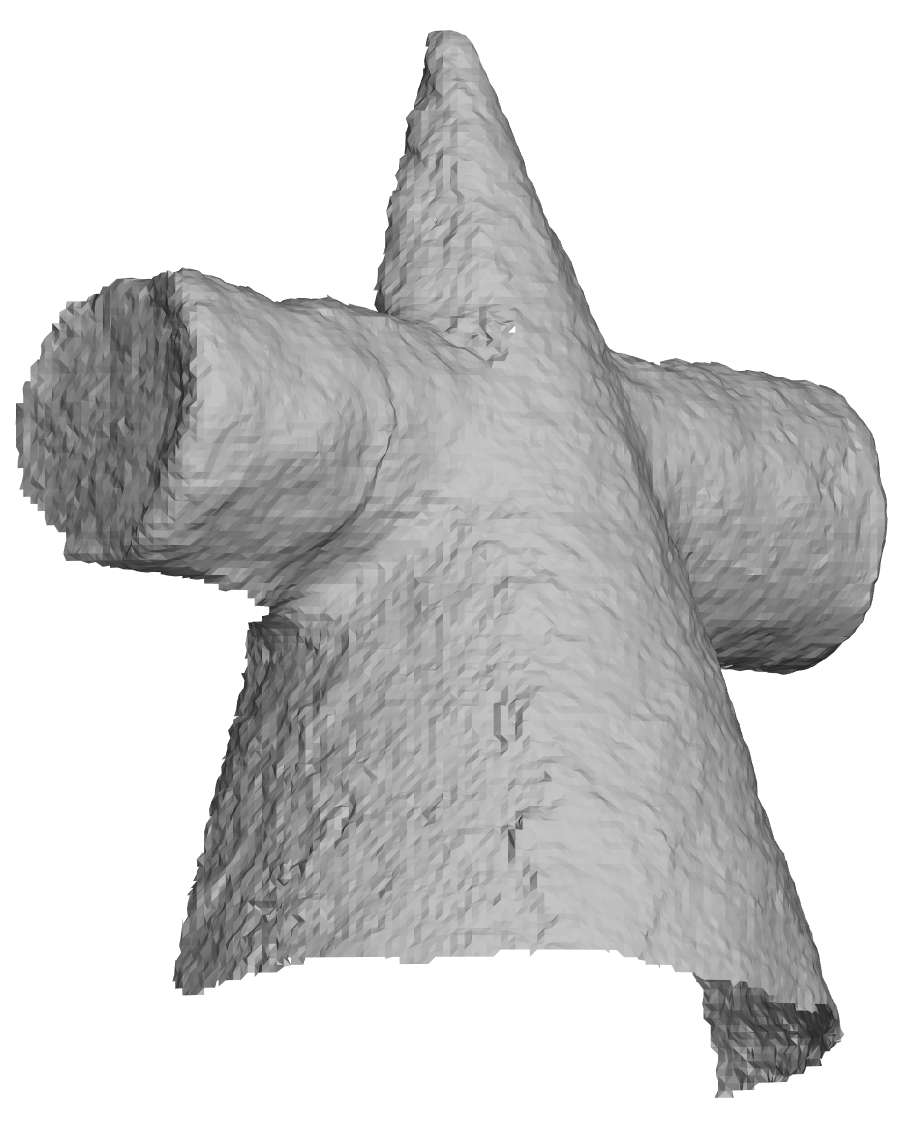} 
    \includegraphics[width=\textwidth]{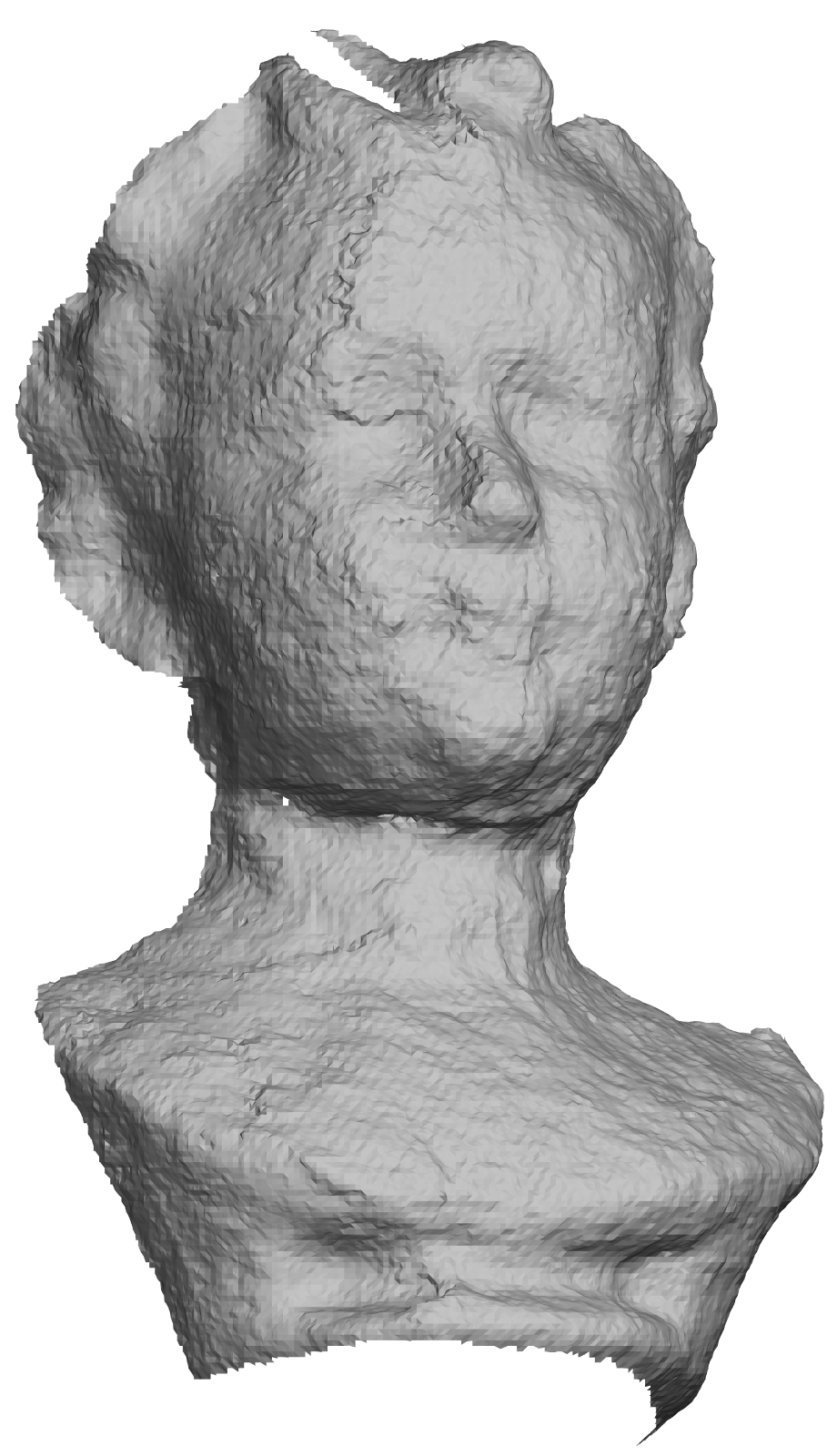}
    \caption{Noisy}
    \label{fig_kinect-a}
\end{subfigure}
\begin{subfigure}{0.12\linewidth}
    \includegraphics[width=\textwidth]{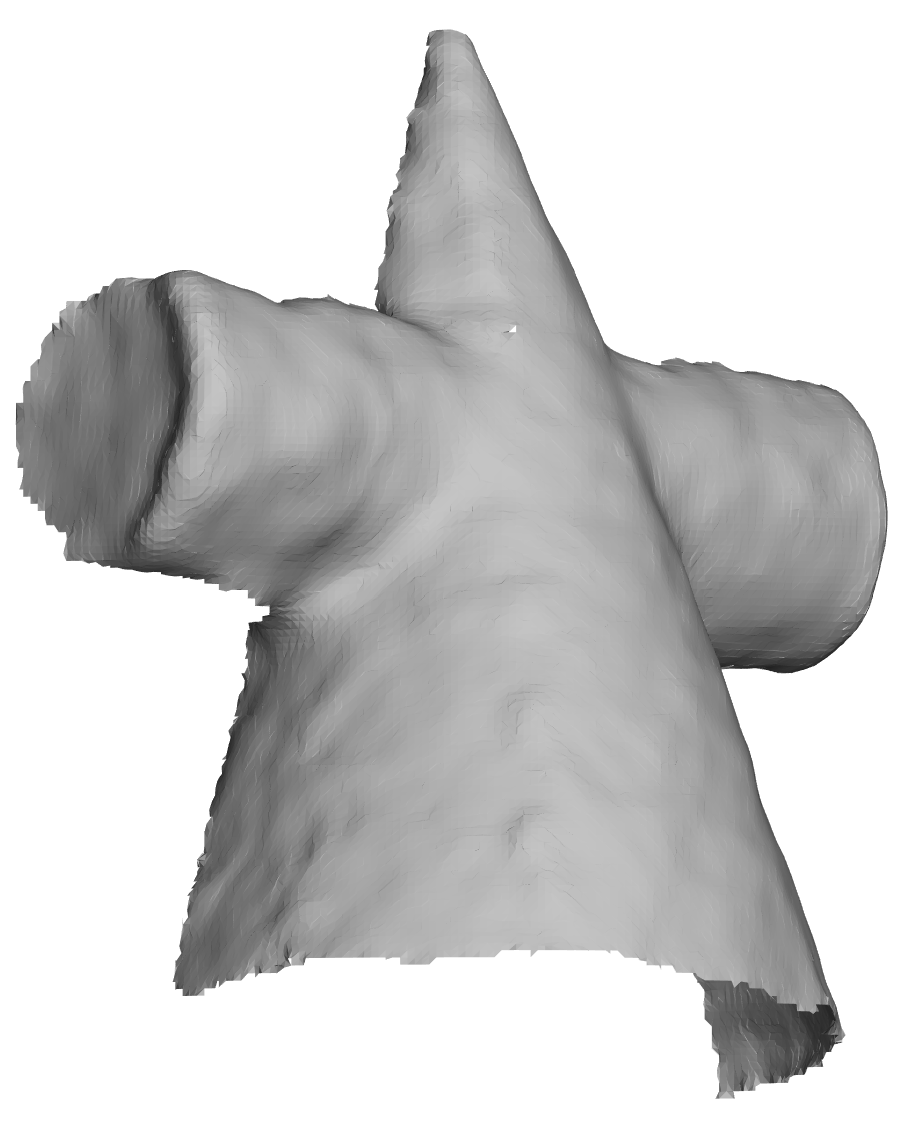}
    \includegraphics[width=\textwidth]{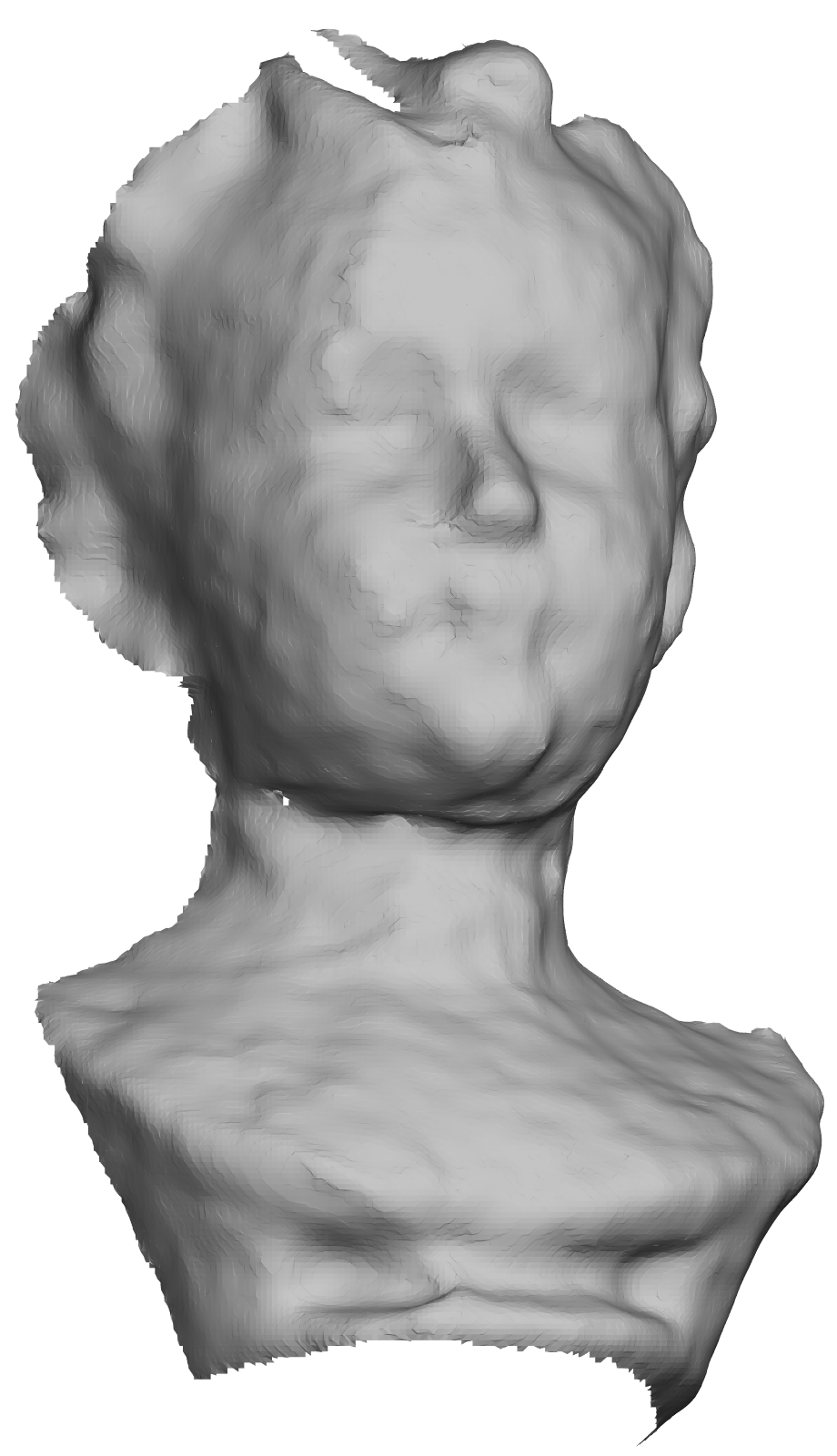} 
    \caption{BF}
    \label{fig_kinect-b}
\end{subfigure}
\begin{subfigure}{0.12\linewidth}
    \includegraphics[width=\textwidth]{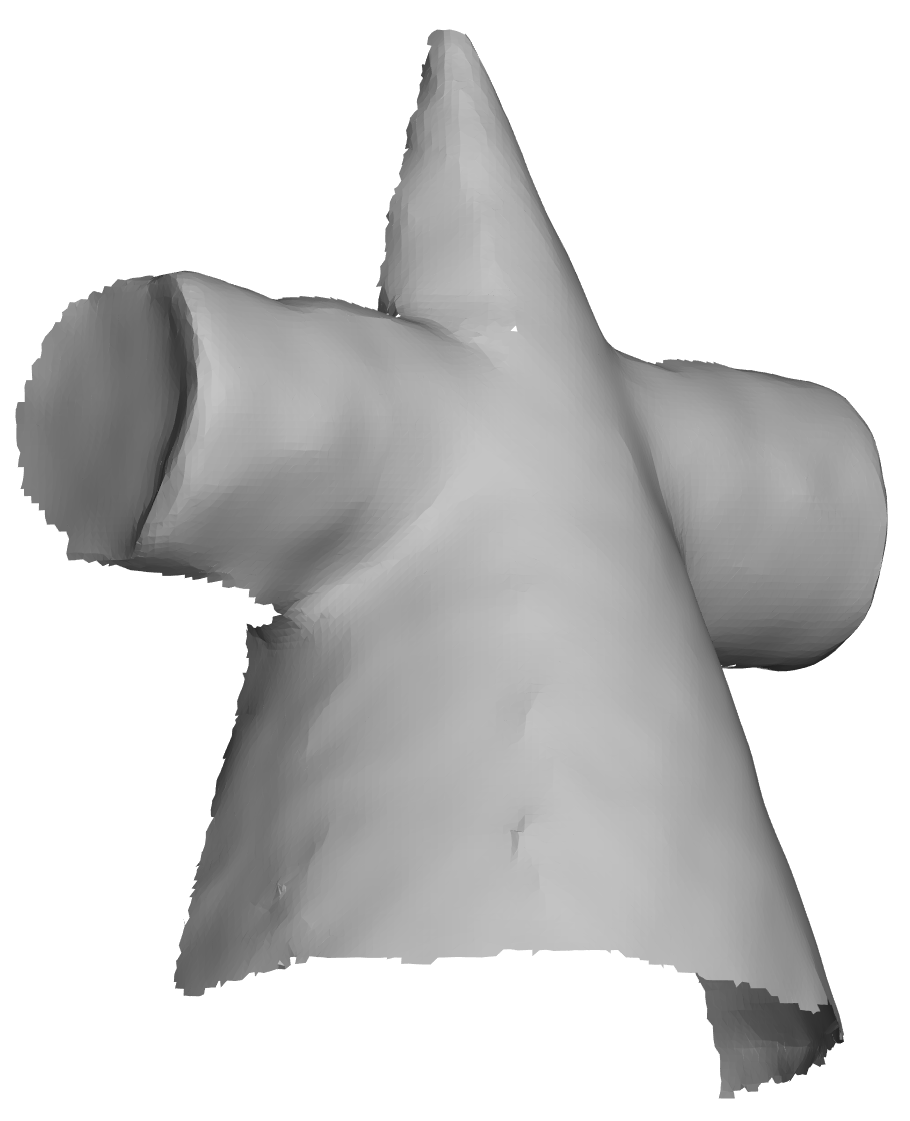}
    \includegraphics[width=\textwidth]{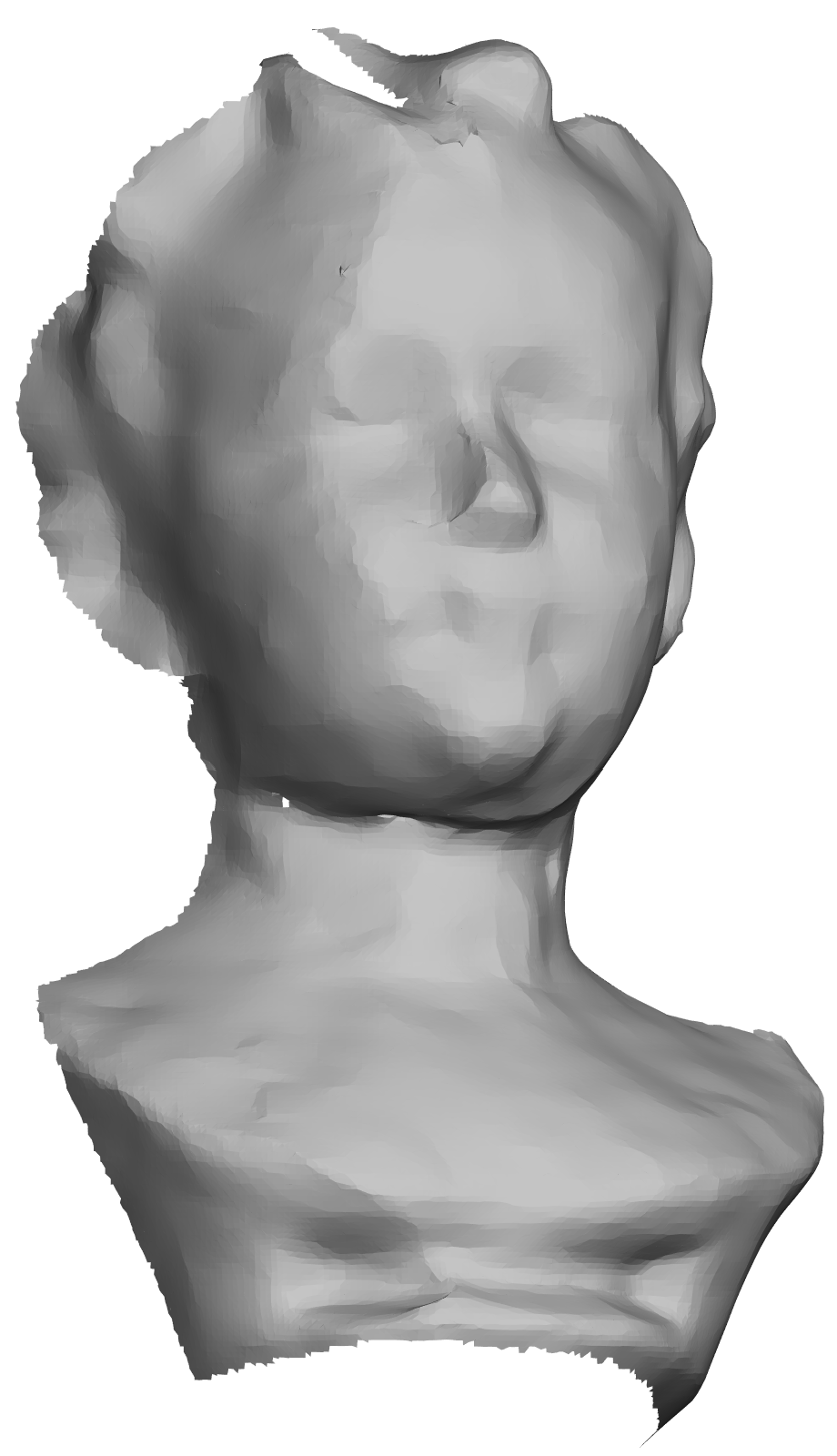}
    \caption{TV}
    \label{fig_kinect-c}
\end{subfigure}
\begin{subfigure}{0.12\linewidth}
    \includegraphics[width=\textwidth]{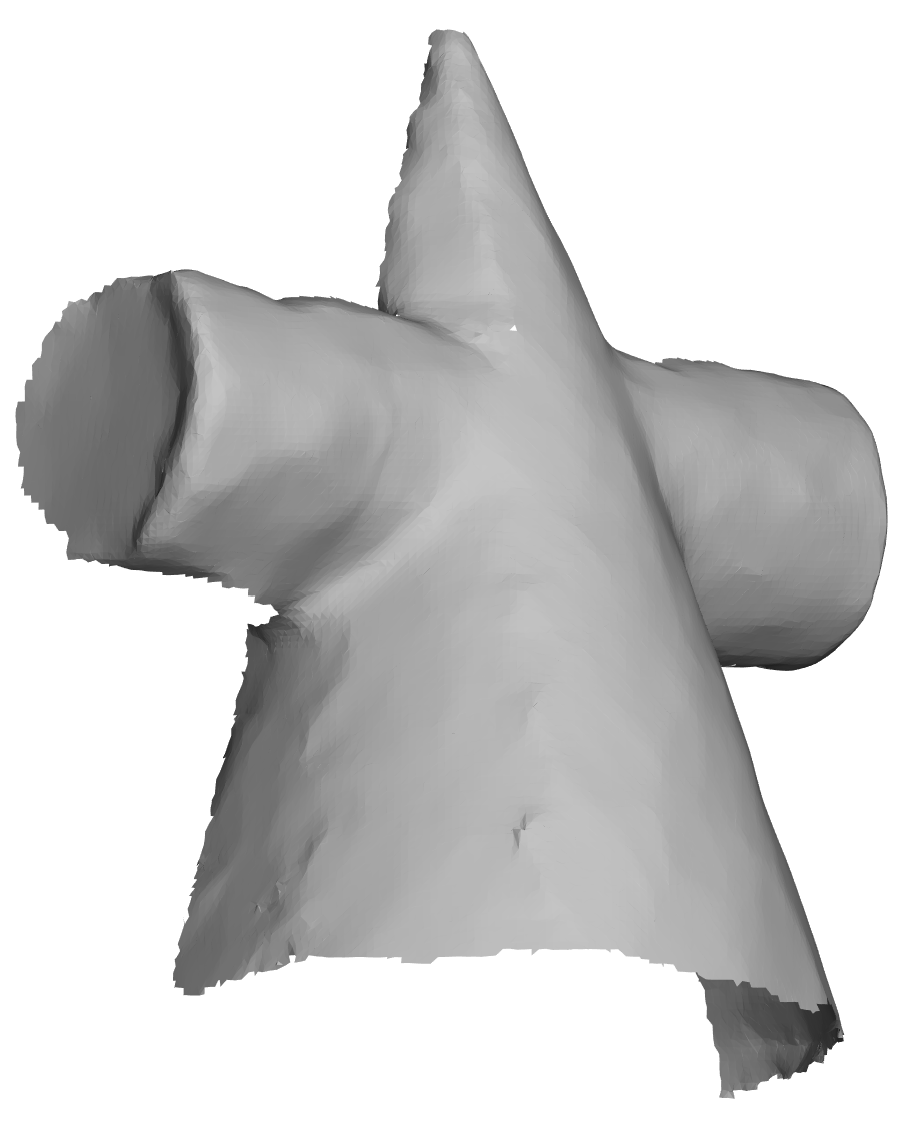}
    \includegraphics[width=\textwidth]{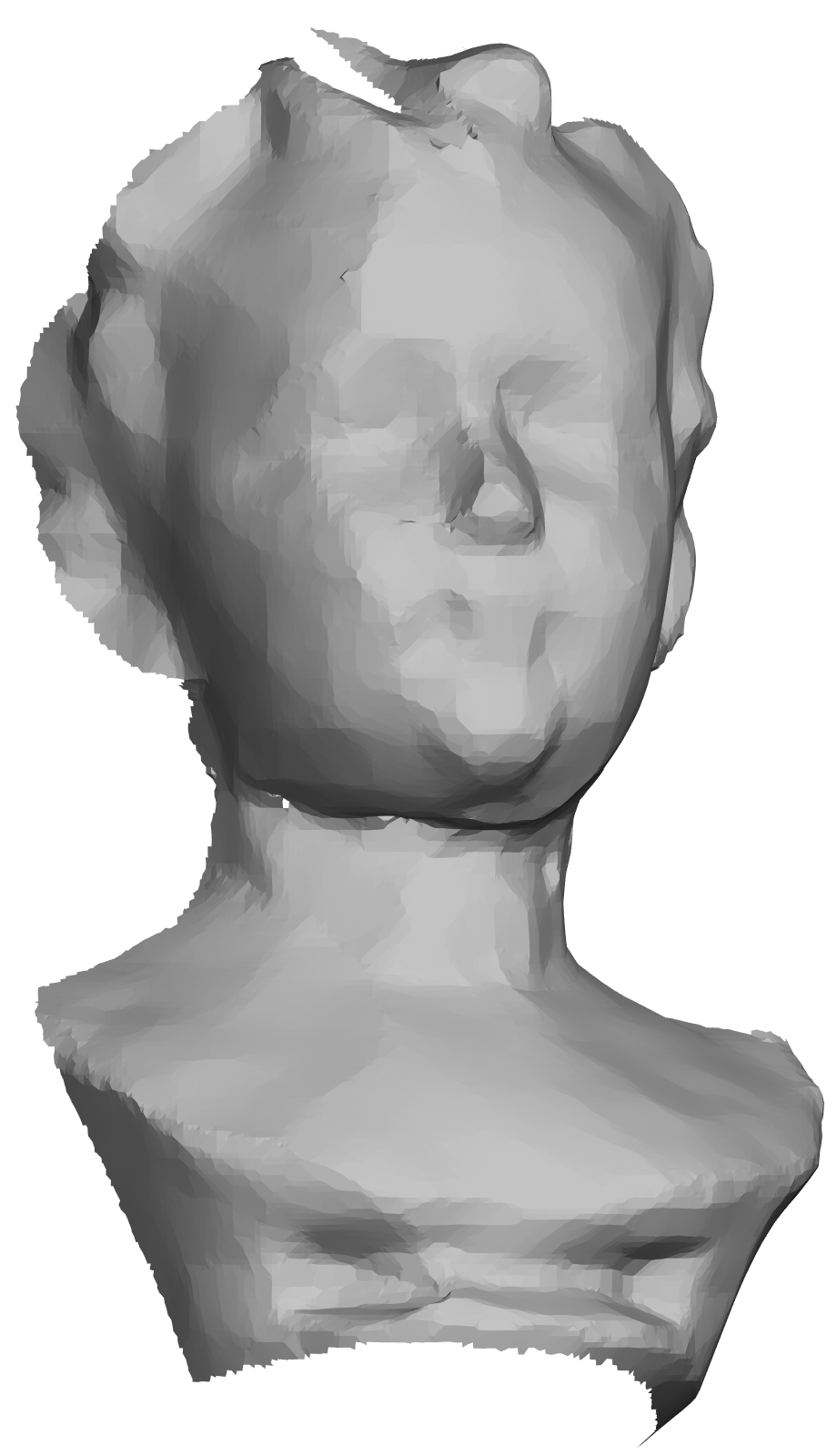}
    \caption{L0}
    \label{fig_kinect-d}
\end{subfigure}
\begin{subfigure}{0.12\linewidth}
    \includegraphics[width=\textwidth]{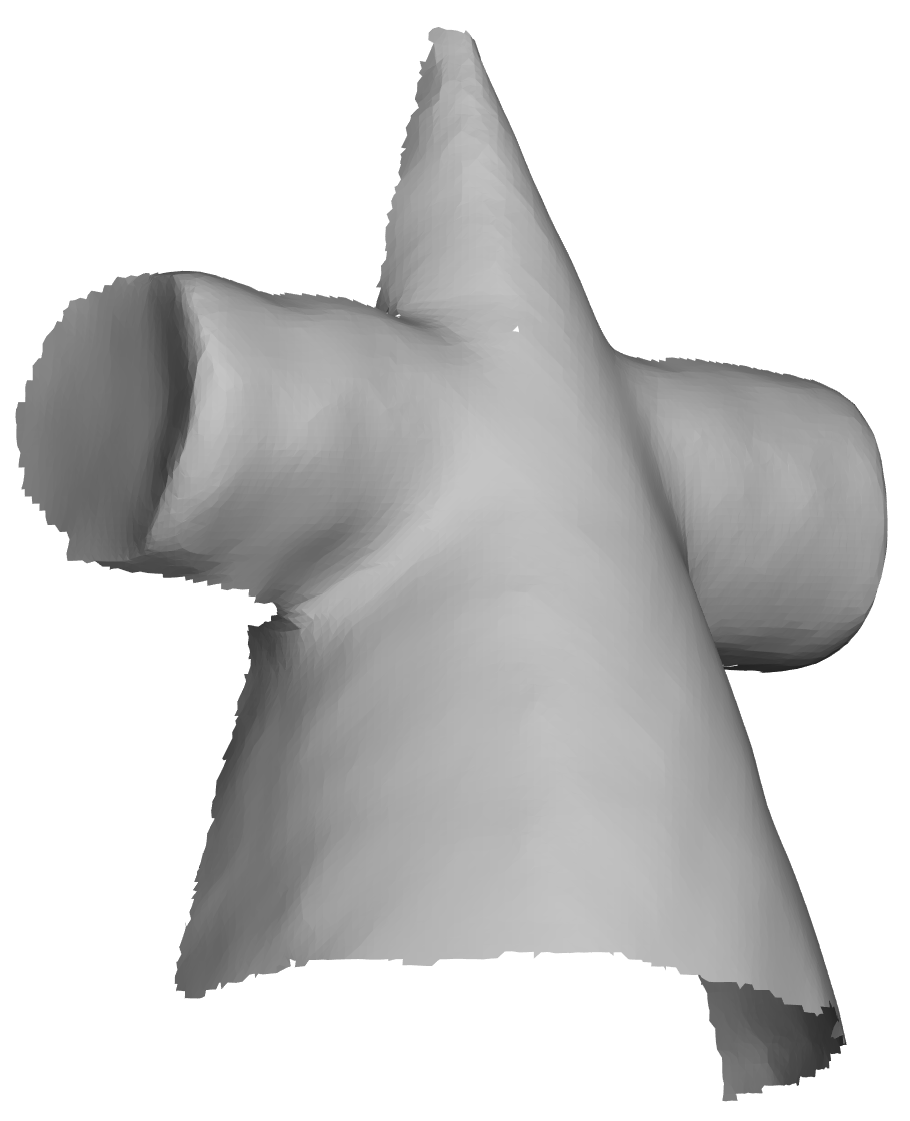}
    \includegraphics[width=\textwidth]{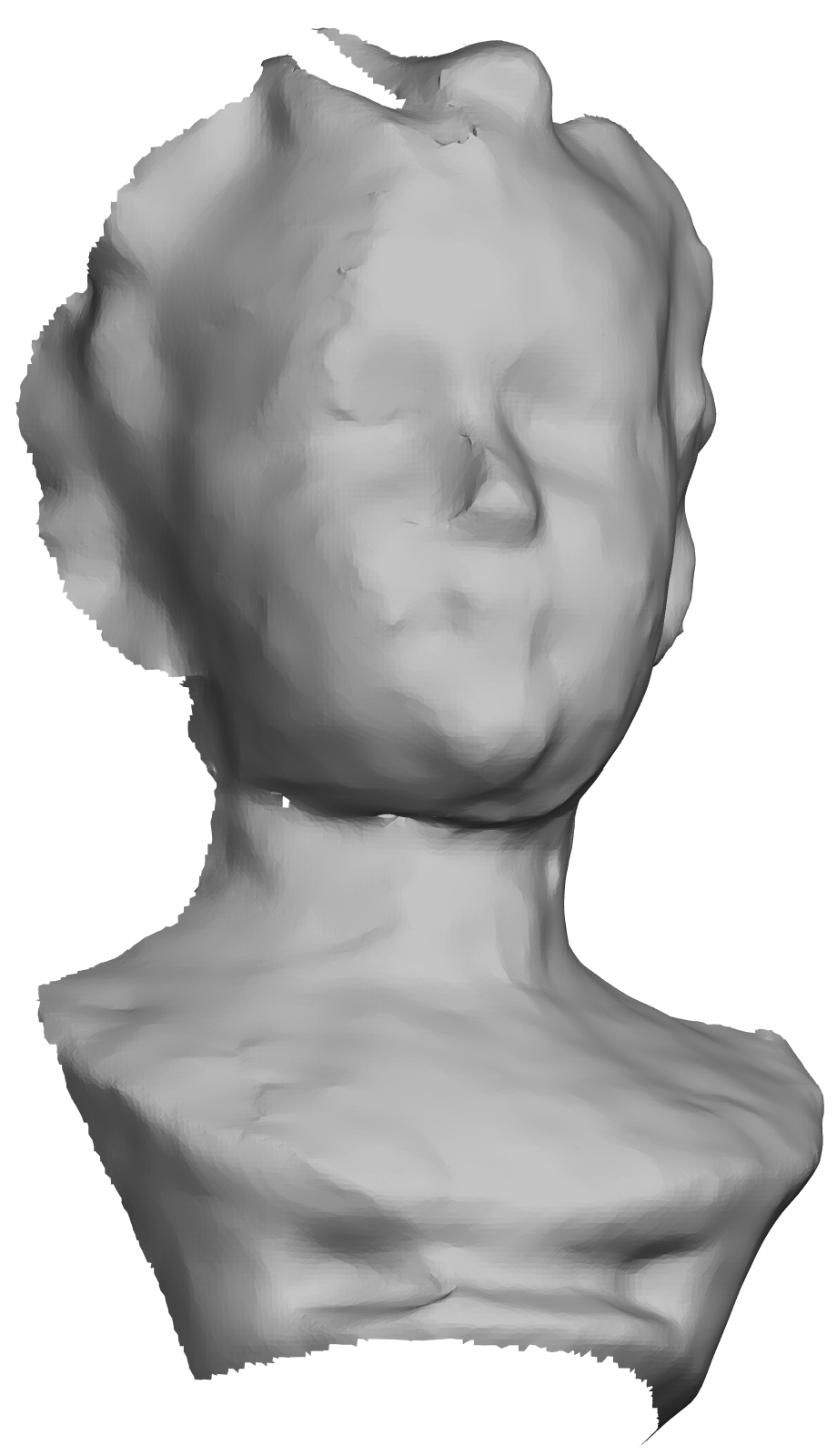}
    \caption{HO}
    \label{fig_kinect-e}
\end{subfigure}
    \begin{subfigure}{0.12\linewidth}
    \includegraphics[width=\textwidth]{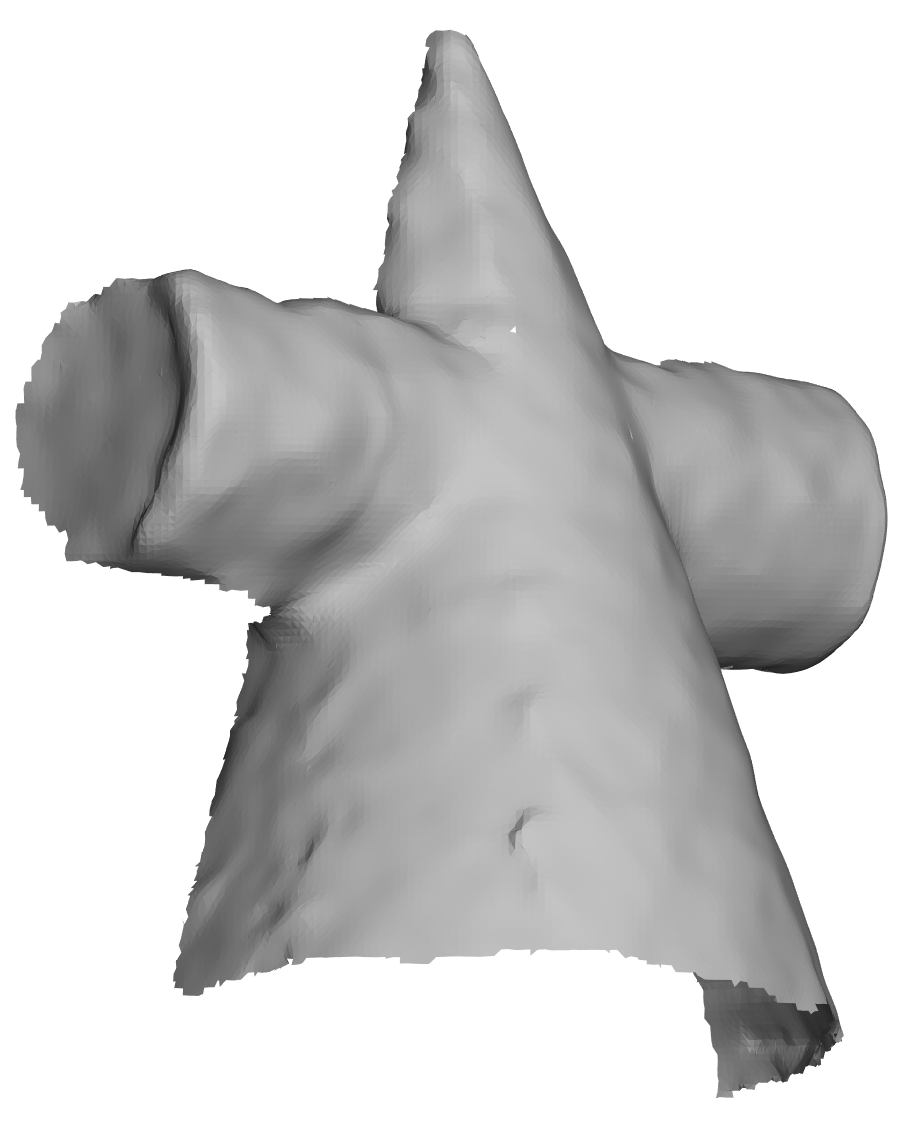}
    \includegraphics[width=\textwidth]{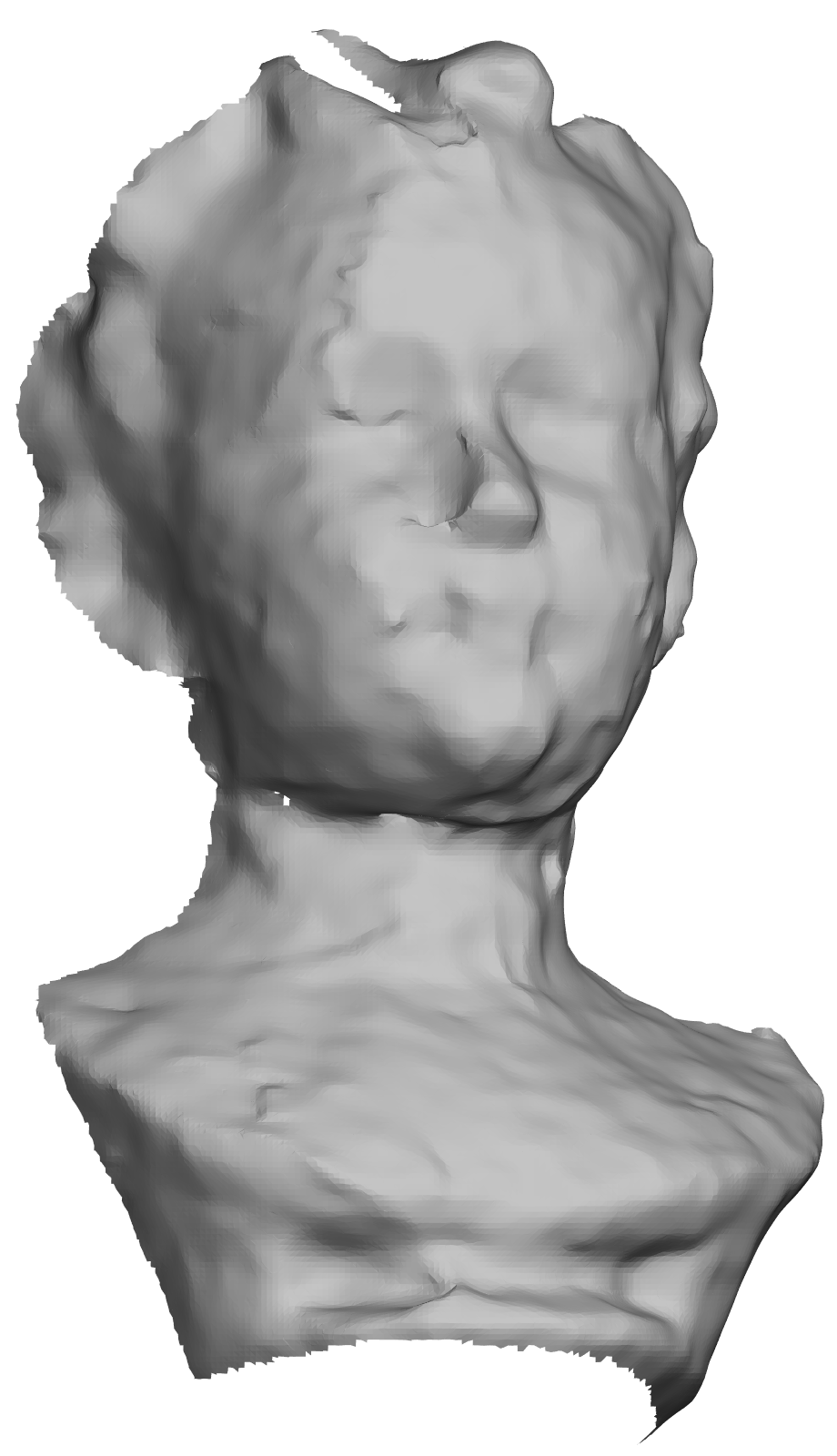}
    \caption{CNR}
    \label{fig_kinect-f}
\end{subfigure}
\begin{subfigure}{0.12\linewidth}
    \includegraphics[width=\textwidth]{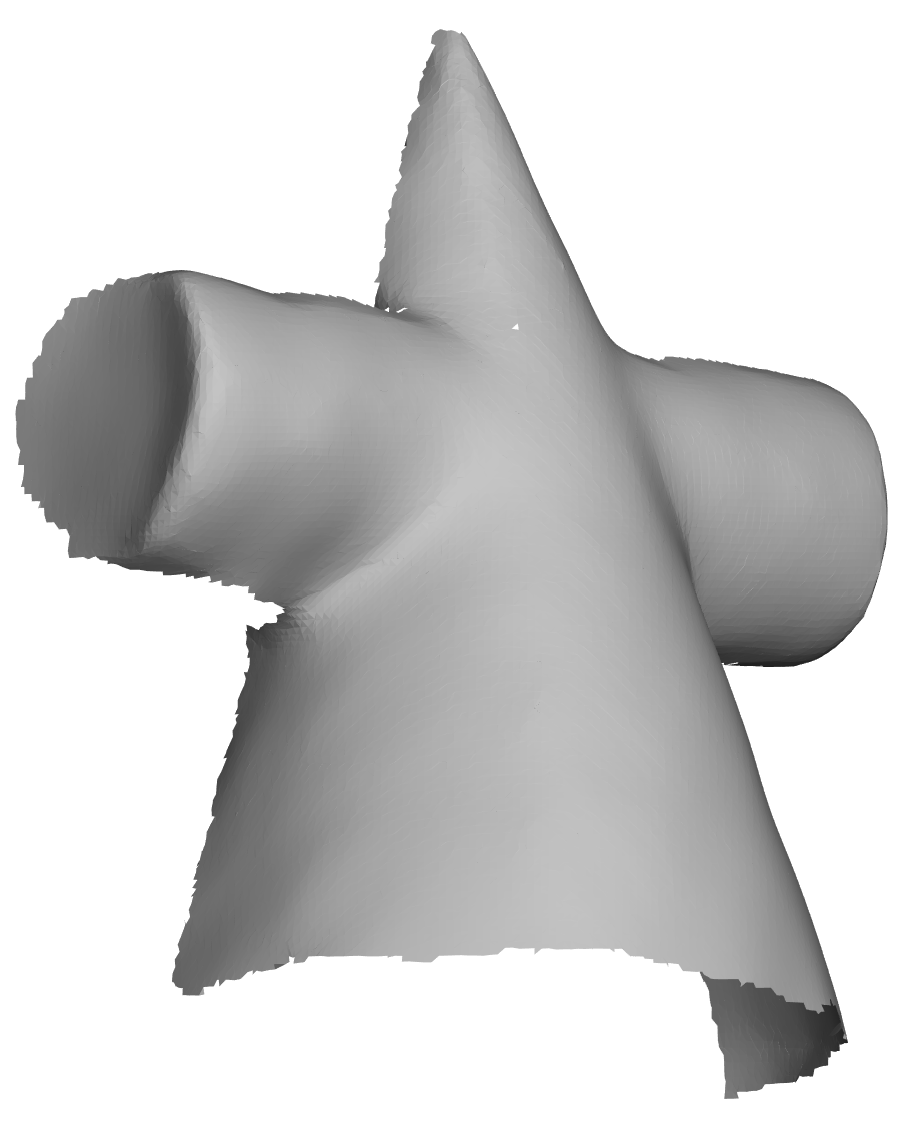}
    \includegraphics[width=\textwidth]{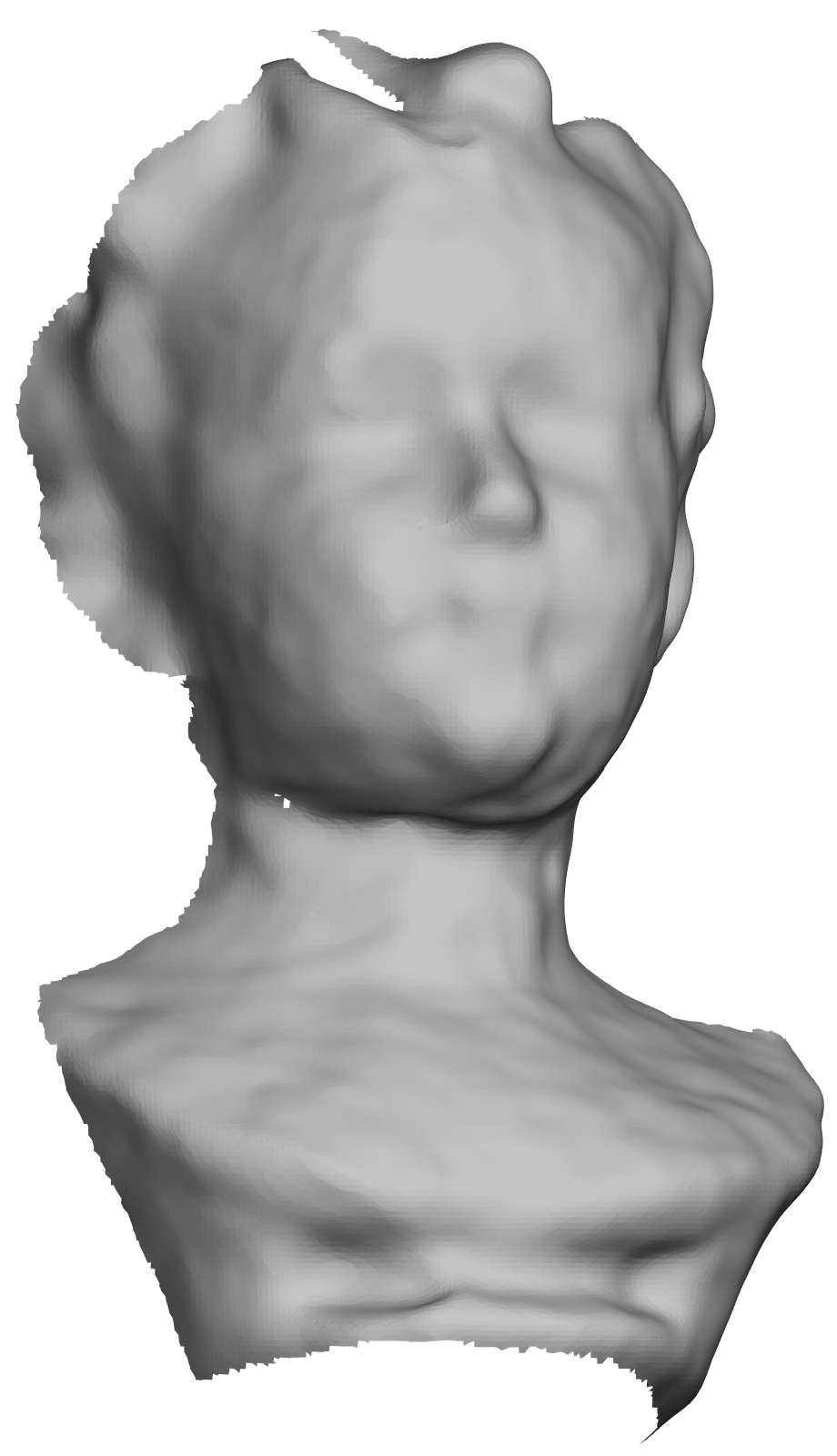}
    \caption{TGV}
    \label{fig_kinect-g}
\end{subfigure}
\begin{subfigure}{0.12\linewidth}
    \includegraphics[width=\textwidth]{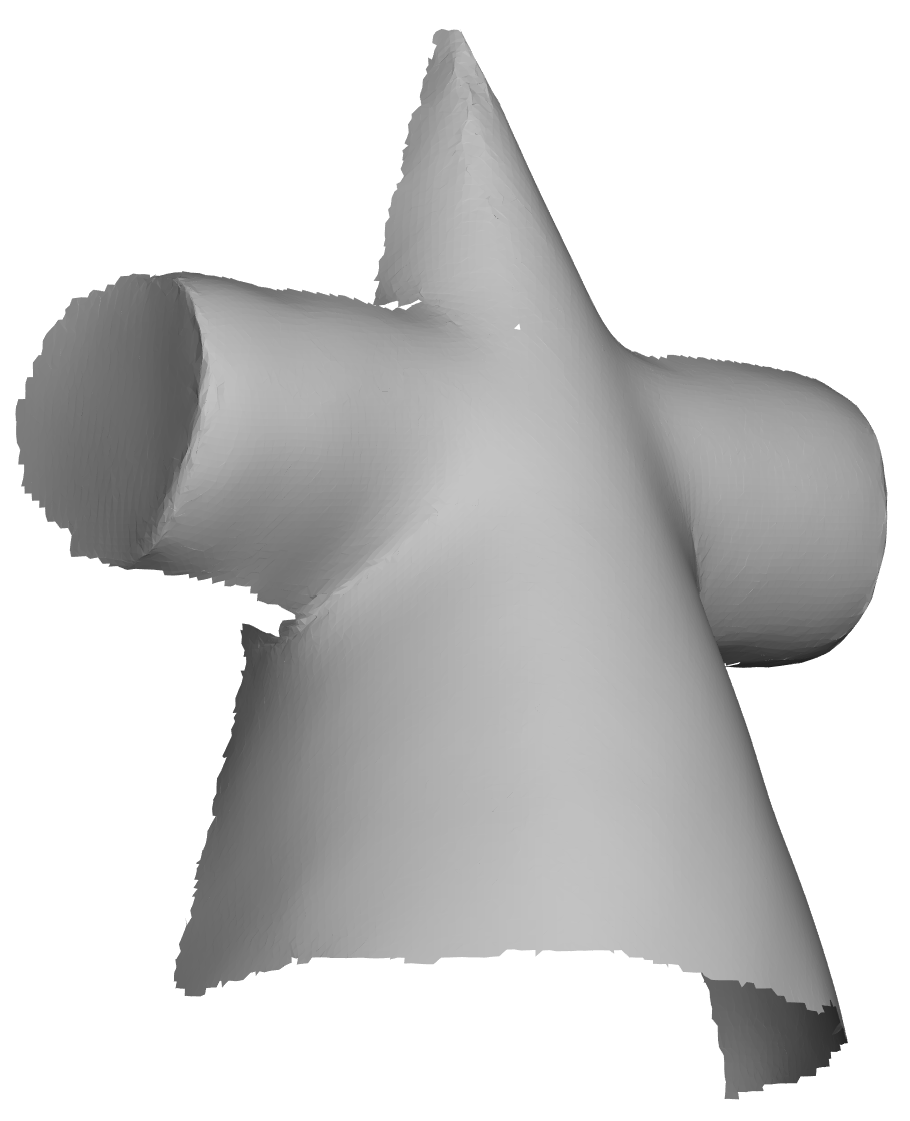}
    \includegraphics[width=\textwidth]{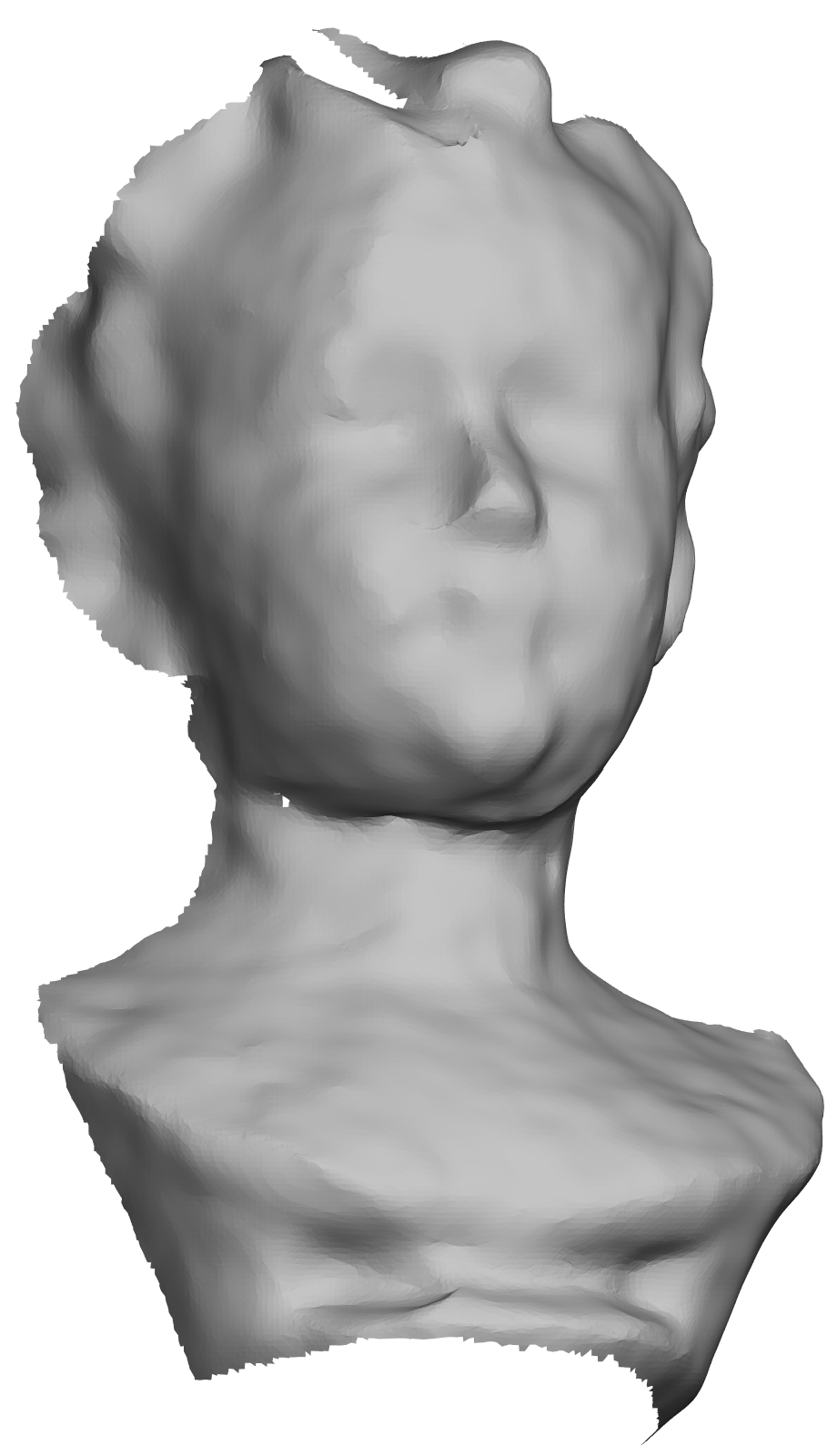}
    \caption{Ours}
    \label{fig_kinect-h}
\end{subfigure}
\caption{Visual comparison of different denoising methods for the scanned data acquired by a laser scanner.}
\label{fig_kinect}
\end{figure*}

A similar trend is observed for data captured with a Kinect sensor, as illustrated in~\cref{fig_kinect}. In this case, the noise level is relatively high (though unknown), and the BF method~\cite{zheng2010bilateral} together with other first-order methods exhibit strong over-smoothing effects due to the need for large regularization parameters to suppress the noise. By contrast, the learning-based CNR method and the higher-order approaches demonstrate greater adaptability to strong noise, yielding more reliable reconstructions.


In summary, the proposed semi-sparsity model consistently yields visually competitive and quantitatively more accurate denoising results across all experiments, including CAD, non-CAD, and real scanned surfaces. Its advantage lies in maintaining a delicate balance between sharp feature preservation and smooth surface recovery, even under strong noise. These results underscore the robustness and versatility of the proposed approach across diverse surface types and noise conditions.

\subsection{Quantitative Evaluation}

To assess the quality of denoising methods, we further adopt two quantitative metrics: the mean angular difference and vertex-based surface-to-surface error that have been widely used in recent studies~\cite{wang2016mesh, li2018non, liu2019triangulated, liu2021mesh}. The first metric computes the average angular deviation between the filtered face normals $\mathbf{N}$ and the ground truth $\mathbf{N}_{gt}$, i.e., 
$$
\theta = \operatorname{mean} \left(\angle(\mathbf{N},\mathbf{N}_{gt})\right),
$$
where $\angle(\mathbf{N},\mathbf{N}_{gt})$ is the angle between the two normals. For fairness, $\theta$ is evaluated after the normal filtering step for each tested method except for the $L_0$ minimization~\cite{he2013mesh}\footnote{This method directly optimizes the vertex positions for mesh denoising, and the face normals are then computed based on updated vertices.}. The second metric is defined as, 
$$
E_{v}
=\sqrt{\frac{1}{3\sum_{\tau} s_{\tau}}
\sum_{i} \left(\sum_{M_{1}(v_i)} s_{\tau}\right)
\operatorname{dist}(v_i, M)^{2}},
$$
where $\operatorname{dist}(v_i, M)$ is the distance between the updated vertex $v_i$ and a triangle of the ground truth that is closest to $v_i$. This metric captures the maximal geometric deviation and offers a direct measure of shape fidelity.

The statistical results are listed in~\cref{tab_tab1}. As indicated, the proposed semi-sparse model produces highly competitive results. In particular, it achieves significantly smaller $\theta$ values in cases of Block, Pyramid, and Cone examples, in which polynomial smoothing surfaces are dominant features and thus admit a strong semi-sparse prior of the face normals. This further demonstrates the validation of using such semi-sparse priors in mesh denoising. In the cases of richer geometric structures, the performance gap becomes smaller yet remains superior to that of the state-of-the-art higher-order TGV method~\cite{liu2021mesh}. A similar trend is evident in the vertex deviation metric $E_{v}$, which consistently favors the proposed model over competing approaches. These quantitative findings align well with the visual comparisons discussed in the previous section.

\begin{table}[!t]
	\footnotesize
	\caption{Quantitative evaluation of different mesh denoising results. The best two results are highlighted with \textbf{bold} and \underline{underline} values.}
    \label{tab_tab1}
	\begin{center}
		\setlength{\tabcolsep}{0.1mm}{
			\begin{tabular}{|c|c|c|c|c|c|c|c|c|}
                \hline
				Index & Mesh & TV~\cite{zhang2015variational} & HO~\cite{liu2019novel} & L0~\cite{he2013mesh} & BF~\cite{zheng2010bilateral} & CNR~\cite{wang2016mesh} & TGV~\cite{liu2021mesh} & Ours \\
				\hline
                \multirow{7}{*}{\rotatebox[origin=c]{90}{\parbox[c]{2cm}{\centering Mean angular difference ($\theta$)}}}
				&Block  & 3.12 & 2.90 & 4.35 & 5.30 & 2.63 & \underline{1.88} & \textbf{0.74}      \\
				&Fandisk  & 2.62 & 3.67 & 3.92 & 5.51 & 2.31 & \underline{2.20} & \textbf{2.15}      \\
				&Lucy  & 9.88 & 9.63 & 13.0 & 8.86 & 7.91 & \underline{7.34} & \textbf{7.30}      \\
				&Gargoyle  & 10.8 & 9.72 & 12.0 & 9.94 & 8.96 & \underline{7.86} & \textbf{7.79}      \\
				&Pyramid  & 6.79 & 7.18 & 6.50 & 8.45 & 6.40 & \underline{6.19} & \textbf{5.84}      \\
				&Cone  & 7.45 & 7.41 & 7.80 & 8.16 & 7.11 & \underline{6.96} & \textbf{6.61}      \\
				&Boy  & 9.16 & 8.98 & 9.42 & 10.0 & 8.97 & \underline{8.91} & \textbf{8.82}      \\
				\hline
                \hline
                \multirow{7}{*}{\rotatebox[origin=c]{90}{\parbox[c]{2cm}{\centering Vertex-based error ($E_{v}$)}}}
				&Block  & 2.01 & 1.70 & 2.15 & 1.80 & \underline{0.86} & 0.95 & \textbf{0.71}      \\
				&Fandisk  & 2.39 & 1.67 & 2.17 & 1.62 & 1.47 & \underline{1.20} & \textbf{1.19}      \\
				&Lucy  & 0.38 & 0.61 & 0.72 & 0.50 & 0.30 & \underline{0.26} & \textbf{0.25}      \\
				&Gargoyle  & 0.59 & 0.67 & 0.63 & 2.22 & 1.58 & \underline{0.54} & \textbf{0.50}      \\
				&Pyramid  & 4.69 & 4.23 & 3.47 & 4.45 & \underline{3.38} & 4.52 & \textbf{3.08}      \\
				&Cone  & 3.98 & 3.33 & 3.45 & \underline{2.69} & 2.78 & 3.57 & \textbf{2.14}      \\
				&Boy  & 6.08 & \underline{5.66} & 6.04 & 6.19 & 6.07 & 5.94 & \textbf{5.54}      \\
				\hline 
		\end{tabular}}
	\end{center}
\end{table}

\begin{table*}[!t]
	\small
	\caption{The running time (seconds) for meshes with different vertices $|V|$ and faces $|F|$. The best two results are highlighted with \textbf{bold} and \underline{underline} values.}
    \label{tab_tab2}
	\begin{center}
		\setlength{\tabcolsep}{1.2mm}{
			\begin{tabular}{|c|c|c|c|c|c|c|c|c|}
                \hline
				Mesh & $|V|$/$|F|$ (K) & TV~\cite{zhang2015variational} & HO~\cite{liu2019novel} & L0~\cite{he2013mesh} & BF~\cite{zheng2010bilateral} & CNR~\cite{wang2016mesh} & TGV~\cite{liu2021mesh} & Ours \\
				\hline
				Block  &8.8/17.6 & 0.928 & 1.235 & 8.470 & \underline{0.725} & \textbf{0.353} & 3.916 & 3.091      \\
				Fandisk & 6.2/12.9  & 0.542 & 1.191 & 4.224 & \underline{0.431} & \textbf{0.388} & 1.956 & 1.519      \\
				Lucy & 143.3/298.5  & 16.614 & 28.060 & 48.557 & \textbf{4.531} & \underline{5.701} & 36.222 & 30.106      \\
				Gargoyle  & 85.6/171.1 & \underline{8.652} & 12.931 & 35.179 & \textbf{2.958} & 19.319 & 31.600 & 20.174      \\
				Pyramid & 6.6/12.6 & 0.832 & 1.245 & 5.198 & \underline{0.468} & \textbf{0.456} & 2.584 & 2.051      \\
				Cone  & 31.2/61.3 & 2.897 & 5.009 & 30.200 & \underline{2.235} & \textbf{1.186} & 19.121 & 14.482      \\
				Boy  & 76.9/152.2 & \underline{6.434} & 10.980 & 56.115 & 7.809 & \textbf{3.169} & 35.388 & 25.901      \\
				\hline
		\end{tabular}}
	\end{center}
\end{table*}

In addition, we present the execution time of each method to evaluate computational efficiency. In general, the computational cost is not only determined by the compared algorithms but also closely related to the scale of problems (the tested meshes), for which the sizes of vertices $(|V|)$ and faces $(|F|)$ are listed in~\cref{tab_tab2}. As we can see, the learning-based CNR approach~\cite{wang2016mesh} is the fastest, benefiting from the small size of the neural network. The BF method~\cite{zheng2010bilateral} is the most efficient among the remaining classical methods. The HO method~\cite{liu2019novel} is slightly slower than TV regularization~\cite{zhang2015variational} due to the additional computation of higher-order differential operators. In contrast, $L_0$ minimization~\cite{he2013mesh}, despite its first-order nature, relies on edge-based differential operators, making it comparatively more computationally expensive. The proposed semi-sparsity scheme is a little bit slower than the HO method~\cite{liu2019novel}, since it also involves the first-order difference, while our semi-sparsity scheme is somewhat faster than the TGV-based regularization~\cite{liu2021mesh}, as it requires the computation of more complex symmetrical discrete higher-order differential operators.

Overall, the quantitative comparisons further indicate that our semi-sparse regularization method is highly effective in recovering fine geometric details—including sharp edges, multi-scale features, and smooth surface regions from noisy input. In most cases, it achieves the lowest reconstruction error among all tested methods. Importantly, the proposed approach demonstrates consistent performance across all categories of surfaces—CAD, non-CAD, and scanned meshes, highlighting its robustness and general applicability rather than being tailored to a specific class of mesh data.

\section{Conclusion and Future Work}
\label{conclusion}

In this work, we have presented a semi-sparsity regularization framework for triangular mesh denoising. Our approach follows a two-step strategy: first restoring triangle face normals, and then updating vertex positions accordingly. In the normal filtering stage, the semi-sparsity model enables effective smoothing of the normal field while preserving critical geometric features. The associated optimization problem is efficiently solved using a multi-block ADMM algorithm. In the subsequent vertex updating step, mesh geometry is refined to align with the denoised normal field. Extensive evaluations across different surface types demonstrate that the proposed method provides significant advantages in preserving sharp features, recovering smooth transition regions, and avoiding common artifacts such as staircase effects, over-smoothing, over-sharpening and noise amplification. Both visual comparisons and quantitative analyses confirm that our method outperforms several classical approaches and achieves results on par with state-of-the-art techniques. Owing to its robustness, it is well-suited for pre- and post-processing of CAD and man-made meshes containing both sharp and smooth structures. There are several promising directions for future work. The proposed discretized semi-sparsity operator may be applied to other geometry processing tasks, such as mesh segmentation, surface reconstruction, mesh simplification, and detection. 

\appendices
\ifCLASSOPTIONcompsoc
  \section*{Acknowledgments}
  We acknowledge the various online resources, including images, codes, and software, that have contributed to our research.
\else
  \section*{Acknowledgment}
\fi

\bibliography{egbib}
\bibliographystyle{abbrv}

%

\vspace{-8mm}
\begin{IEEEbiography}[{\includegraphics[width=1in,height=1.25in,clip,keepaspectratio]{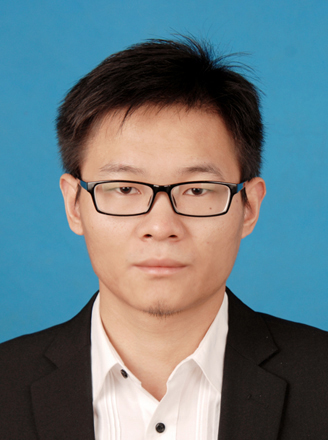}}]{Junqing Huang}
	received the BS degree in Automation from the School of Electrical Engineering, Zhengzhou University, Zhengzhou, China, in 2011, and the MS degree in Mathematics from the School of Mathematical Sciences, Beihang University (BUAA), Beijing, China, in 2015. He is currently a Ph.D. candidate of Department of Mathematics: Analysis, Logic and Discrete Mathematics, Ghent University, Belgium. His research interests include deep learning, image processing, optimal transport and optimization.
\end{IEEEbiography}

\vspace{-8mm}
\begin{IEEEbiography}[{\includegraphics[width=1in,height=1.25in,clip,keepaspectratio]{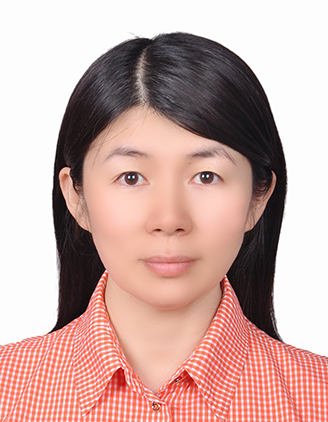}}]{Haihui Wang}	
	received the PhD degree in Mathematics from Peking University, Beijing, China, in 2003. She is currently a full Professor in the School of Mathematical Sciences, Beihang University (BUAA), Beijing, China, and a Professor in the Health Science Center, Beijing University, Beijing, China. Her research interests include artificial intelligence, machine learning, signal and image processing, wavelet analysis and application, and so on.
\end{IEEEbiography}
\vspace{-8mm}
\begin{IEEEbiography}[{\includegraphics[width=1in,height=1.25in,clip,keepaspectratio]{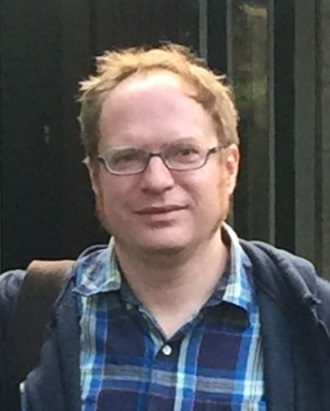}}]{Michael Ruzhansky}
	is a senior full Professor in Department of Mathematics, Ghent University, Belgium, a Professorship in School of Mathematical Sciences, Queen Mary University of London, and Honorary Professorship in Department of Mathematics, Imperial College London, UK.	He was awarded by FWO (Belgium) the prestigious Odysseus 1 Project in 2018, he was recipient of several Prizes and Awards: ISAAC Award in 2007,
	Daiwa Adrian Prize in 2010 and Ferran Sunyer I Balaguer Prizes in 2014 and 2018. His research interests include different areas of analysis, in particular, theory of PDEs, microlocal analysis, and harmonic analysis, and so on.
\end{IEEEbiography}

\end{document}